\documentclass[a4paper,11pt]{article}\usepackage[]{graphicx}\usepackage[]{color}
\makeatletter
\def\maxwidth{ %
  \ifdim\Gin@nat@width>\linewidth
    \linewidth
  \else
    \Gin@nat@width
  \fi
}
\makeatother

\definecolor{fgcolor}{rgb}{0.345, 0.345, 0.345}
\newcommand{\hlnum}[1]{\textcolor[rgb]{0.686,0.059,0.569}{#1}}%
\newcommand{\hlstr}[1]{\textcolor[rgb]{0.192,0.494,0.8}{#1}}%
\newcommand{\hlcom}[1]{\textcolor[rgb]{0.678,0.584,0.686}{\textit{#1}}}%
\newcommand{\hlopt}[1]{\textcolor[rgb]{0,0,0}{#1}}%
\newcommand{\hlstd}[1]{\textcolor[rgb]{0.345,0.345,0.345}{#1}}%
\newcommand{\hlkwb}[1]{\textcolor[rgb]{0.69,0.353,0.396}{#1}}%
\newcommand{\hlkwc}[1]{\textcolor[rgb]{0.333,0.667,0.333}{#1}}%
\newcommand{\hlkwd}[1]{\textcolor[rgb]{0.737,0.353,0.396}{\textbf{#1}}}%

\usepackage{framed}
\makeatletter
\newenvironment{kframe}{%
 \def\at@end@of@kframe{}%
 \ifinner\ifhmode%
  \def\at@end@of@kframe{\end{minipage}}%
  \begin{minipage}{\columnwidth}%
 \fi\fi%
 \def\FrameCommand##1{\hskip\@totalleftmargin \hskip-\fboxsep
 \colorbox{shadecolor}{##1}\hskip-\fboxsep
     \hskip-\linewidth \hskip-\@totalleftmargin \hskip\columnwidth}%
 \MakeFramed {\advance\hsize-\width
   \@totalleftmargin\z@ \linewidth\hsize
   \@setminipage}}%
 {\par\unskip\endMakeFramed%
 \at@end@of@kframe}
\makeatother

\definecolor{shadecolor}{rgb}{.97, .97, .97}
\definecolor{messagecolor}{rgb}{0, 0, 0}
\definecolor{warningcolor}{rgb}{1, 0, 1}
\definecolor{errorcolor}{rgb}{1, 0, 0}
\newenvironment{knitrout}{}{} 

\usepackage{alltt}
\usepackage[charter,expert]{mathdesign}
\usepackage[margin=1.4in]{geometry}
\usepackage{natbib}
\usepackage{linguex}
\usepackage{setspace}
\usepackage{caption}
\usepackage{microtype}
\usepackage[colorlinks=true,
            linkcolor=red,
            urlcolor=blue,
            citecolor=black]{hyperref}

\newcommand{\sscr}[1]{\textrm{\scriptsize #1}}

\SetTracking
  [ no ligatures  = {},
    spacing       = {600*,-100*,
 },
    outer spacing = {450,250,150},
    outer kerning = {*,*} ]
  { encoding = * }
  { 100 }
\IfFileExists{upquote.sty}{\usepackage{upquote}}{}
\begin{document}

\noindent{\textls{\MakeUppercase{Generalised Additive Mixed Models for dynamic analysis\\ in linguistics: A practical introduction}}\footnote{
This introduction has been updated a few times. Thanks to Martijn Wieling, Bodo Winter and Donald Derrick for helpful comments and suggestions. Details of the changes are shown on the associated GitHub page (see end of section \ref{sec:intro} for the link).
}\vspace{1ex}\\\textit{M{\'a}rton S{\'o}skuthy\\ University of York}}\vspace{1ex}\\
\noindent [last updated: \today]

\section{Introduction}\label{sec:intro}

\noindent This is a hands-on introduction to Generalised Additive Mixed Models (GAMMs; \citealt{wood06}) in the context of linguistics with a particular focus on dynamic speech analysis. Dynamic speech analysis is a term used to refer to analyses that look at measureable quantities of speech that vary in space and/or time. Temporal variation can be short-term (e.g.\ formant contours and pitch tracks) or long-term (e.g.\ diachronic change or change over the life-span). Similarly, spatial variation can sometimes be measured in millimetres (e.g.\ tongue contours), and sometimes in kilometres (e.g.\ the acoustic realisation of a sound category as a function of location on a dialect map). The focus of this introduction is mostly on short-term temporal variation in phonetics, and more specifically on formant trajectories (though one of the examples also involves diachronic change). This choice is mostly practical: I chose to illustrate GAMMs through formant trajectories simply because I'm comfortable talking about them. However, most of the concepts and techniques discussed here are more general and are also applicable to other types of short-term and long-term temporal trajectories.

The main goal of this introduction is to explain some of the main ideas underlying GAMMs, and to provide a practical guide to frequentist significance testing using these models. Some of the suggestions below are based on simulation-based work presented in \citet{soskuthy16labphon} and \citeauthor{soskuthy16gam} (in prep), but this introduction is meant as a standalone guide. 

The following discussion is divided into two parts, which can be read in slightly different ways. The first part (section \ref{sec:theory}) looks at what GAMMs actually are, how they work and why/when we should use them. Although the reader can replicate some of the example analyses in this section, this is not essential -- reading the section should be enough. The second part (section \ref{sec:tutorial}) is a tutorial introduction that illustrates the process of fitting and evaluating GAMMs in the R statistical software environment \citep{R}, and the reader is strongly encouraged to work through the examples on their own machine.

Since a lot of the research in GAMM theory is closely intertwined with software development in R  (e.g.\ the author of one of the main textbooks on GAMM, \citealt{wood06} is also the maintainer of one of the main GAMM software packages), it is difficult to talk about GAMMs without using some of the terminology and conventions of R. Therefore, although the discussion in the first part is mostly conceptual, I do rely on R code to illustrate the structure of different models. However, I've tried not to use too much code, and it should be possible to follow the discussion without a strong background in R. The second part relies much more heavily on R and will be mostly of interest to readers who want to fit their own models using R. Readers who want to learn more about R before reading this introduction may want to consult \citet{baayen08} and \citet{johnson08}, who both provide thorough introductions to R for beginners using examples from linguistics. 
GAMMs are a type of regression model and they are closely related to mixed effects regression. This tutorial assumes some background in regression modelling and it will help to be familiar with mixed effects models as well. \citet{winter13} is a short but excellent introduction to mixed effects modelling, while \citet{gelmanandhill07} and \citet{baayen08} provide more in-depth treatments of the topic.

The examples in this introduction rely on two R packages: {\tt mgcv} \citep{wood06} and {\tt itsadug} \citep{itsadug}. These should be installed and loaded before trying to run the analyses that follow. I've also put together a script with a few `GAMM hacks' ({\tt gamm\_hacks.r}), which help to keep the code in the tutorial tidier and may also be useful for the reader's own analyses. This should be sourced after loading {\tt itsadug}, as it overrides some of the functions in that package. Finally, the data sets for the tutorial are in two separate files called {\tt words\_50.csv} and {\tt glasgow\_r.csv}, which should be imported into R as {\tt words.50} and {\tt gl.r}.

The data sets, the GAMM hacks script, the slides for \citet{soskuthy16labphon} and the source code for the PDF are all available from my GitHub page: \\ \url{https://github.com/soskuthy/gamm_intro}.

I relied on a number of sources for this introduction, but the main body of the text is light on references to make the discussion easier to follow. There is a more detailed list in the final section that also includes links and brief summaries of each source.

\section{A gentle introduction to GAMM theory} \label{sec:theory}

\subsection{GAMs}\label{sec:gams}

Before discussing GAMMs, let's start with a slightly simpler type of model called Generalised Additive Models (that's GAMM without the `mixed' part). The easiest way to understand GAMs is through a comparison with linear regression models. We will work through a specific example, where our goal will be simply to fit a regression line/curve to an F2 trajectory. The formant measurements are in Hz and the trajectory is represented by 11 measurement points taken at equal intervals (going from the very beginning to the very end). The figure below shows the trajectory (available as {\tt traj.csv} from the GitHub page):
\begin{knitrout}
\definecolor{shadecolor}{rgb}{0.969, 0.969, 0.969}\color{fgcolor}

{\centering \includegraphics[width=0.5\textwidth]{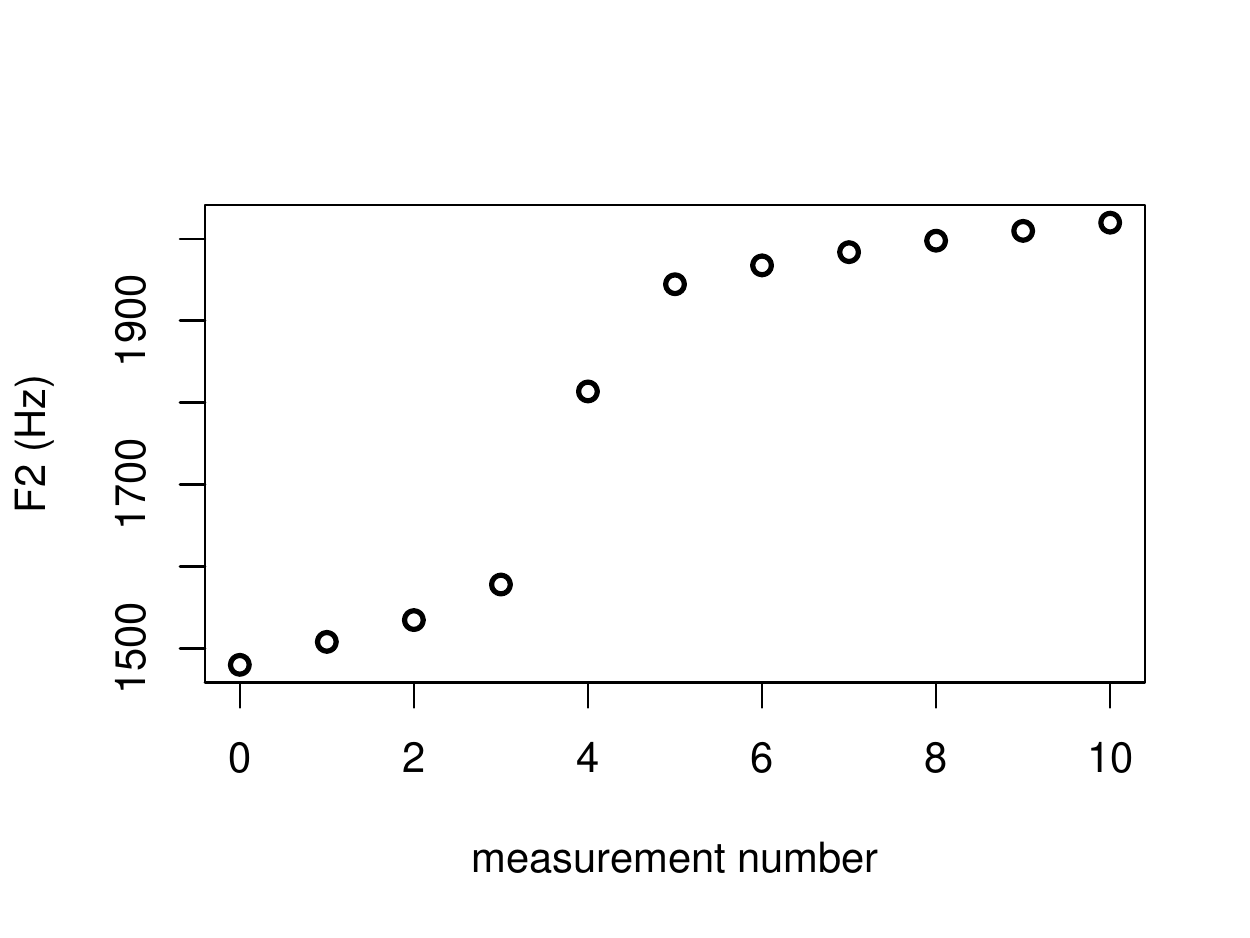} 

}

\end{knitrout}

\noindent Here is a linear model that fits a line to the trajectory in R:

\begin{knitrout}\small
\definecolor{shadecolor}{rgb}{0.969, 0.969, 0.969}\color{fgcolor}\begin{kframe}
\begin{alltt}
\hlstd{traj} \hlkwb{<-} \hlkwd{read.csv}\hlstd{(}\hlstr{"traj.csv"}\hlstd{)}  \hlcom{# importing the data}
\hlstd{demo.lm} \hlkwb{<-} \hlkwd{lm}\hlstd{(f2} \hlopt{~} \hlstd{measurement.no,} \hlkwc{data} \hlstd{= traj)}
\end{alltt}
\end{kframe}
\end{knitrout}

\noindent Though I'll try to use mathematical notation sparingly in this tutorial, it is useful to write out the model as a formula, as it will help with the transition to GAMs:%
\footnote{I've left out the so-called error term to keep things simple, though this should technically be part of the model specification.}
\begin{equation}
Y_{\sscr{f2}} = \alpha + \beta_{1} X_{\sscr{measurement.no}}
\end{equation}
This is simply an equation for a straight line: $X_{\sscr{measurement.no}}$ stands for the values along the \textit{x}-axis (measurement number), while $Y_{\sscr{f2}}$ stands for corresponding values along the \textit{y}-axis (F2 in Hz). $\alpha$ and $\beta_{1}$ specify the \textit{intercept} (i.e.\ the height) and the \textit{slope} of the regression line that represents the relationship between F2 and {\tt measurement.no}. When we fit a regression line to the actual trajectory, the result looks like this:
\begin{knitrout}
\definecolor{shadecolor}{rgb}{0.969, 0.969, 0.969}\color{fgcolor}

{\centering \includegraphics[width=0.5\textwidth]{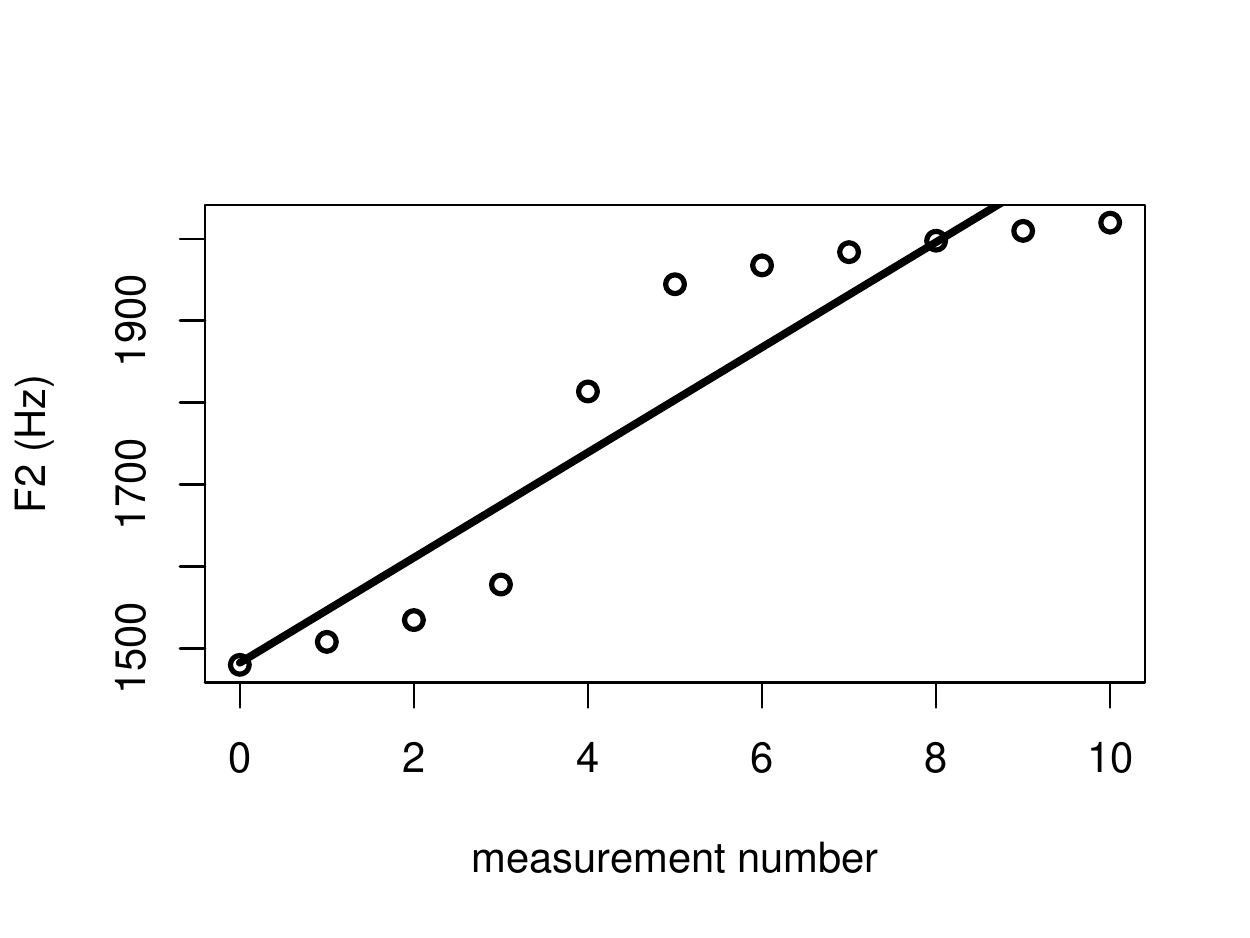} 

}

\end{knitrout}

\noindent The regression line is unable to capture the non-linear nature of the trajectory. This is bad news for models that seek to make causal inferences about such trajectories. Such discrepancies between the data and the model create systematic patterns in the \textit{residuals} of the model (i.e.\ the difference between the predicted vs.\ the actual value of the outcome variable; in this case, the distances between the prediction line and the individual data points along the \textit{y}-axis), which makes confidence intervals and \textit{p}-values unreliable. This is illustrated by the following two figures. The figure on the left shows how the residuals are calculated by explicitly indicating the distance between the predicted and the observed values of the outcome variable. The figure on the right plots the raw residuals (the lengths of the red dashed lines on the left) against measurement number:

\begin{knitrout}
\definecolor{shadecolor}{rgb}{0.969, 0.969, 0.969}\color{fgcolor}

{\centering \includegraphics[width=0.495\textwidth]{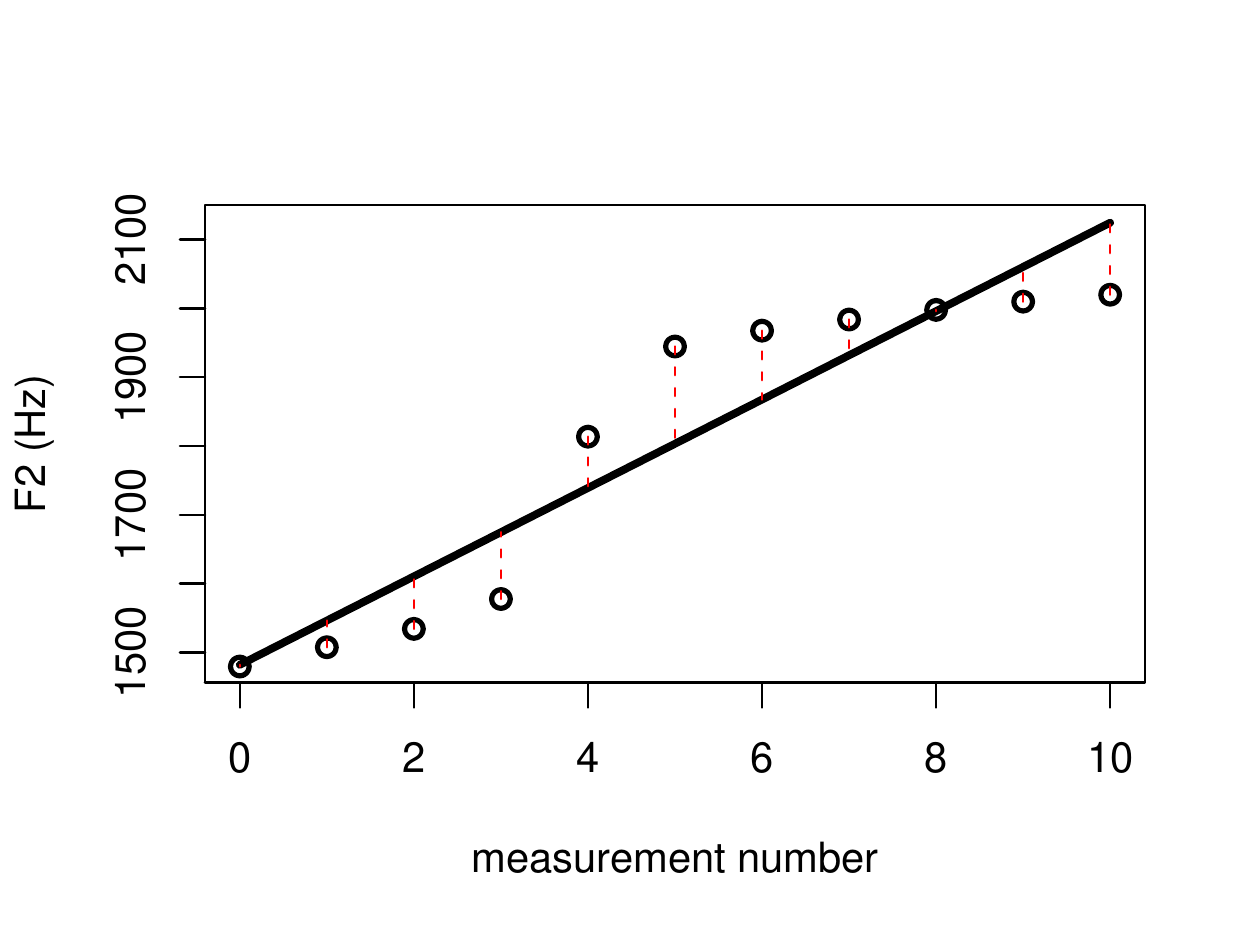} 
\includegraphics[width=0.495\textwidth]{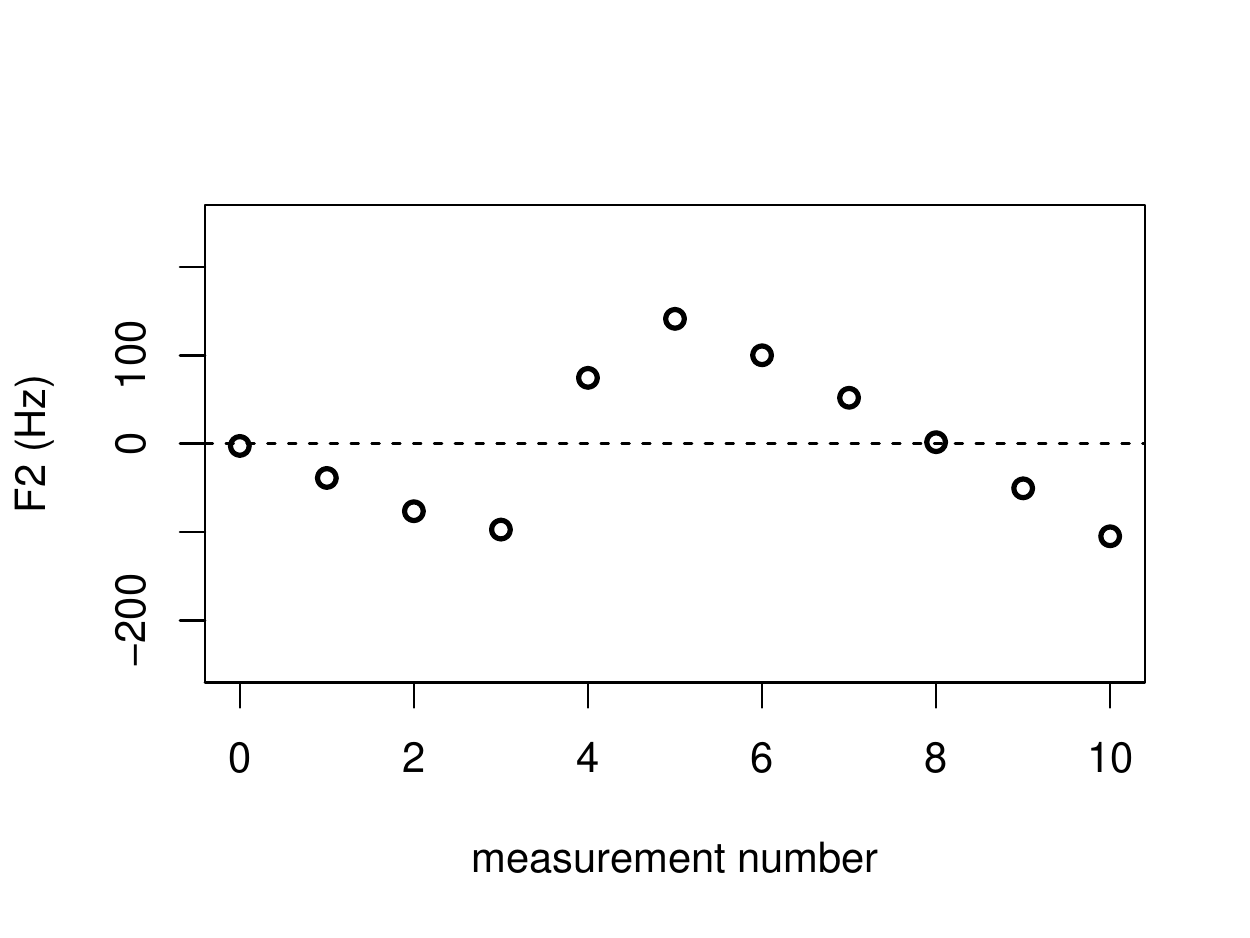} 

}

\end{knitrout}

\noindent The residuals shouldn't vary systematically as a function of other predictors, yet there is a clear pattern in the figure above, which -- in this case -- results from the inability of the model to capture non-linearities. If we want to account for non-linear relationships, we'll need to take a different approach.

GAMs provide one way of getting around this problem. Let us refer to the slopes and intercepts of linear regression models as \textit{parametric terms}. GAMs differ from traditional linear regression models by allowing so-called \textit{smooth terms} alongside parametric terms. Smooth terms are extremely flexible, so much so that their mathematical representation in model specifications is simply `some function of $x$':
\begin{equation}
Y_{\sscr{f2}} = \alpha + f_1 (X_{\sscr{measurement.no}})
\end{equation}
This model specification does not say anything about the shape of the function linking $X_{\sscr{measurement.no}}$ and $Y_{\sscr{f2}}$. The only requirement on the smooth term $f_1 (X_{\sscr{measurement.no}})$ is that it should be a smooth function of one or more predictor variables. Since the shape of this smooth function is not constrained in the same way as it is for regression lines, GAMs can deal with non-linearity. The following R function can be used to fit a GAM with the structure above to the F2 trajectory:\footnote{Throughout this tutorial, we'll use the {\tt bam()} function from the {\tt mgcv} package to fit GAMs and GAMMs. An alternative is the {\tt gam()} function from the same package, but {\tt gam()} is less flexible and often slower than {\tt bam()}, so we won't use it here.}

\begin{knitrout}\small
\definecolor{shadecolor}{rgb}{0.969, 0.969, 0.969}\color{fgcolor}\begin{kframe}
\begin{alltt}
\hlstd{demo.gam} \hlkwb{<-} \hlkwd{bam}\hlstd{(f2} \hlopt{~} \hlkwd{s}\hlstd{(measurement.no,} \hlkwc{bs} \hlstd{=} \hlstr{"cr"}\hlstd{),} \hlkwc{data} \hlstd{= traj)}
\end{alltt}
\end{kframe}
\end{knitrout}

\noindent The \verb+s()+ notation is used to distinguish smooth terms from parametric ones. The \verb+bs="cr"+ parameter tells R to use a so called `cubic regression spline' as the smooth term (\verb+bs+ actually stands for `basis', a concept that will be discussed in more detail below). The model fit is shown below.
\begin{knitrout}
\definecolor{shadecolor}{rgb}{0.969, 0.969, 0.969}\color{fgcolor}

{\centering \includegraphics[width=0.5\textwidth]{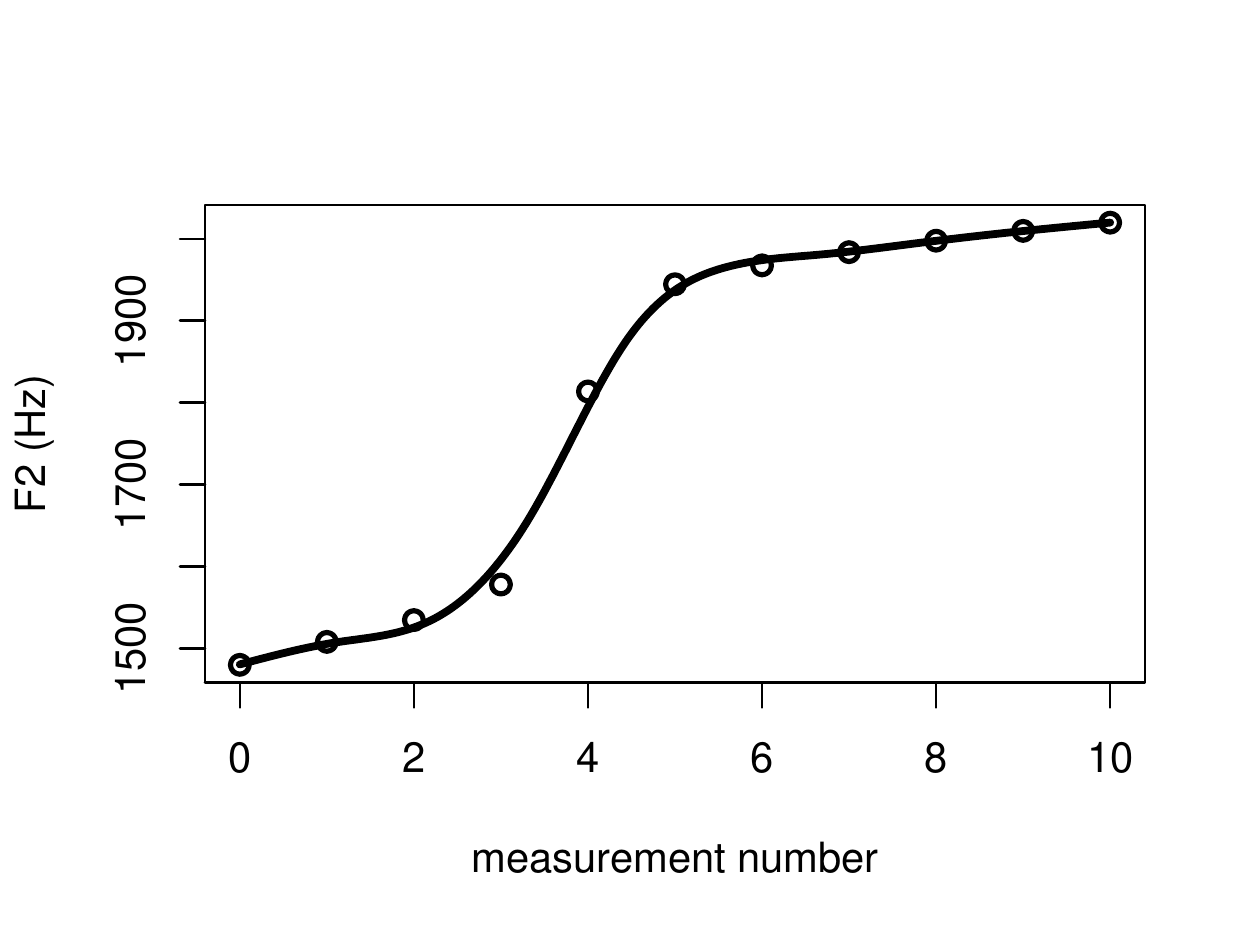} 

}

\end{knitrout}

\noindent So how are such wiggly smooth terms created in practice? We are not going to discuss the mathematical underpinnings of GAMs, but there are two fundamental and relatively straightforward concepts that need to be understood if one wants to work with these models: \textit{basis functions} and the \textit{smoothing parameter}. Basis functions are simple functions that add up to a (potentially) more complex curve. For instance, consider the following model:

\begin{knitrout}\small
\definecolor{shadecolor}{rgb}{0.969, 0.969, 0.969}\color{fgcolor}\begin{kframe}
\begin{alltt}
\hlstd{demo.poly} \hlkwb{<-} \hlkwd{lm}\hlstd{(f2} \hlopt{~} \hlstd{measurement.no} \hlopt{+} \hlkwd{I}\hlstd{(measurement.no}\hlopt{^}\hlnum{2}\hlstd{)} \hlopt{+}
                     \hlkwd{I}\hlstd{(measurement.no}\hlopt{^}\hlnum{3}\hlstd{),} \hlkwc{data}\hlstd{=traj)}
\end{alltt}
\end{kframe}
\end{knitrout}

\noindent Or, in mathematical notation:
\begin{equation}
Y_{\sscr{f2}} = \alpha + \beta_1 X_{\sscr{measurement.no}} + \beta_2 X_{\sscr{measurement.no}}^2 + \beta_3 X_{\sscr{measurement.no}}^3
\end{equation}
And the model fit:
\begin{knitrout}
\definecolor{shadecolor}{rgb}{0.969, 0.969, 0.969}\color{fgcolor}

{\centering \includegraphics[width=0.5\textwidth]{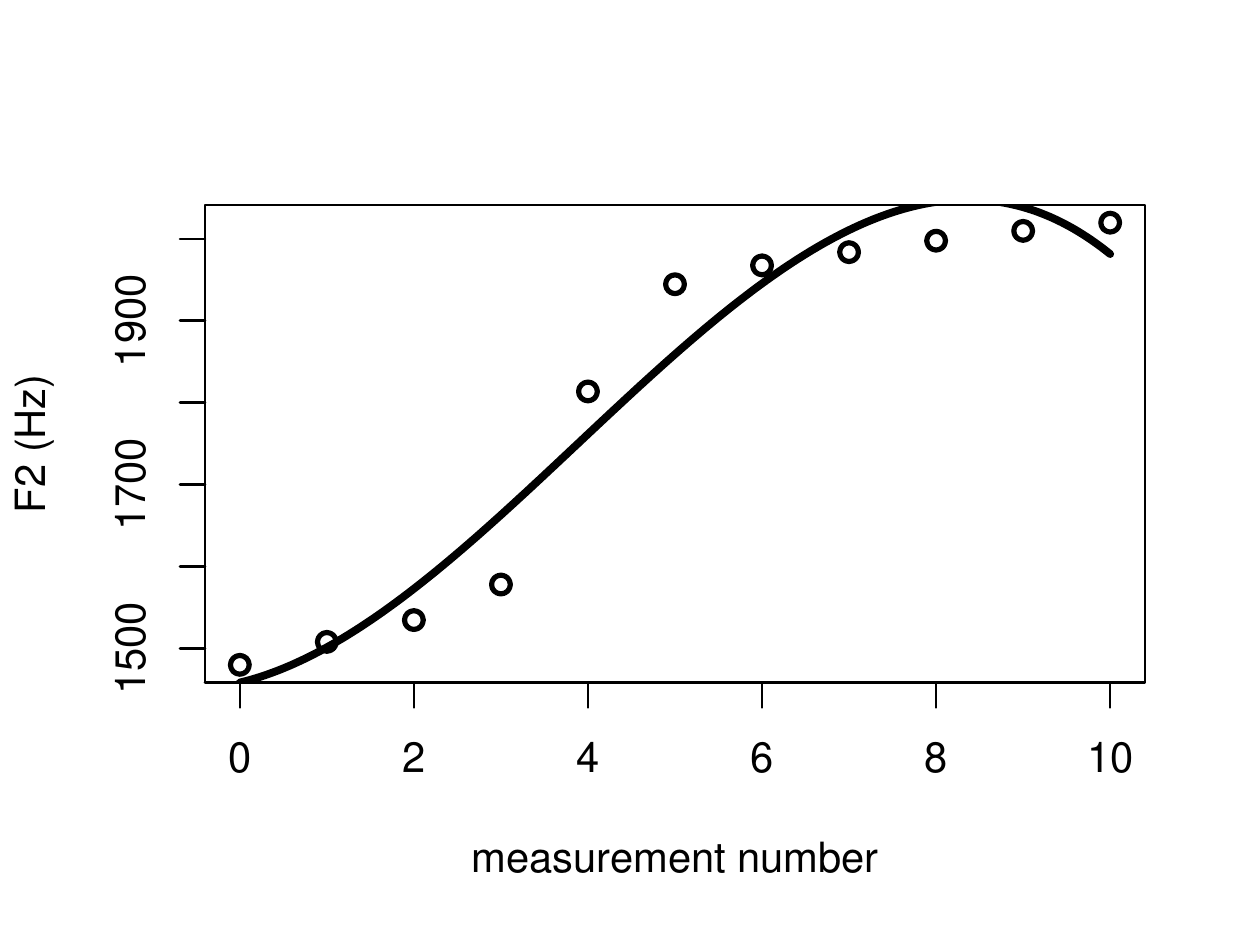} 

}

\end{knitrout}

\noindent This is an example of polynomial regression, where both a variable $x$ and some of its powers ($x^2$, $x^3$, \ldots) are included as predictors. The terms $x$, $x^2$, $x^3$ are referred to as basis functions. The fitted curve is obtained by multiplying each of the basis functions by the corresponding coefficient and then adding them up. The plot on the left below shows the basis functions of the GAM smooth for the formant trajectory before multiplication by the coefficients. The plot on the right shows the same basis functions after multiplication, and their sum (i.e.\ the predicted formant trajectory minus the intercept term).

\begin{figure}[h!]
\begin{knitrout}
\definecolor{shadecolor}{rgb}{0.969, 0.969, 0.969}\color{fgcolor}

{\centering \includegraphics[width=0.495\textwidth]{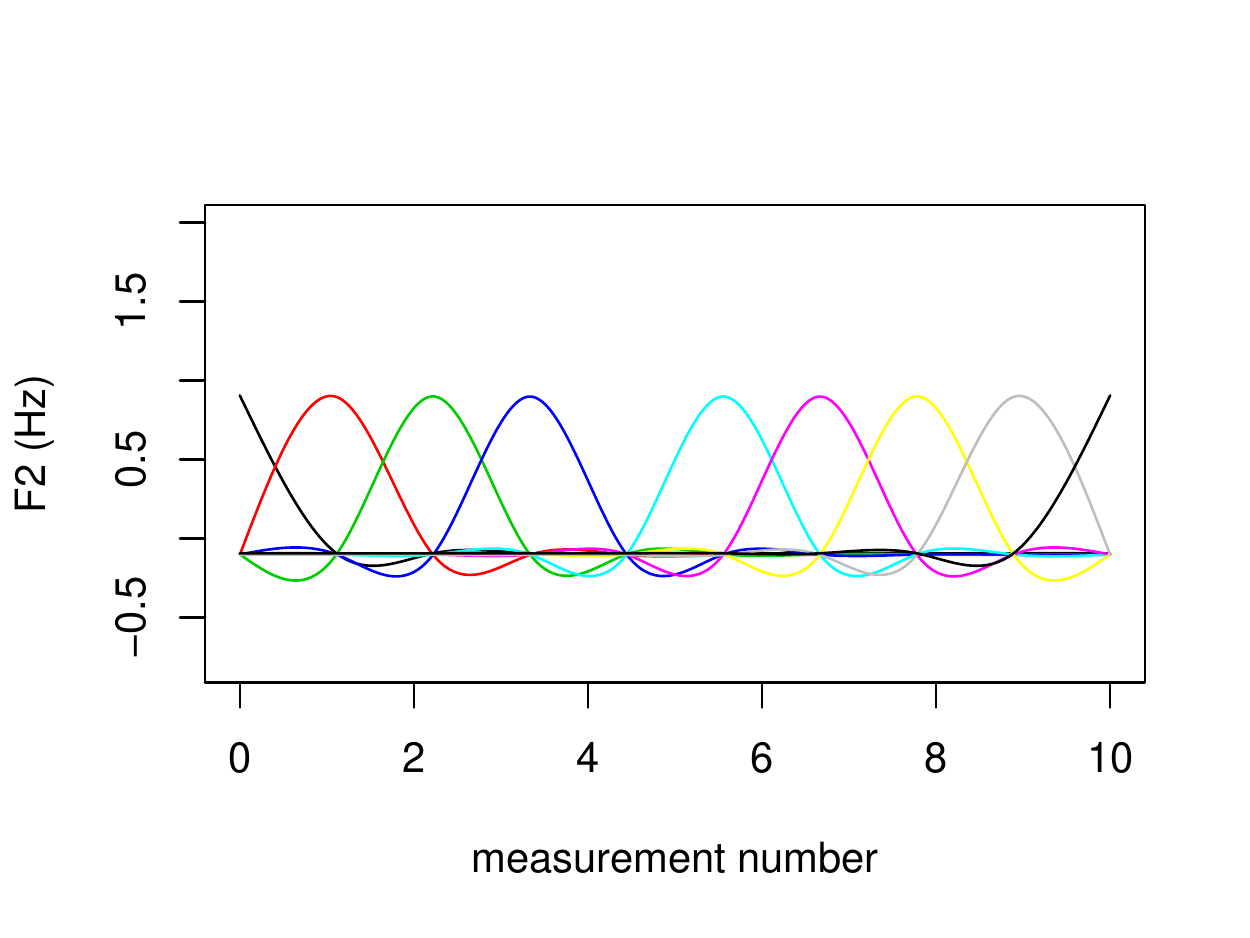} 
\includegraphics[width=0.495\textwidth]{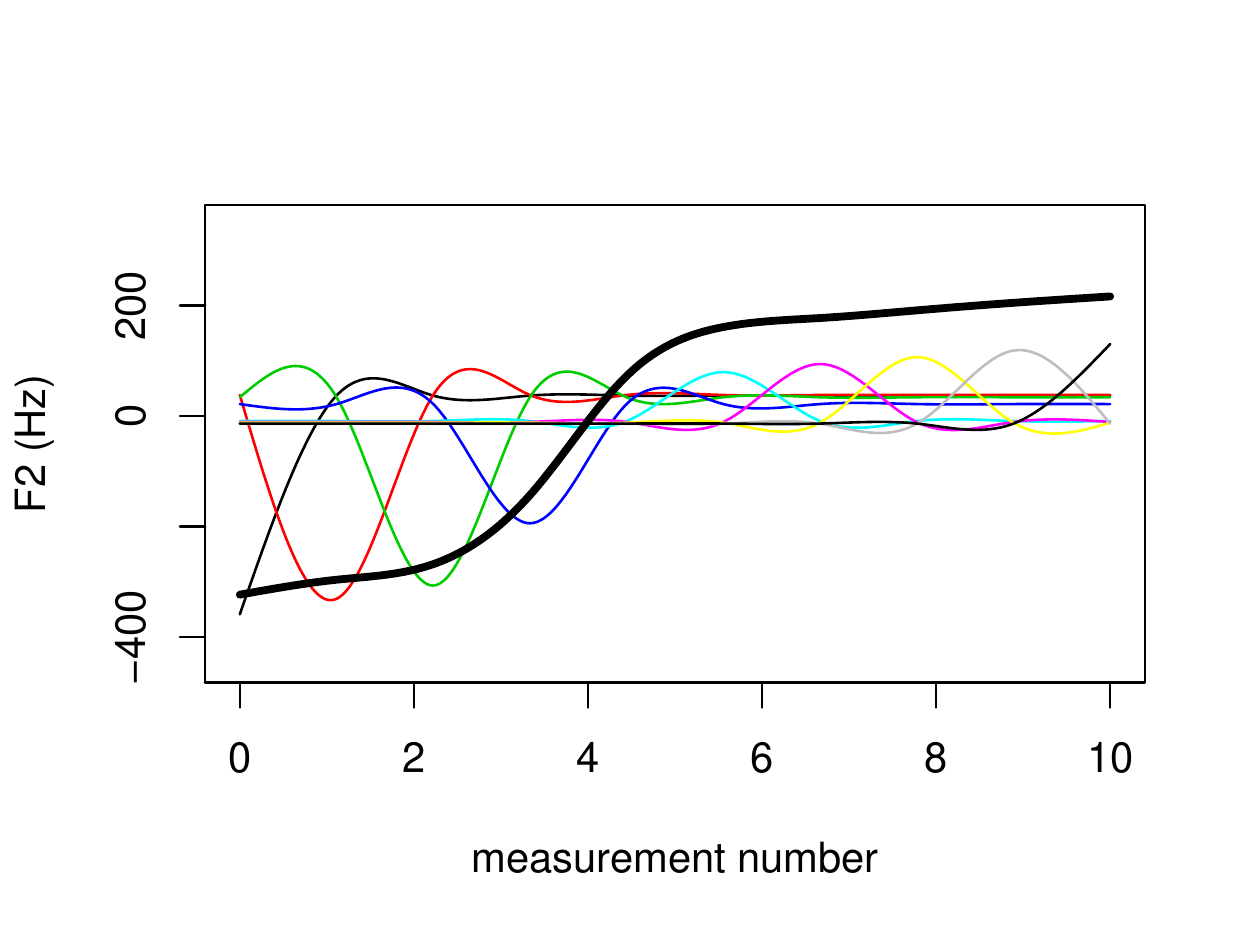} 

}

\end{knitrout}
\caption*{}\label{smooths}\vspace*{-1.3cm}
\end{figure}

\noindent The above smooth is made up of 9 basis functions, which is a default setting for certain types of smooths in R. Note that the basis functions are placed at regular intervals, and that they converge on each other at certain points (this is clearer in the plot on the left-hand side). These `convergence points' are called \textit{knots}, and there are 10 of them (the number of basis functions + 1): 2 at the edges of the plot and 8 in the middle. There is a simple intuition linked to the number of knots / basis functions: increasing this number allows for more wiggliness in the smooth, while decreasing it makes the smooth\ldots{} well, smoother. 

In linear regression models where the basis functions are included manually (e.g.\ polynomial regression), the number of basis functions (or knots) has to be chosen by the modeller. This can be tricky, as using too few basis functions can lead to oversmoothing (i.e.\ missing some of the non-linearity in the data), while using too many can lead to overfitting (i.e.\ missing the real trend in the data by fitting a curve to random noise). This is where GAMs really shine. GAMs rely on a value called the smoothing parameter. The coefficients for the individual basis functions contained in a GAM smooth are estimated in such a way that the resulting curve has a controlled degree of wiggliness determined by the smoothing parameter. The higher the smoothing parameter, the smoother ( = less wiggly) the estimated curve. This is illustrated below, where the number of basis functions is always the same, but the value of the smoothing parameter is varied.\footnote{As before, the curve is shown without the intercept, which means that it is centred around 0 along the \textit{y}-axis. This is why some of the values are negative. The graphs also include shifted versions of the actual data points to make the degree of smoothing clearer.}
\begin{knitrout}
\definecolor{shadecolor}{rgb}{0.969, 0.969, 0.969}\color{fgcolor}

{\centering \includegraphics[width=0.495\textwidth]{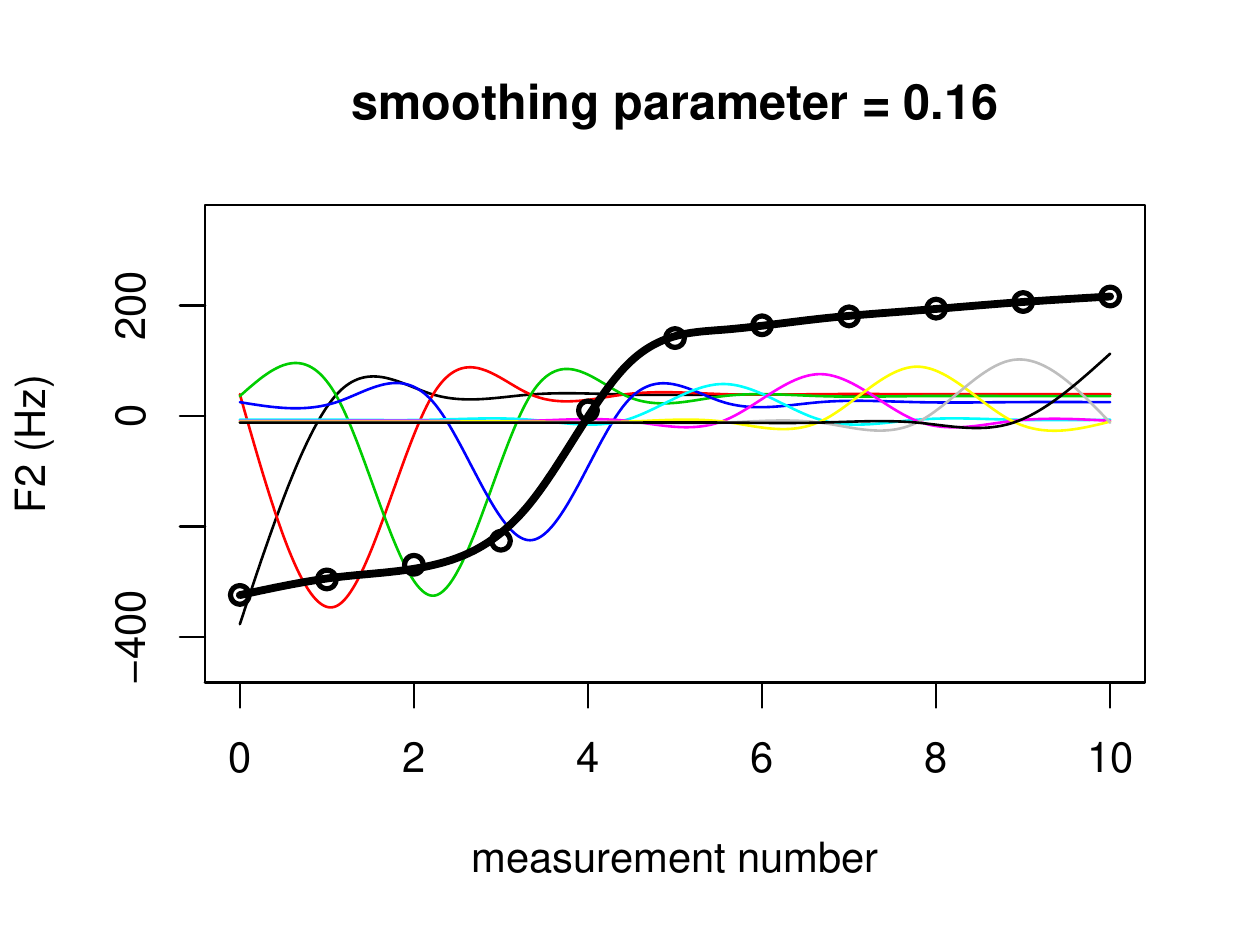} 
\includegraphics[width=0.495\textwidth]{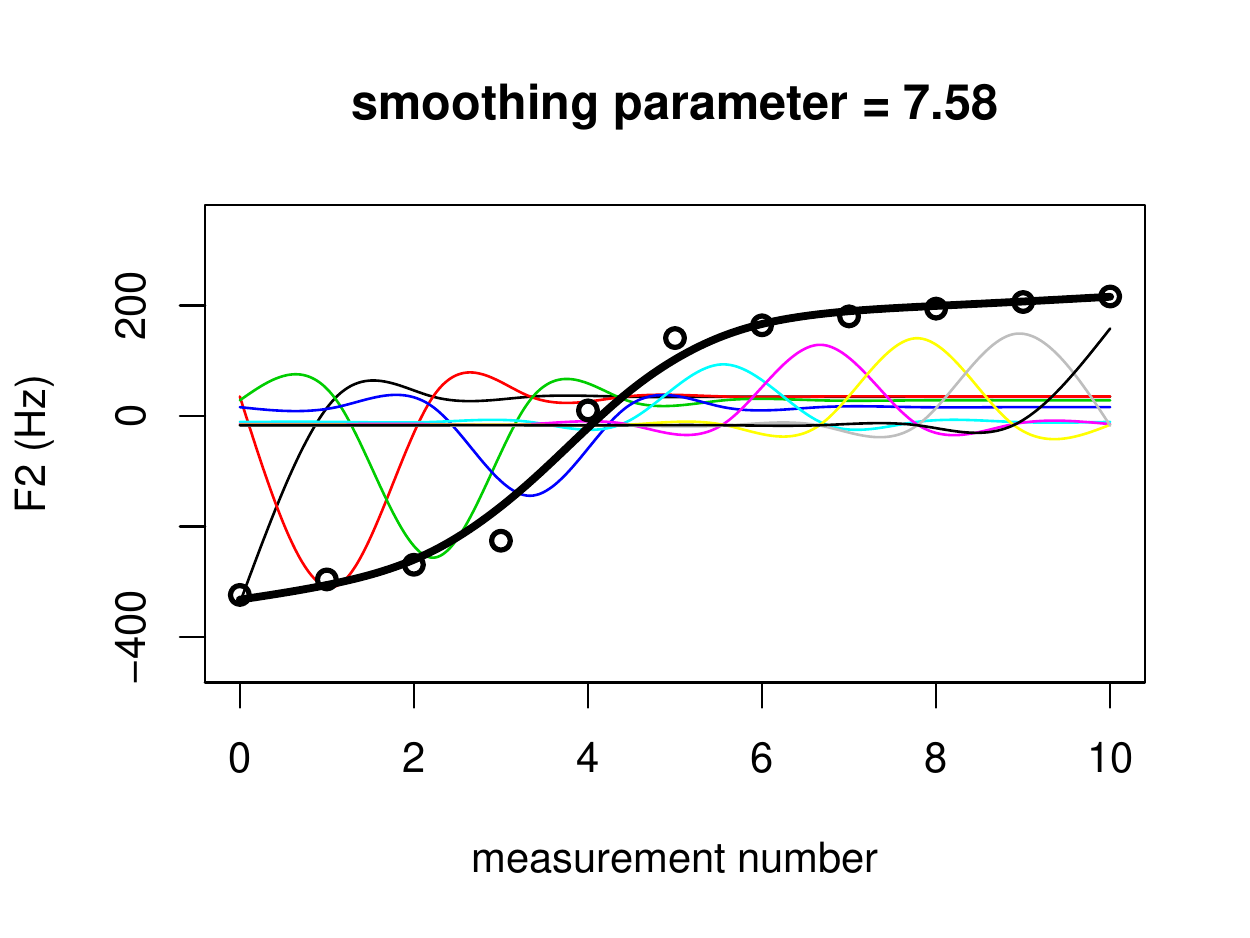} 
\includegraphics[width=0.495\textwidth]{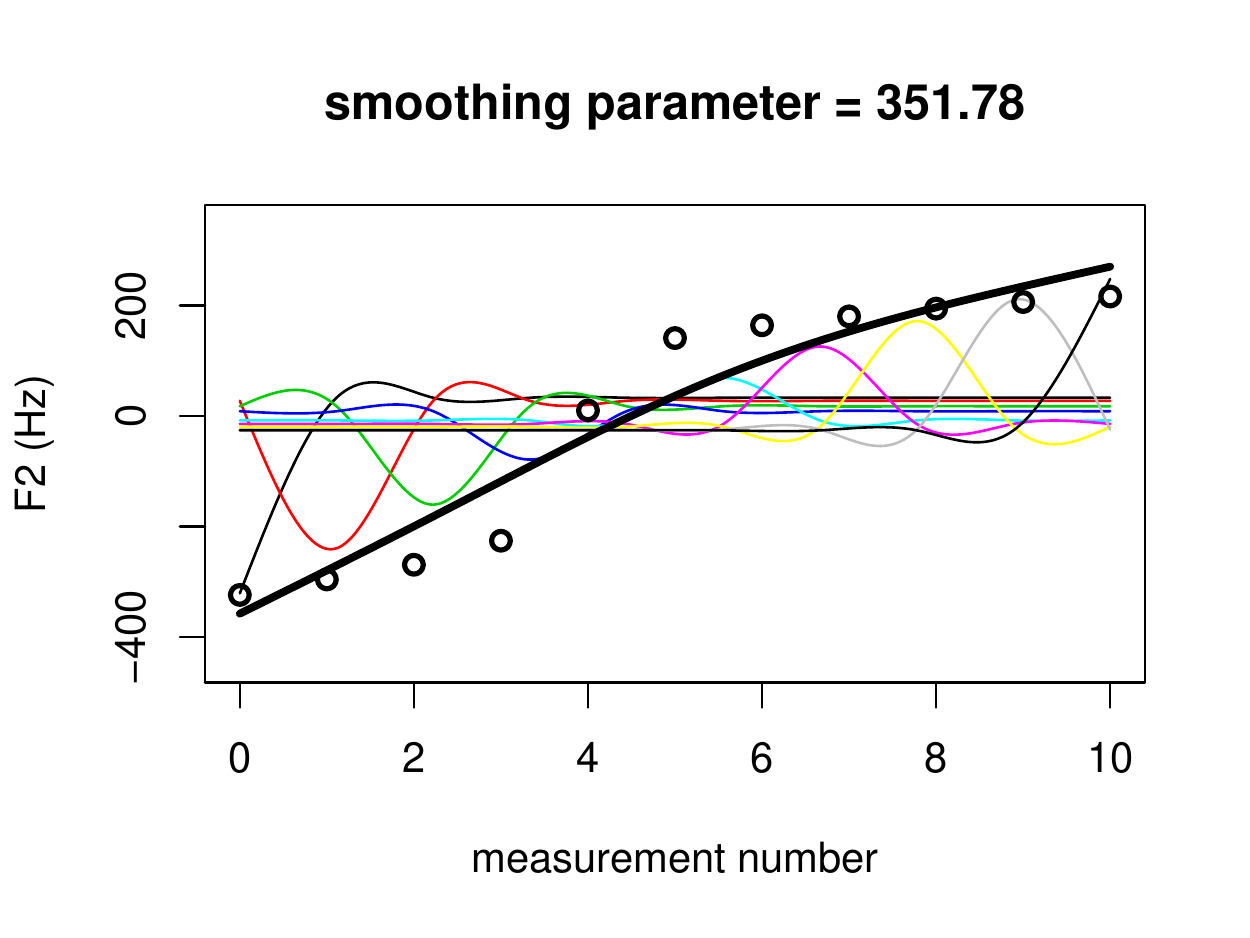} 
\includegraphics[width=0.495\textwidth]{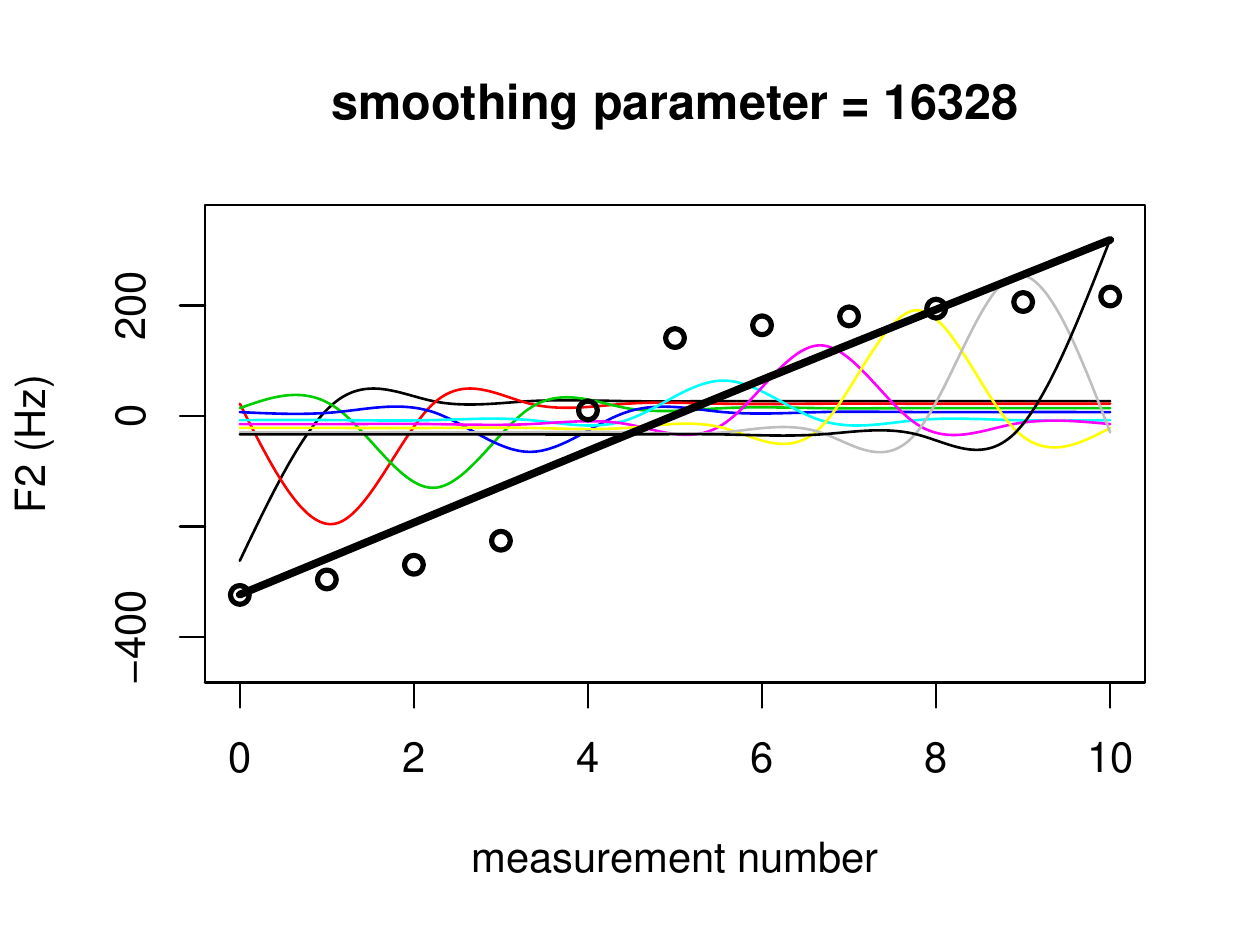} 

}

\end{knitrout}

\noindent In other words, the degree of smoothness / wiggliness in GAMs is mostly determined by the smoothing parameter. The role of the number of basis functions / knots is (mostly) reduced to setting an upper limit on the degree of wiggliness: having 3 knots ( = 2 basis functions) obviously allows for much less wiggliness than 10 or 100 knots, even if we set the smoothing parameter extremely low.

The GAMs that we use in this introduction (i.e.\ the ones implemented in the {\tt mgcv} package) do not require the modeller to decide on a value for the smoothing parameter: they estimate it directly from the data. This can be done using methods based on cross-validation or maximum likelihood estimation. We won't discuss these in detail, but a brief overview will be helpful (see {\tt ?gam.selection} from \verb+mgcv+ for more detail and references). Both sets of methods aim at choosing a value for the smoothing parameter that makes the resulting curve generalisable beyond the specific sample under investigation. This is especially clear in the case of cross-validation. Cross-validation works by (i) creating subsets of the full data set that each exclude a single data point, (ii) refitting the model to each of these subsets and (iii) checking how well the fitted models predict the excluded data points. Choosing a value for the smoothing parameter that is too low will result in a curve that is overly wiggly. This excess wiggliness will be used to capture idiosyncratic variation in the data set, which leads to bad performance on out-of-sample observations (and therefore a high error-rate in cross-validation). If the smoothing parameter is too high, the curve will fail to capture non-linear patterns both in within-sample and out-of-sample observations, which, again, leads to a high error-rate in cross-validation. Note that {\tt mgcv} actually only fits the model once, and uses a mathematical trick to calculate the cross-validation score (see \citealt{wood06} for more detail). Maximum likelihood estimation relies on a different technique that works by treating smooths as random effects for the purposes of estimating their smoothing parameters (but the outcome is the same: a curve that -- all things being equal -- avoids under/overfitting and is generalisable beyond the sample).

As a result of smoothing parameter estimation, the number of basis functions has little bearing on the shape of the smoother provided that there are enough of them to represent the degree of wiggliness in the data, so a smooth with a high number of basis functions will often look very similar to one with a lower number. This is illustrated below using a slightly modified version of our formant trajectory with 50 measurement points instead of 11 and a small amount of added noise. Three models were fit to this trajectory with different values for {\tt k} (10, 20 and 50). The plots show the data, the fitted curves and the so-called \textit{estimated degrees of freedom}, or EDF (see below). All three curves look very similar.

\begin{knitrout}\small
\definecolor{shadecolor}{rgb}{0.969, 0.969, 0.969}\color{fgcolor}\begin{kframe}
\begin{alltt}
\hlstd{demo.gam.k.10} \hlkwb{<-} \hlkwd{bam}\hlstd{(f2} \hlopt{~} \hlkwd{s}\hlstd{(measurement.no,} \hlkwc{bs} \hlstd{=} \hlstr{"cr"}\hlstd{,} \hlkwc{k} \hlstd{=} \hlnum{10}\hlstd{),}
                    \hlkwc{data} \hlstd{= traj.50)}
\hlstd{demo.gam.k.20} \hlkwb{<-} \hlkwd{bam}\hlstd{(f2} \hlopt{~} \hlkwd{s}\hlstd{(measurement.no,} \hlkwc{bs} \hlstd{=} \hlstr{"cr"}\hlstd{,} \hlkwc{k} \hlstd{=} \hlnum{20}\hlstd{),}
                    \hlkwc{data} \hlstd{= traj.50)}
\hlstd{demo.gam.k.50} \hlkwb{<-} \hlkwd{bam}\hlstd{(f2} \hlopt{~} \hlkwd{s}\hlstd{(measurement.no,} \hlkwc{bs} \hlstd{=} \hlstr{"cr"}\hlstd{,} \hlkwc{k} \hlstd{=} \hlnum{50}\hlstd{),}
                    \hlkwc{data} \hlstd{= traj.50)}
\end{alltt}
\end{kframe}
\end{knitrout}
\begin{knitrout}
\definecolor{shadecolor}{rgb}{0.969, 0.969, 0.969}\color{fgcolor}

{\centering \includegraphics[width=0.32\textwidth]{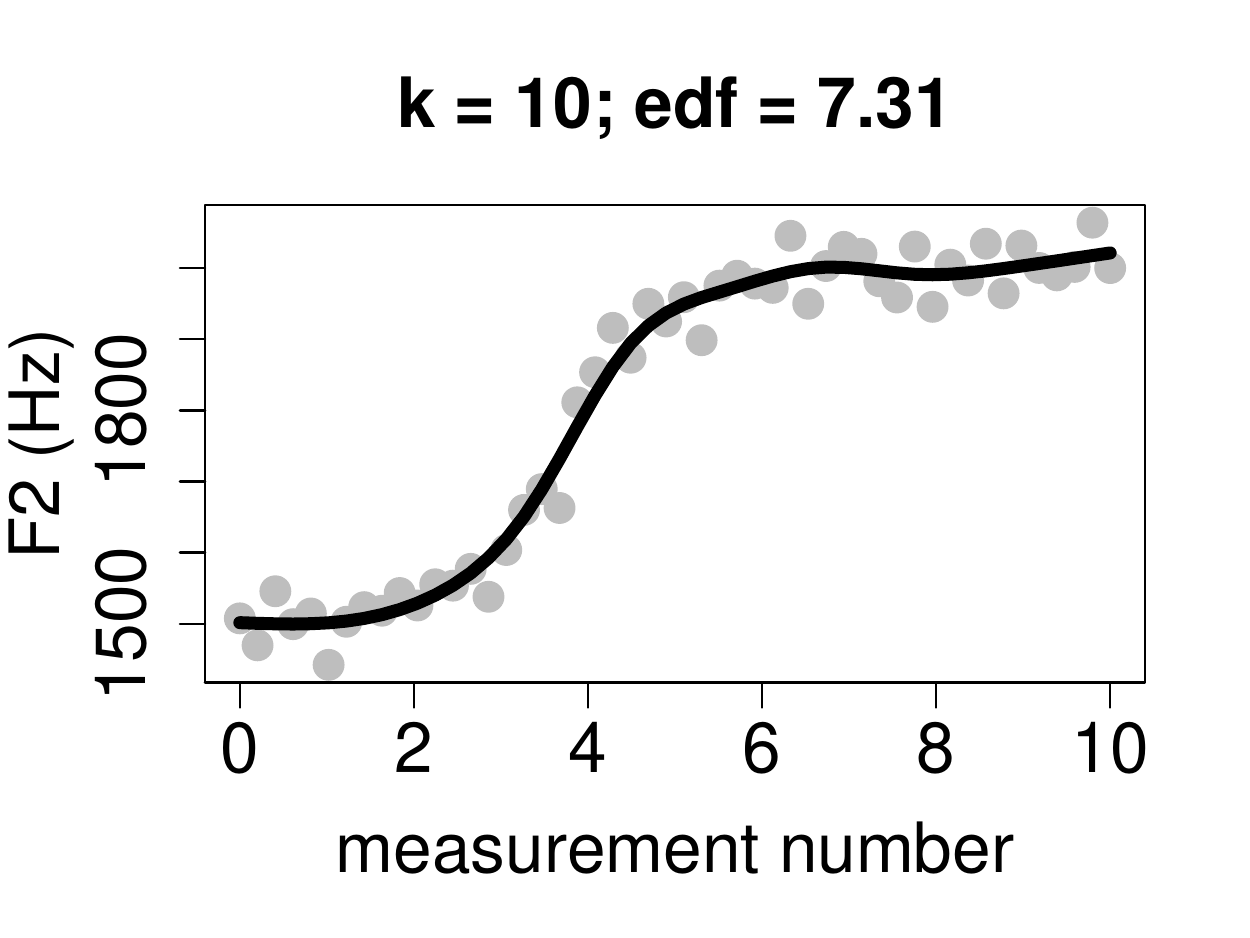} 
\includegraphics[width=0.32\textwidth]{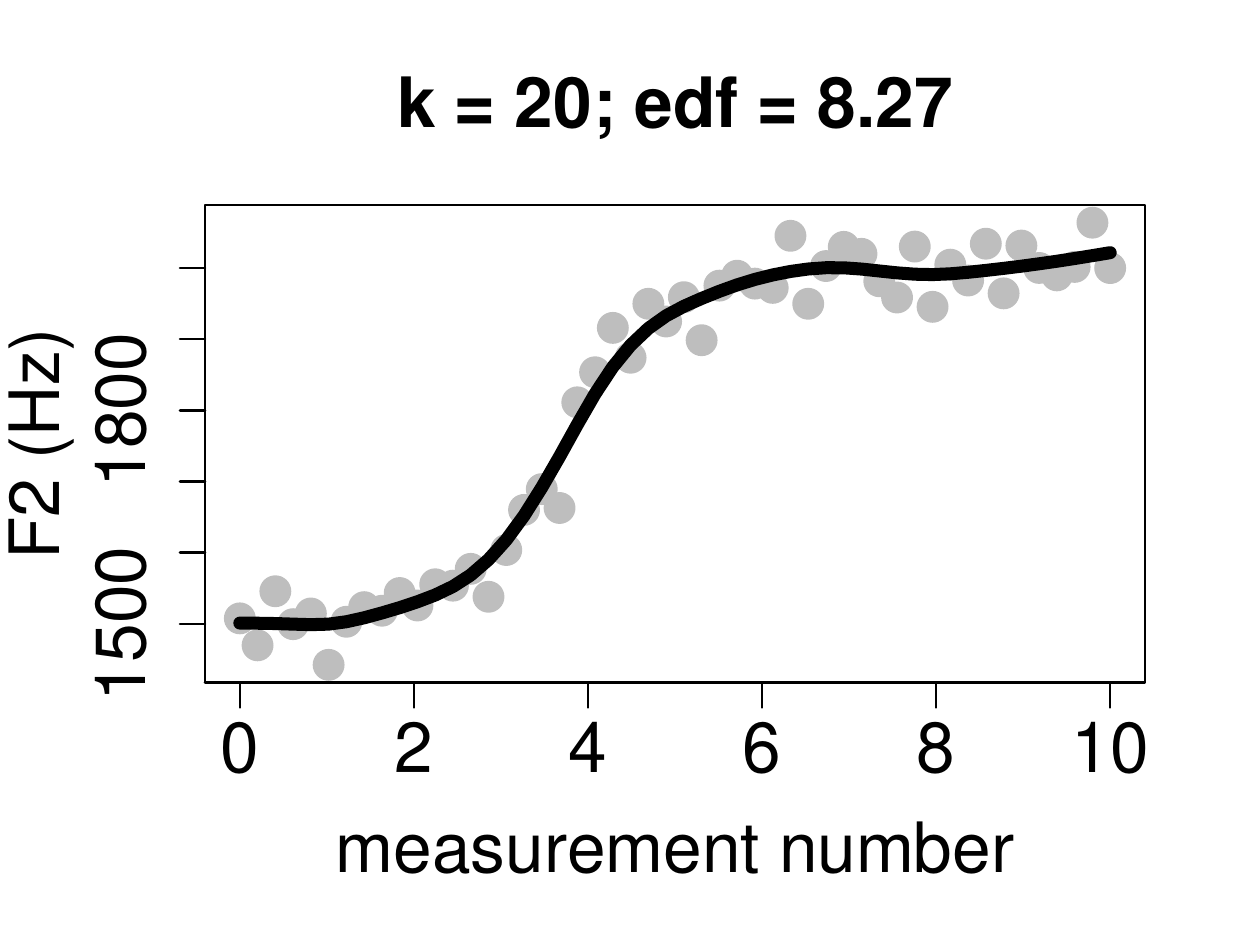} 
\includegraphics[width=0.32\textwidth]{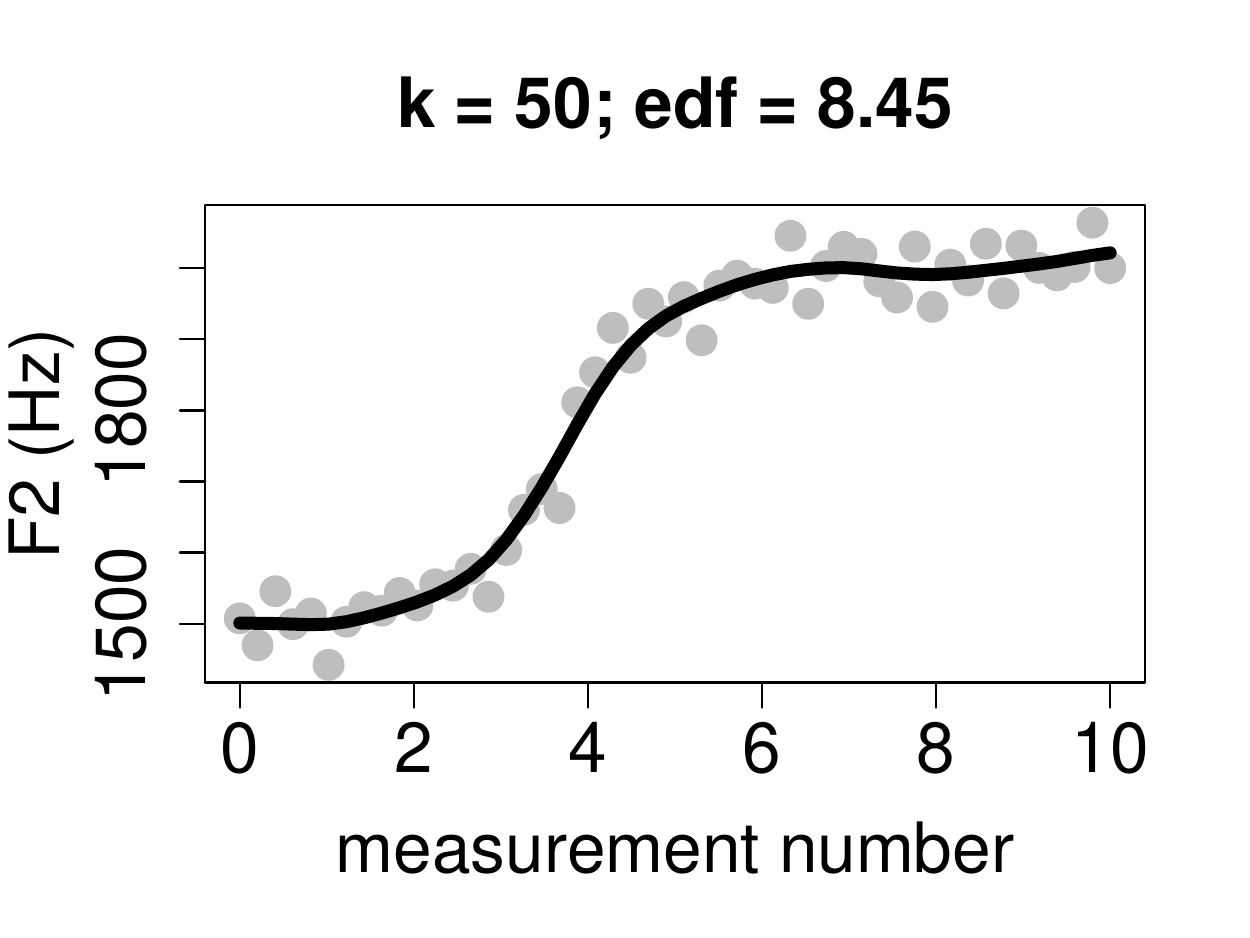} 

}

\end{knitrout}

\noindent On the basis of the above discussion, it would seem that the best strategy is to set {\tt k} to a high value so that model is not restricted in terms of how much wiggliness it can attribute to the underlying curve. And once we've done that, we could forget about the whole thing, basically trusting {\tt mgcv} to do the right thing for us.

Unfortunately, that's not a good strategy. As is often the case with statistical methods, knowing only a little about GAMs and setting up models without understanding what they do is probably worse than using less advanced methods in an informed way. First of all, although cross-validation and other estimation methods attempt to avoid overfitting, they are not always successful at doing so, which means that an unrealistically high value of {\tt k} does sometimes lead to an unrealistically wiggly curve. Second, there are several situations where the number of basis functions needs to be changed, and these are almost impossible to avoid while working with dynamic speech data:
\begin{itemize}
  \item \textit{too few measurements to support} \verb+k+ \textit{knots}: If our trajectories only have 11 measurements, the maximum number of knots is also 11 (that is, the maximum number of basis functions is 10). Since the default value for \verb+k+ is 10 (although this number may be different depending on the type of smooth), it needs to be lowered if there are less than 10 unique values for a given variable.
  \item \textit{not enough wiggliness allowed}: The default value of \verb+k+ can only support a certain amount of wiggliness in the data. If the actual trajectories show a greater degree of non-linearity, \verb+k+ needs to be increased.
  \item \textit{computational inefficiency due to high} \verb+k+: The higher the value of \verb+k+, the longer it will take to fit the model. Therefore, it is a good idea not to increase \verb+k+ any further than necessary. In realistic scenarios, the modeller may be forced to choose a \verb+k+ that is actually lower than would be ideal. 
\end{itemize}
My recommendation is therefore to try and make an informed choice about {\tt k} rather than attempting to rely on a one-size-fits-all strategy (e.g.\ always using the default value or always specifying a high value for {\tt k}). For instance, formant trajectories for monophthongs and diphthongs tend to be relatively smooth and they don't typically show any meaningful `high-frequency' wiggliness, so there's no point in setting {\tt k} to a very high value (I'm reluctant to suggest specific values here, but e.g.\ formant trajectories for single vowels will hardly ever require more than 10 basis functions). An intonation contour for a longer chunk of speech will likely show much more wiggliness, so {\tt k} ought to be set a bit higher too. Setting {\tt k} to a reasonable value requires a bit of experience, so do have a play around with different values of {\tt k} in the examples in the second part of this introduction. {\tt mgcv} also provides some functions and advice for evaluating {\tt k}; see {\tt ?gam.check} and {\tt ?choose.k} from \verb+mgcv+.

The concepts of basis function and smoothing parameter are closely related to the so-called \textit{estimated degrees of freedom} (or EDF). When the coefficients for basis functions are estimated without any constraints, as in the case of polynomial regression, a smooth with \textit{p} basis functions uses up exactly \textit{p} degrees of freedom. However, when the coefficients are constrained by a smoothing parameter, the effective degrees of freedom taken up by the smooth go down (this is because the coefficients for the individual basis functions become dependent on each other). For instance, at extremely high values of the smoothing parameter, the smooth becomes a straight line, which only uses up a single degree of freedom, even if it is represented by more than one basis function. Conversely, at extremely low values of the smoothing parameter, the smooth will use all the potential wiggliness provided by the \verb+k+ $-$ 1 basis functions, which corresponds to \verb+k+ $-$ 1 degrees of freedom. At intermediate values, the smooth uses an intermediate number of degrees of freedom. The EDF is an estimate of the degrees of freedom that are actually used by a smooth with a given number of basis functions and a given smoothing parameter. Significance tests of smooth terms in GAMs rely on the EDF and not the number of basis functions -- and, as a result, model summaries for GAMs always include the EDF. Note that the EDFs for the three models with different {\tt k} values above are very similar (see the plots for the EDF values).

\subsection{GAMMs}\label{section:gamms}

We are now in a position to move on to GAMMs, that is, generalised additive \textit{mixed} models. This tutorial assumes that the reader is already familiar with random intercepts and slopes from linear mixed effects models and knows how to implement them in R. Just as a reminder: random intercepts in linear mixed models capture by-group random variation in the outcome variable (e.g.\ between-speaker differences in average F2 values); random slopes capture by-group random variation in the effect of a predictor variable on the outcome variable (e.g.\ between-speaker differences in the effect of style on F2 -- certain speakers exhibit more stylistic adaptation than others). GAMMs are to GAMs as linear mixed effects models are to linear models. That is, GAMMs incorporate random effects alongside parametric and smooth terms. Similar to linear mixed effects models, these random effects can be random intercepts and random slopes. However, GAMMs also offer a third option: random smooths. 

Random smooths are similar to random slopes, but they are more flexible than the latter: while random slopes can only capture by-group variation in linear effects, random smooths can also deal with by-group variation in non-linear effects. An example will help to make this point clearer. Let's assume that the trajectory that we've looked at earlier is an example of a specific vowel, and that we have four
further tokens for this vowel. The four trajectories are shown below (this data set is available as {\tt traj$\_$random.csv}):
\begin{knitrout}
\definecolor{shadecolor}{rgb}{0.969, 0.969, 0.969}\color{fgcolor}

{\centering \includegraphics[width=0.495\textwidth]{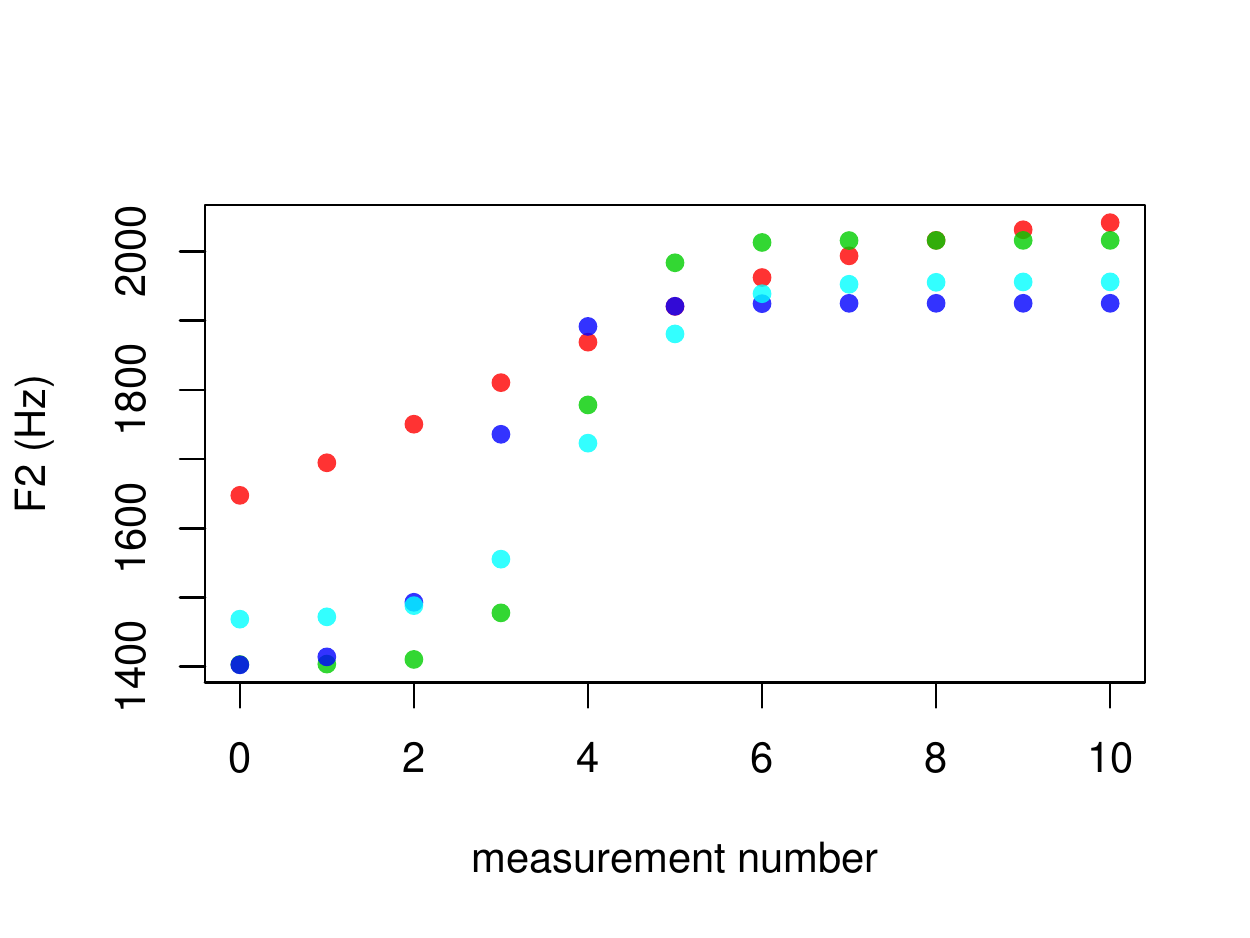} 

}

\end{knitrout}
\noindent First, let's use linear mixed effects models to fit straight line approximations to the trajectories. We'll fit two versions of the same model: one with random intercepts only, and a second one with random intercepts and random slopes. The random intercepts model is shown on the left below, while the random intercepts $+$ slopes model is shown on the right.

\begin{figure}[!h]
\vspace*{-0.5cm}
\begin{knitrout}
\definecolor{shadecolor}{rgb}{0.969, 0.969, 0.969}\color{fgcolor}

{\centering \includegraphics[width=0.495\textwidth]{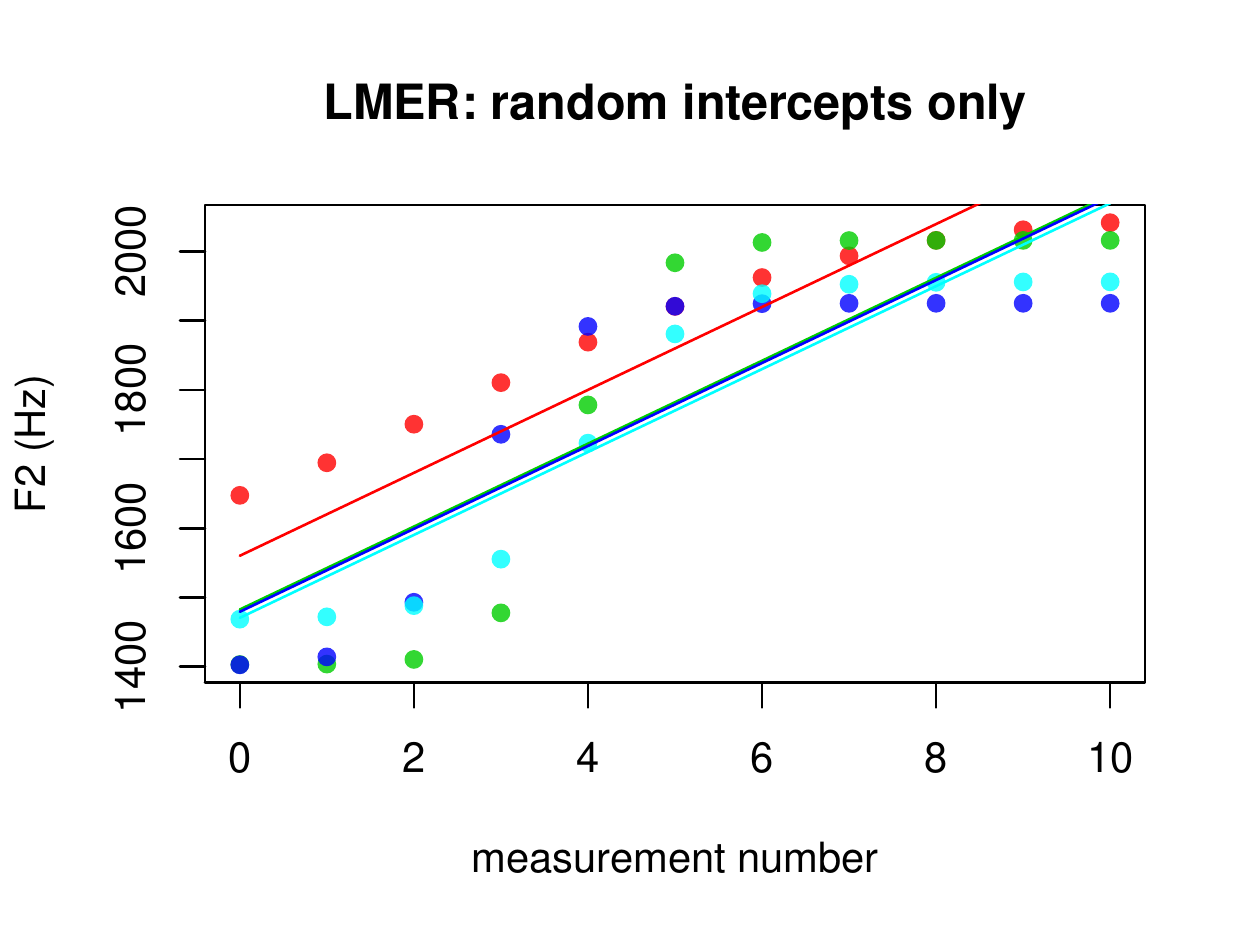} 
\includegraphics[width=0.495\textwidth]{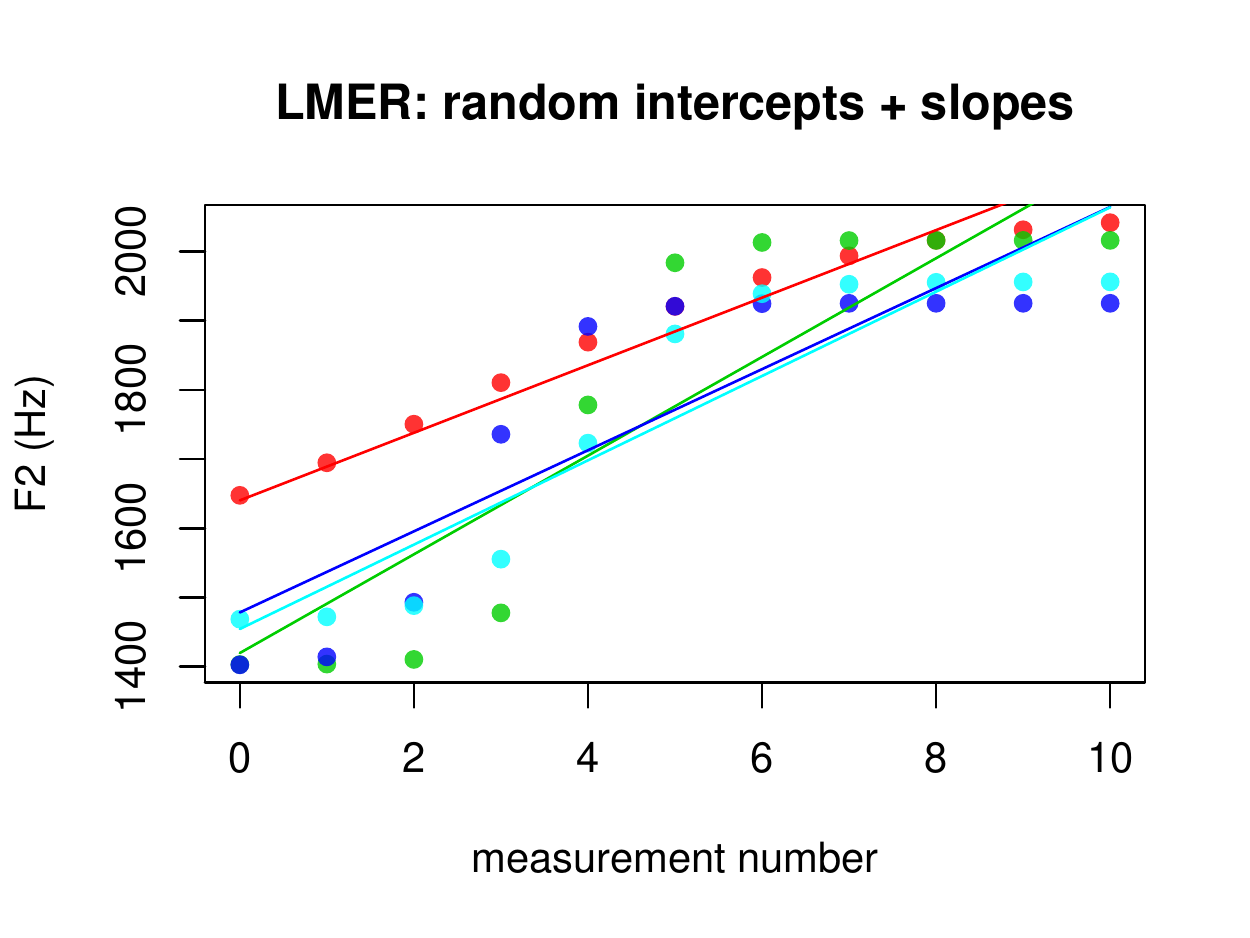} 

}

\end{knitrout}
\caption*{} \label{lmers}\vspace*{-1cm}
\end{figure}

\noindent Unsurprisingly, the model fit is not great. In the random intercepts only model, the fitted lines are parallel to each other, as the slopes are not allowed to vary. In the random intercepts + slopes model, both the height and the slope of the lines vary. Now let's fit three GAMMs to the same trajectories: one with random intercepts only, a second one with random intercepts $+$ slopes and a third one with random smooths.
\begin{knitrout}
\definecolor{shadecolor}{rgb}{0.969, 0.969, 0.969}\color{fgcolor}

{\centering \includegraphics[width=0.495\textwidth]{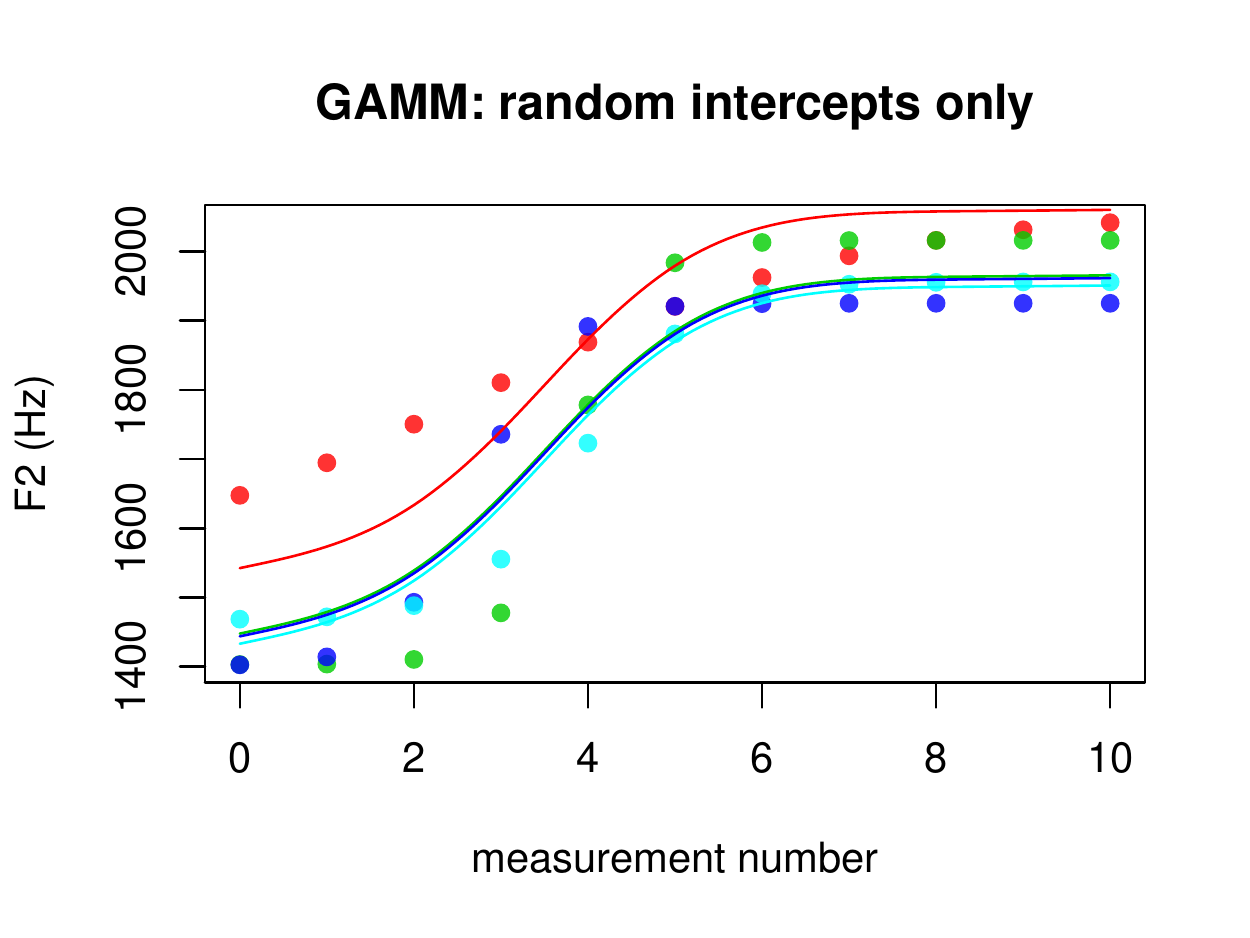} 
\includegraphics[width=0.495\textwidth]{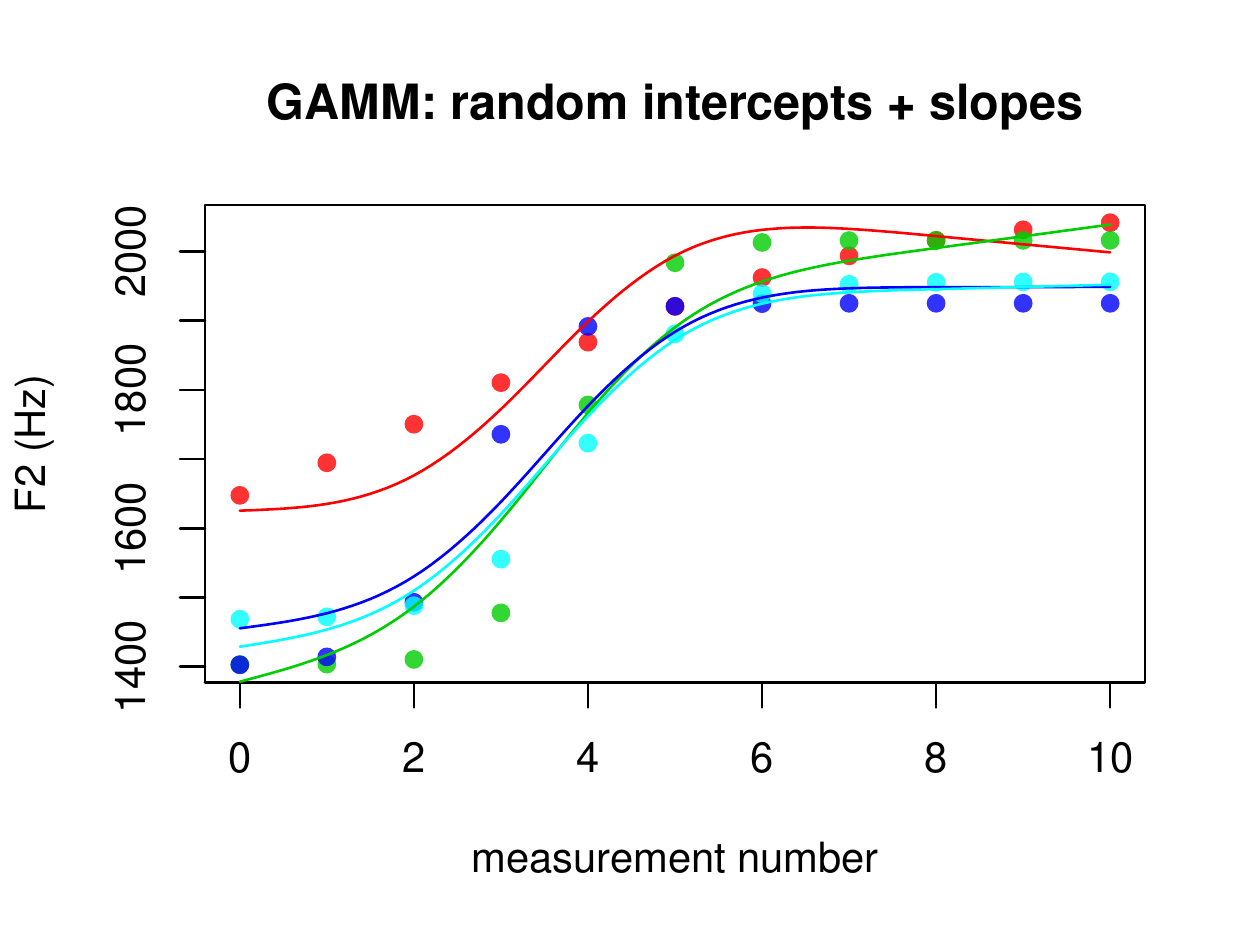} 
\includegraphics[width=0.495\textwidth]{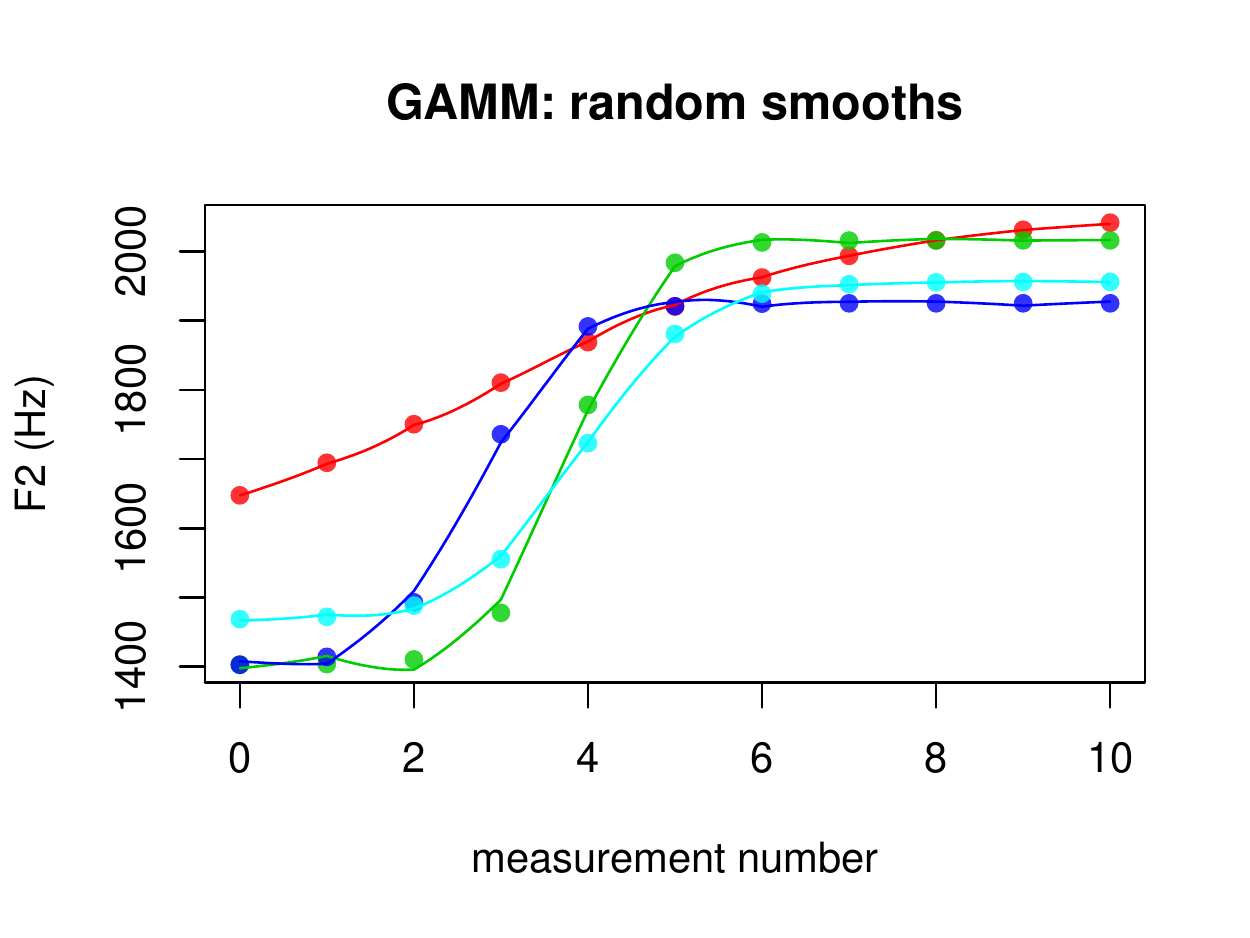} 

}

\end{knitrout}
\noindent The model with random intercepts simply varies the height of the lines, but does not yield a particularly good fit. The one with slopes does slightly better: in this case, the same curve is essentially rotated and stretched to match the actual trajectories. Random smooths clearly provide the best fit by fitting individual curves to each trajectory. Note, however, that random smooths are also extremely resource intensive: fitting four separate random smooths to the data requires $4 \times \textrm{\tt k}$ basis functions, and the same number of coefficients need to be estimated.\footnote{The separate random smooths each use up {\tt k} rather than $\textrm{\tt k} - 1$ basis functions (as opposed to the smooths that we've looked at before). The additional basis function plays the role of a random intercept. Since there are four smooths (one for each trajectory), the overall number of basis functions is $4 \times \textrm{\tt k}$.}

\subsection{Residual autocorrelation}

Another important concept in GAMM theory is that of \textit{residual autocorrelation}. Consider the example of a linear model fitted to the wiggly trajectory from section \ref{sec:gams}. As explained above, the model fit leaves systematic patterns in the residuals, which are reproduced below for convenience. One way of conceptualising these patterns is through the notion of autocorrelation: the correlation between observed values at fix intervals within a time series, where the size of the interval is usually referred to as the \textit{lag}.%
\footnote{
See \citet[390]{venablesandripley02} for the formula for calculating autocorrelation, which is slightly different from the formula for Pearson's correlation coefficient.
} %
For instance, the autocorrelation at lag 1 in the residuals below is calculated by taking the f2 values at measurement points 0, 1, 2, \ldots{} and correlating them with the f2 values at measurement points 1, 2, 3, \ldots{} The autocorrelation at lag 2 is calculated by correlating the f2 values at measurement points 0, 1, 2, \ldots{} with those at measurement points 2, 3, 4, \ldots{} These autocorrelation values are usually shown in a plot where the horizontal axis shows different lag values and the vertical axis shows the autocorrelation at each lag value. These plots also often include two horizontal lines, which are `approximate 95\% confidence limits' (\citealt{venablesandripley02}; i.e.\ autocorrelation values outside these lines can be regarded as significant at $\alpha = 0.05$, though significance isn't of key importance in this case). The figure on the left below shows the residuals from the linear model, while the figure on the right is an autocorrelation plot based on these residuals.

\begin{knitrout}
\definecolor{shadecolor}{rgb}{0.969, 0.969, 0.969}\color{fgcolor}

{\centering \includegraphics[width=0.495\textwidth]{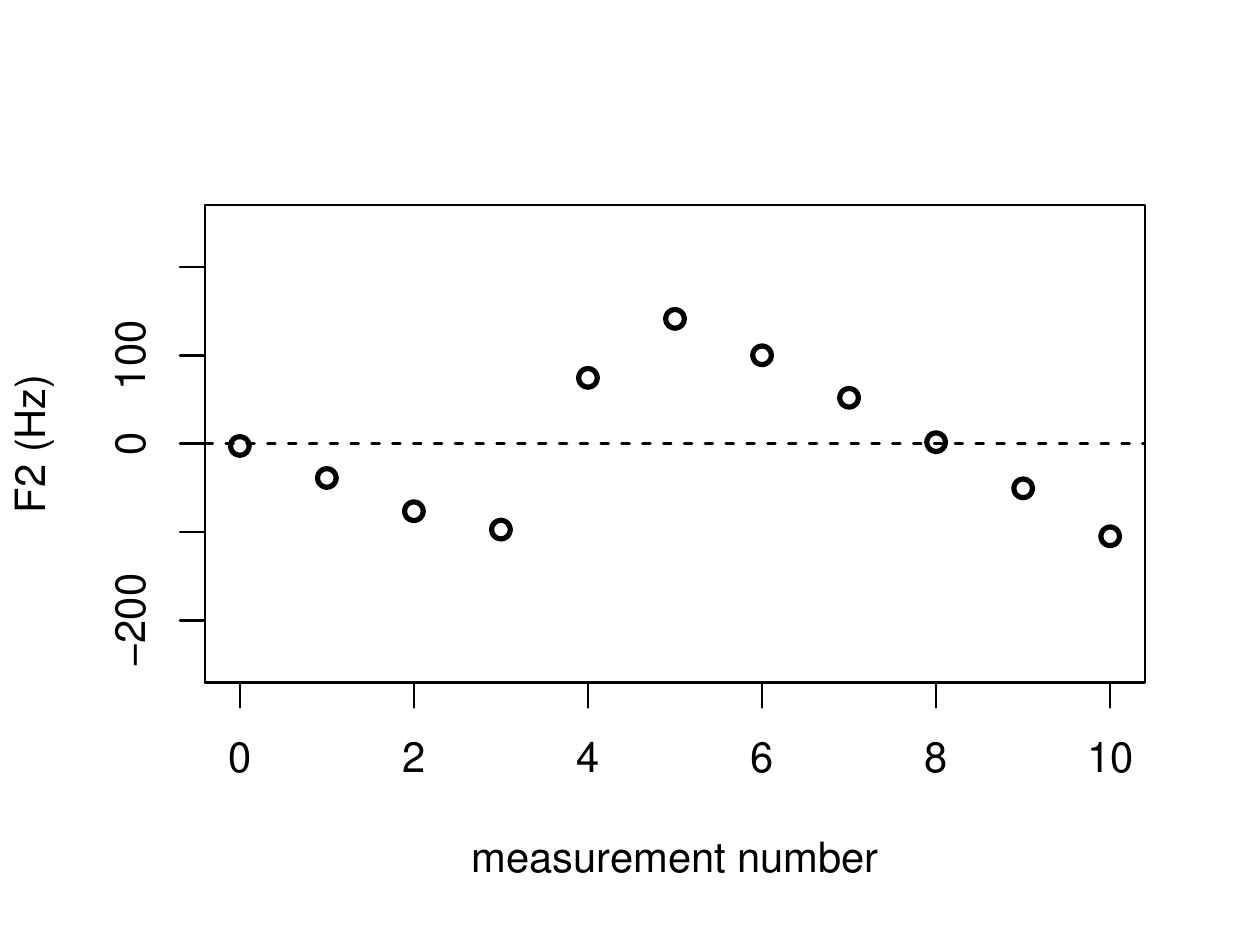} 
\includegraphics[width=0.495\textwidth]{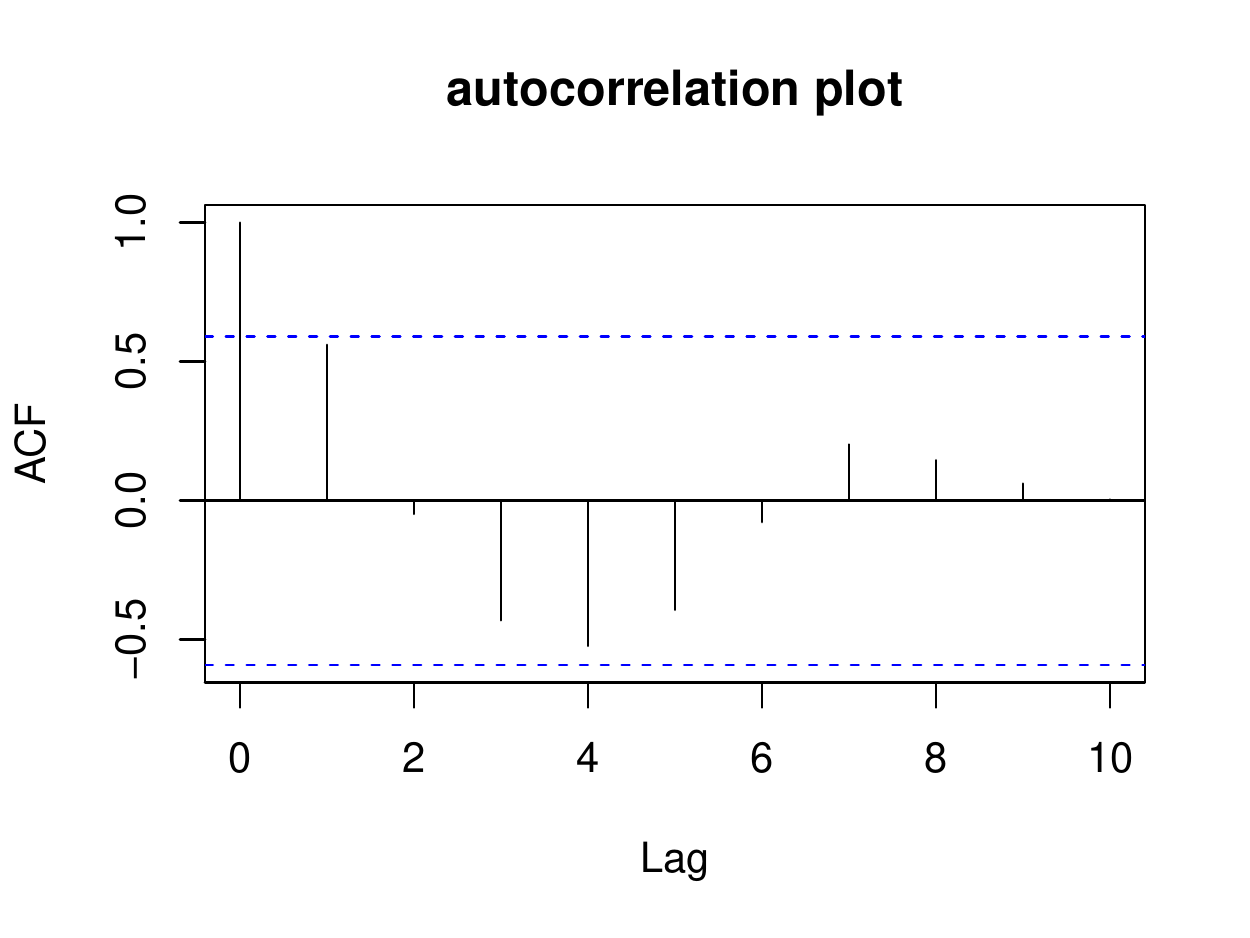} 

}

\end{knitrout}

\noindent The autocorrelation at lag 0 is simply the correlation of the residuals with themselves, which, of course, always yields a value of 1. The autocorrelation at lag 1 is slightly over 0.5, which is a moderately high value. The formant trajectory is non-linear, but it moves smoothly over time, which means that neighbouring measurements are always relatively close to each other. Since this non-linear movement is not appropriately captured by a linear model, this pattern remains in the residuals, which leads to the observed autocorrelation value. The autocorrelation at lag 2 is very close to 0, but at lags 3, 4 and 5 we observe higher negative values. Negative values indicate a zigzag pattern in the residuals. The fact that negative autocorrelations are only observed at higher lag values indicates that the zigzag pattern occurs not between adjacent measurements but over a slightly longer period. This is indeed what we observe in the residuals: we start with negative values, which become positive at measurement point 4 and then go back to negative at measurement point 9. Note that the zigzag pattern in the current set of residuals is due to the relatively sudden formant transition near the middle of the trajectory, which is entirely smoothed over by the linear model.

Another example of a residual autocorrelation plot is presented below. The residuals come from another linear model, which was fitted to the 50-point trajectory from the previous section (which was introduced in the discussion of different {\tt k} values). The residuals look slightly messier as this trajectory included a small amount of added noise.

\begin{knitrout}
\definecolor{shadecolor}{rgb}{0.969, 0.969, 0.969}\color{fgcolor}

{\centering \includegraphics[width=0.495\textwidth]{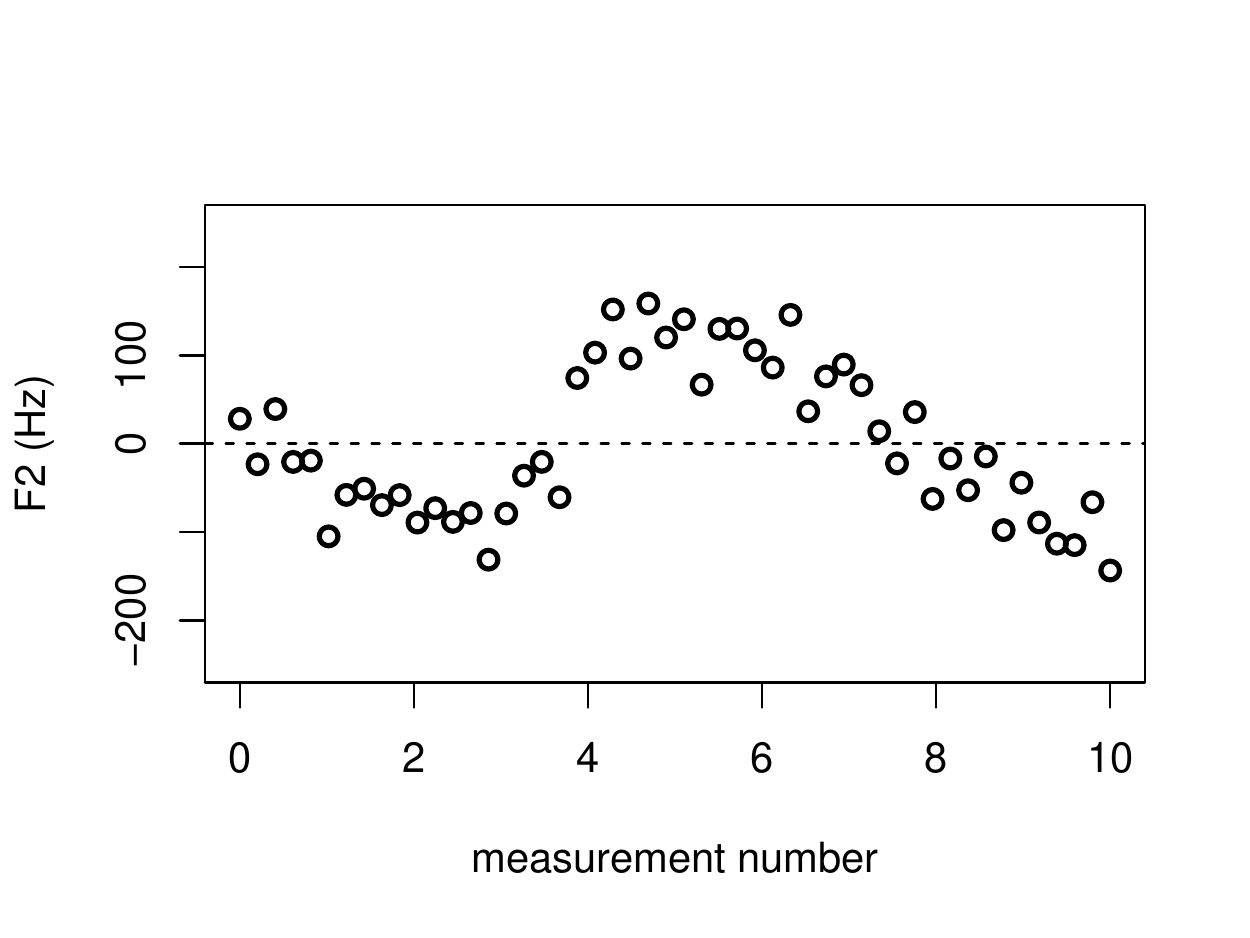} 
\includegraphics[width=0.495\textwidth]{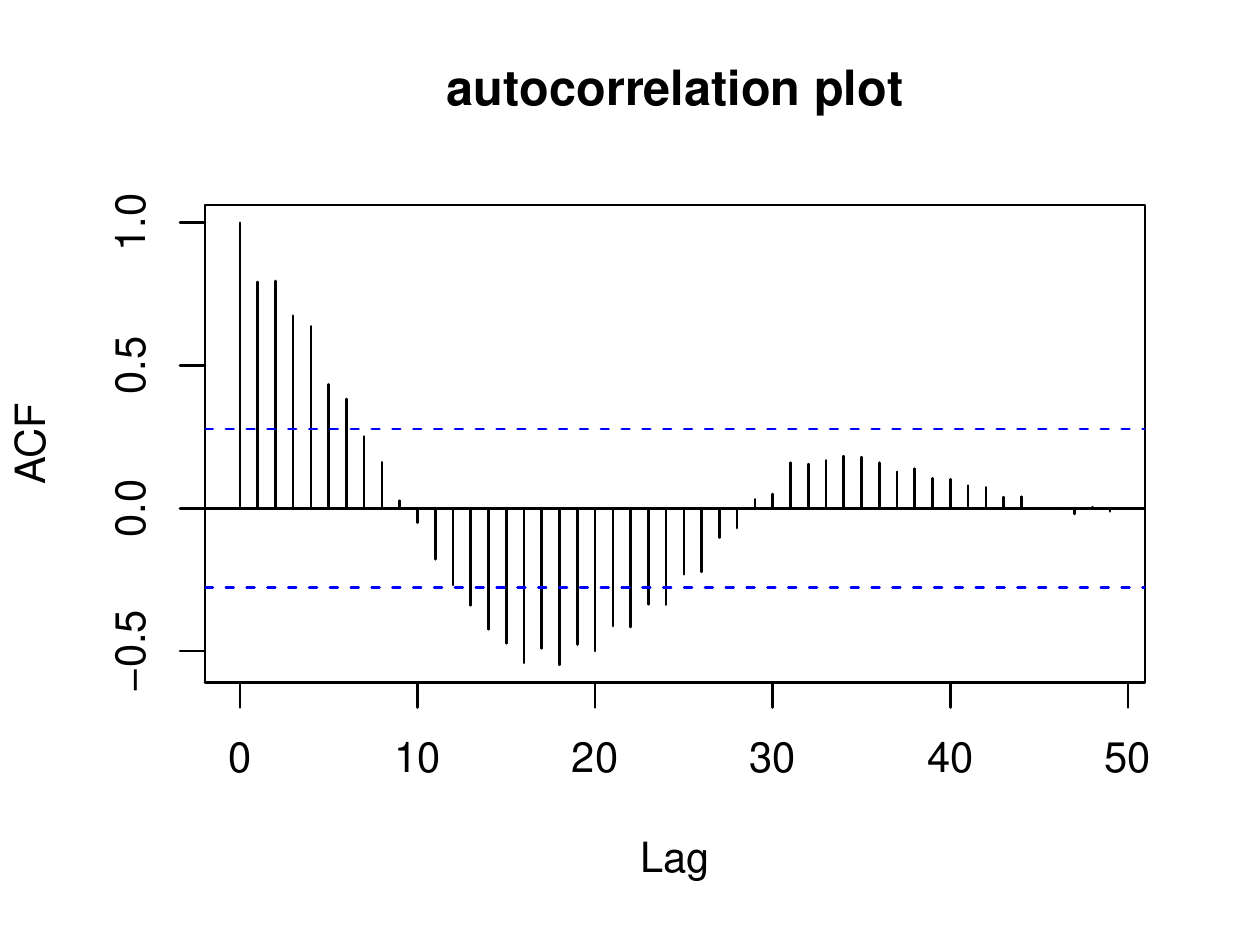} 

}

\end{knitrout}

\noindent Though the overall shape of the autocorrelation plot is similar to that of the previous one, the actual autocorrelation values are not the same. This is especially clear for the first few lag values, where the autocorrelation is really high (close to 0.8). The reason for these high values is that the trajectory moves very slowly when viewed as a function of measurement number -- on average, the change between two neighbouring points will be very small. If the trajectory didn't include any noise, the autocorrelation at low lag values would likely be even higher.

Autocorrelation in the residuals is problematic in that it can lead to inaccurate -- and often downward biased -- standard errors, confidence intervals, and \textit{p}-values. There are two ways of dealing with this issue (cf.\ \citealt{baayenetal16}): (i) fitting more accurate models that capture all patterns in the residuals (often through using random smooths) or (ii) including a so-called \textit{error model} in the regression, which essentially adjusts the model output to control for the biasing effect of residual autocorrelation. Both of these methods will be discussed in more detail below.

\subsection{Why do we need random smooths and/or error models?}

Why are patterns in the residuals such a serious issue for regression models? And why do we need to add random effects and/or error models to our GAM(M)s? These questions are partly answered in standard texts dealing with mixed effects regression models \citep{baayen08,gelmanandhill07,zuuretal09}. More detailed treatments are given in \citet{barretal13} and \citet{batesetal15}, while \citet{winter13} provides a particularly clear explanation that should be fairly easy to follow for readers with no background in statistics. The focus in many of these texts is on how models without random effects may violate the underlying assumptions of regression models, in particular the assumptions of \textit{linearity} and \textit{independence of errors}. What I'd like to do here is offer a slightly different approach to the question, and discuss the importance of random effects in more intuitive terms.

Regression models attempt to make educated guesses about `hidden' underlying parameters (which correspond to properties of interest in the real world) based on observable data. Each of the data points in a data set provides some information about these parameters, increasing our confidence in the guesses the model makes about them (we will refer to these guesses as estimates). However, it turns out that individual data points don't always provide the same amount of information about the underlying parameters. In other words, two data sets with the same number of observations may provide different amounts of information about the same parameters. These differences are often dependent on the presence or absence of grouping structure and temporal/spatial structure in the data.

In terms of the amount of information per data point, the best case scenario is a data set where there is no grouping or temporal/spatial structure (beyond the variables whose effects we want to estimate). In such data sets, a regression model can assume that individual data points are determined solely by the underlying parameters plus a bit of random noise, so all the information in a given data point can be used towards estimating the underlying parameters. Let's illustrate this point using the formant trajectories from section \ref{section:gamms}. Our data set would have this structure (no grouping/temporal/spatial structure) if each of the measurements (i.e.\ each of the dots in the graphs above) came from a separate vowel. This would mean that we would have to sample 44 different vowel tokens, and only take a single measurement at a random time point for each of them (this would be admittedly a rather weird data set). If we were to, say, double the number of sampled vowels, our confidence in our estimates would also increase -- each new data point would contribute additional information about the underlying parameters.

However, that is not the way our data set is structured. The 44 measurements actually come from only 4 vowels, and measurements taken from a single vowel are not independent of each other. Each of the data points is determined by a combination of factors: the underlying parameters, which trajectory it is in and what values the neighbouring points take on (the last point reflects the fact that there is almost always some autocorrelation in time-series data). Therefore, only part of the information in a given data point can be used towards estimating the underlying parameters -- the rest of the information is actually about other stuff that we might not be interested in. An increase in the size of this data set wouldn't necessarily increase our confidence in our estimates of the underlying parameters. For instance, if we doubled the size of the data set by taking another 11 measurements along the \textit{same} 4 trajectories at different time points, we wouldn't gain much additional information about the parameters that define the underlying curve -- instead, we would gain more information about the individual trajectories themselves. If, on the other hand, we added measurements from 4 additional vowels, that would contribute more information towards the underlying parameters.

When we fit a regression model to our data set without random effects or an appropriate error model, we essentially ignore the grouping/temporal/spatial structure in the data: we pretend that the data set consists of independent measurements, each of them taken from a separate vowel. As a result, we will also erroneously assume that all the information in the individual data points can be used towards estimating the underlying parameters, even though some of the information is actually about other things like the separate trajectories. Since we think that we have more information that we actually do, we become overconfident about our estimates, which leads to anti-conservative results: \textit{p}-values that are biased downwards and overly narrow confidence intervals. When the random effects / error model are correctly specified, the model uses the right amount of information towards estimating the underlying parameters, and the overconfidence issue disappears.

We run into the same problem when we use random intercepts and slopes to fit straight lines to non-linear trajectories. The straight lines correctly `soak up' some of the trajectory-specific information, but not all of it: they can't deal with non-linear dependencies, so those remain in the data. As a result, some of the information that we use towards estimating the underlying parameters will still be about the individual trajectories, so our model remains overconfident. 

The three graphs below illustrate this point by plotting 95\% confidence intervals for three models: one without random effects, one with random intercepts only and one with random smooths.
\begin{knitrout}
\definecolor{shadecolor}{rgb}{0.969, 0.969, 0.969}\color{fgcolor}

{\centering \includegraphics[width=0.495\textwidth]{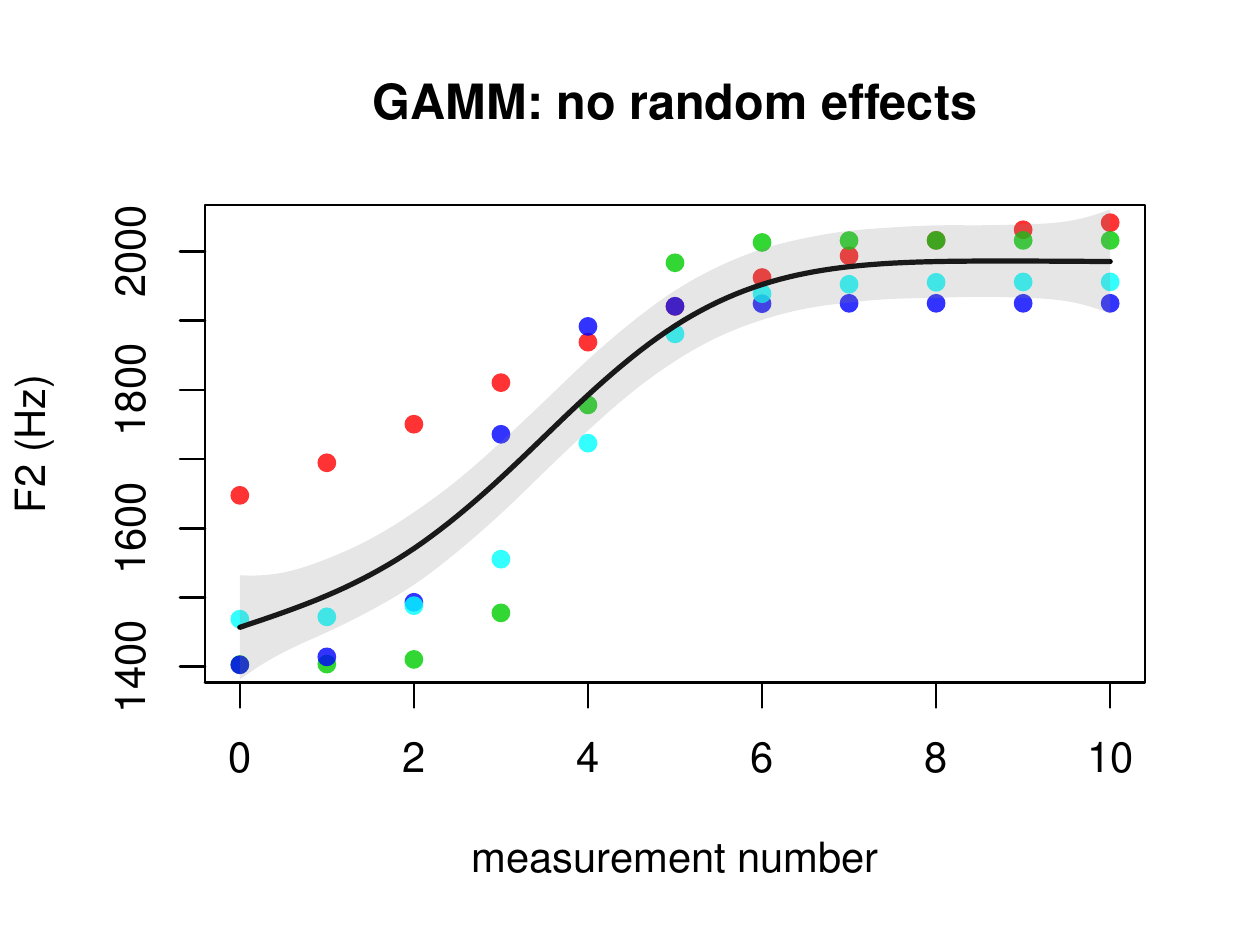} 
\includegraphics[width=0.495\textwidth]{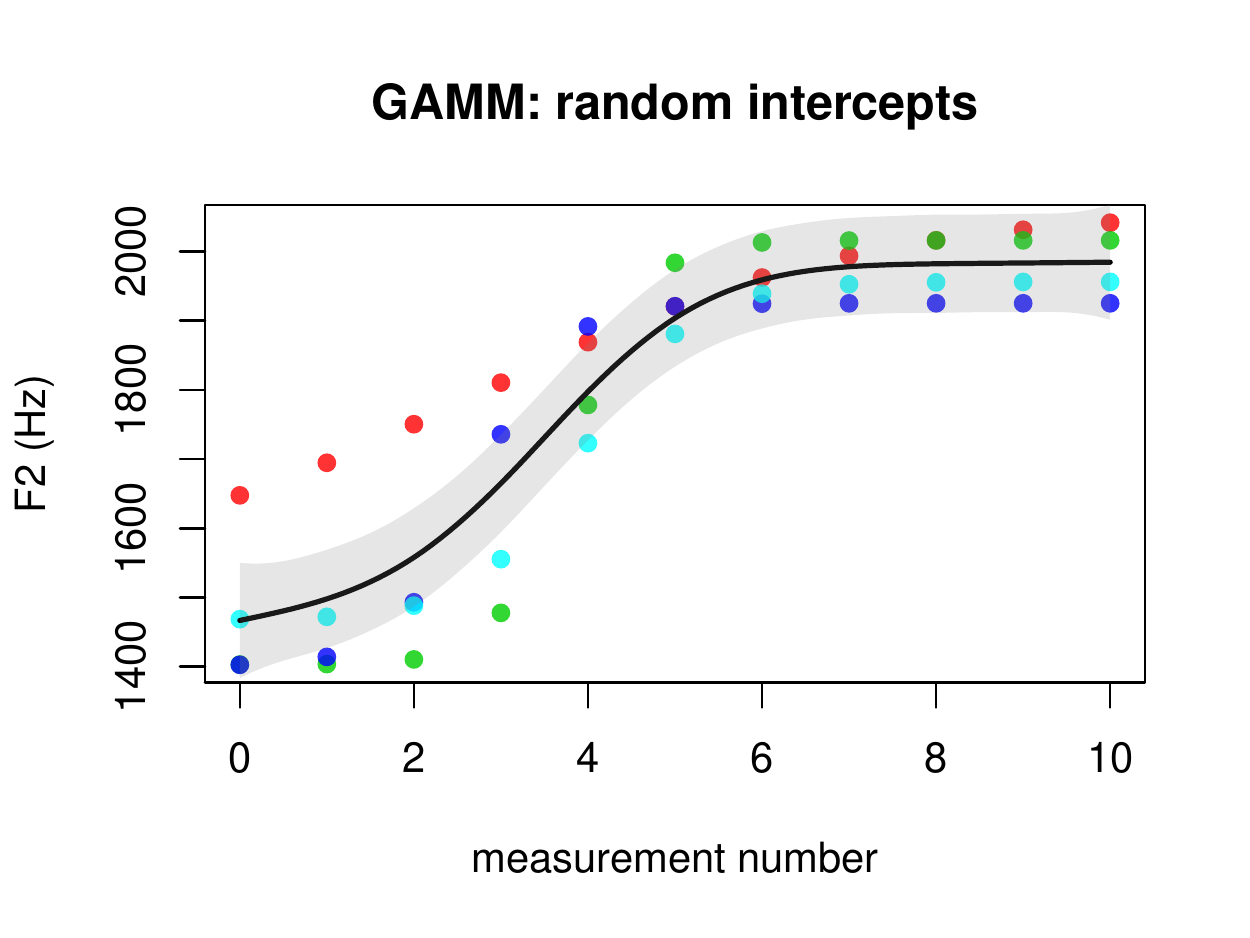} 
\includegraphics[width=0.495\textwidth]{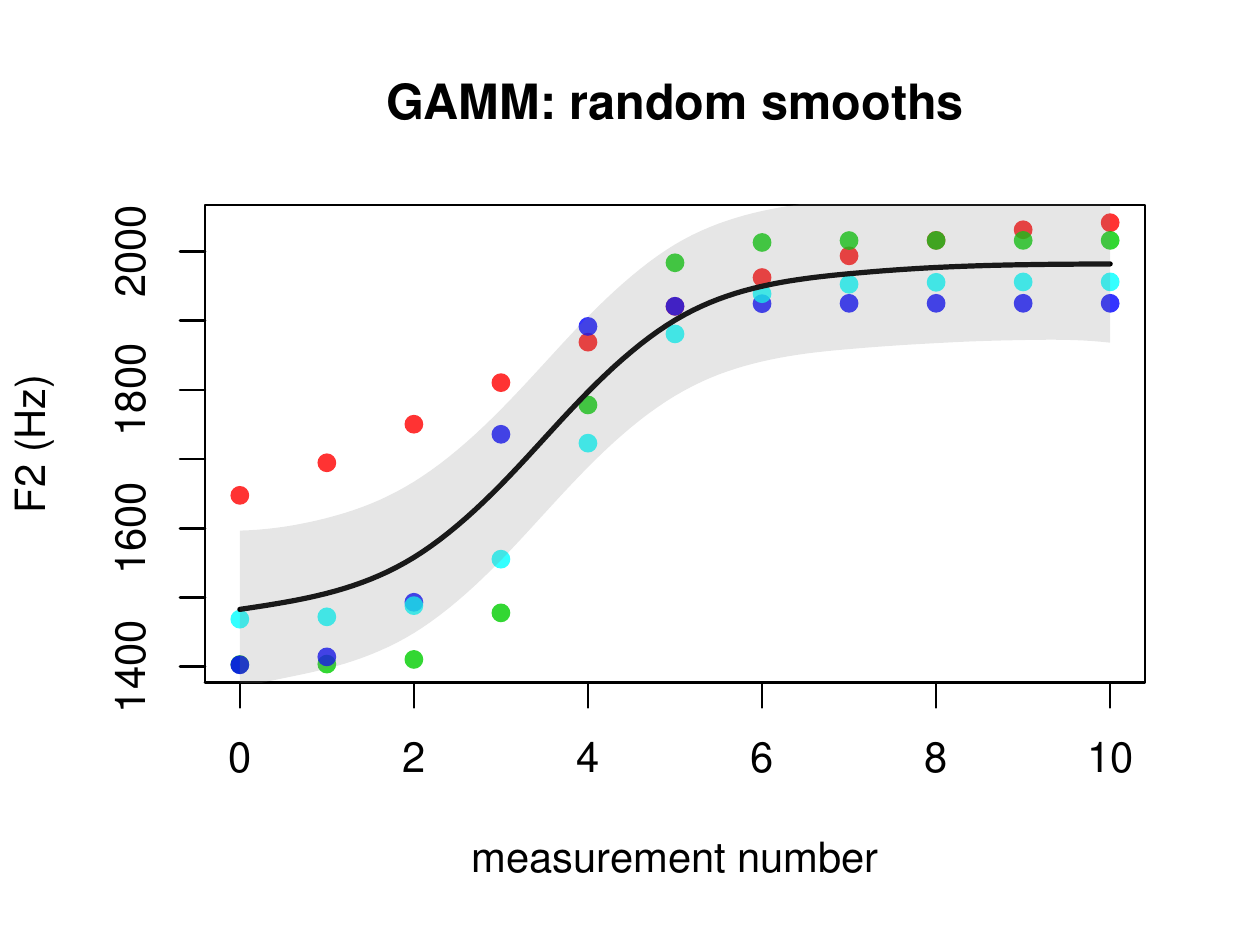} 

}

\end{knitrout}
\noindent The confidence interval becomes slightly wider when we move from the no random effects model to the random intercepts one, and substantially wider for the random smooths model. Although the confidence interval for the final model may seem too wide, it is probably more accurate than the other two. After all, we need to bear in mind that this estimate is really only based on 4 vowels. For comparison, how confident would you be about a group mean based on 4 measurements?

\subsection{Different smooth classes}

Now that we know what GAMMs are, how they are different from GAMs and why random smooths / error models are necessary, it is time to introduce the concept of different smooth classes. We've actually already seen a range of different smooth classes in action, but only one of these has been mentioned explicitly: cubic regression splines. In what follows, we'll look at what distinguishes smooth classes from each other and consider a few examples.

Smooth classes are mainly defined by the basis functions used to generate the smoothers (and also by the type of \textit{smoothing penalty} applied to them, but we won't discuss this). The graphs on page \pageref{smooths} show the basis functions for a smooth that belongs to the class of cubic regression splines. The models below exemplify two further smooth classes: thin plate regression splines (\verb+bs="tp"+) and P-splines (\verb+bs="ps"+; P-splines is short for `penalised B-splines' or `penalised basis splines').

\begin{knitrout}\small
\definecolor{shadecolor}{rgb}{0.969, 0.969, 0.969}\color{fgcolor}\begin{kframe}
\begin{alltt}
\hlstd{demo.gam.tp} \hlkwb{<-} \hlkwd{bam}\hlstd{(f2} \hlopt{~} \hlkwd{s}\hlstd{(measurement.no,} \hlkwc{bs} \hlstd{=} \hlstr{"tp"}\hlstd{),} \hlkwc{data} \hlstd{= traj)}
\hlstd{demo.gam.ps} \hlkwb{<-} \hlkwd{bam}\hlstd{(f2} \hlopt{~} \hlkwd{s}\hlstd{(measurement.no,} \hlkwc{bs} \hlstd{=} \hlstr{"ps"}\hlstd{),} \hlkwc{data} \hlstd{= traj)}
\end{alltt}
\end{kframe}
\end{knitrout}

\noindent While the exact differences among smooth classes are not so important for us, it is instructive to compare their basis functions and the fitted curves. The figures below show the basis functions for thin plate regression splines and P-splines before and after multiplication by the model coefficients, and should be compared with the cubic regression splines on page \pageref{smooths}.

\begin{knitrout}
\definecolor{shadecolor}{rgb}{0.969, 0.969, 0.969}\color{fgcolor}

{\centering \includegraphics[width=0.495\textwidth]{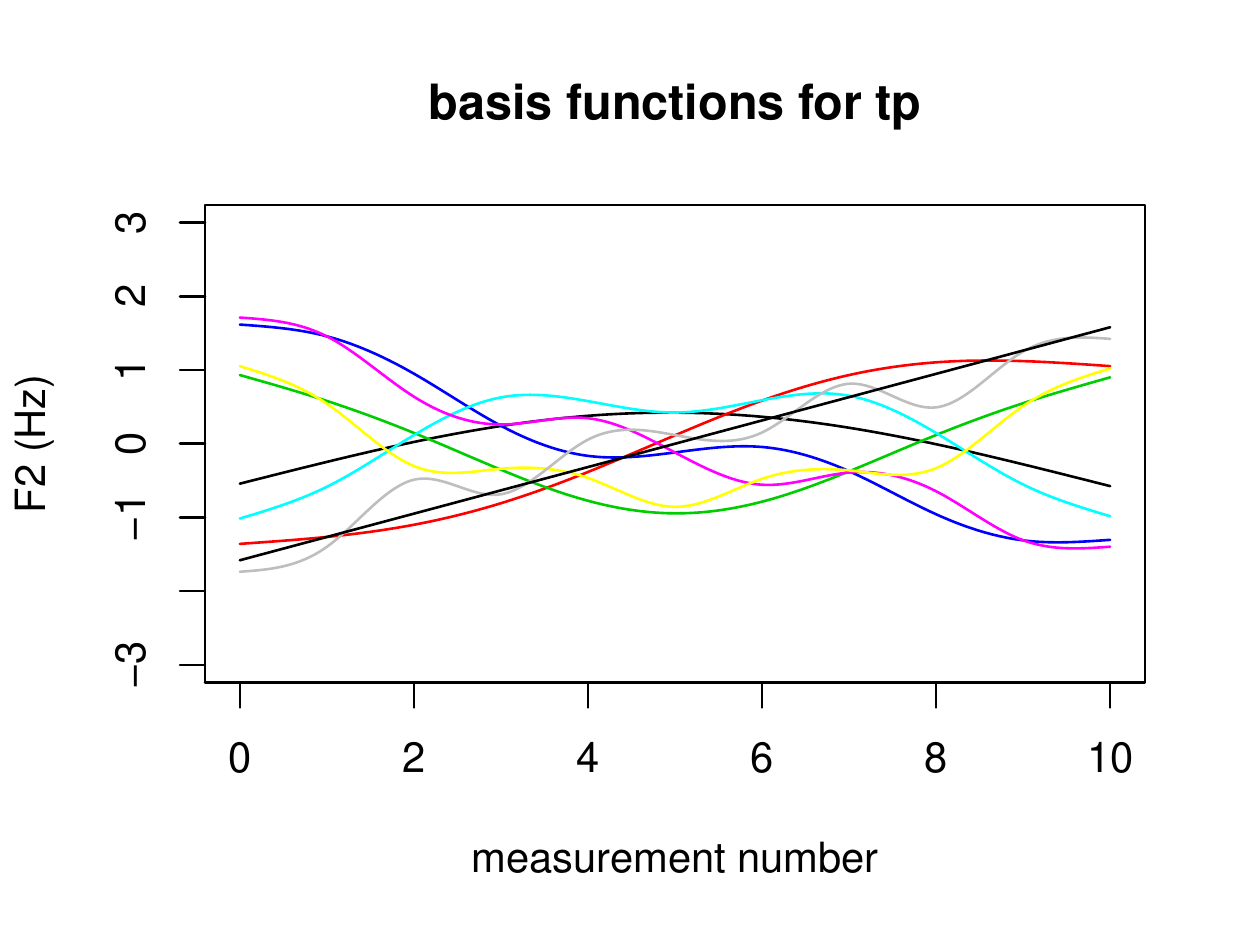} 
\includegraphics[width=0.495\textwidth]{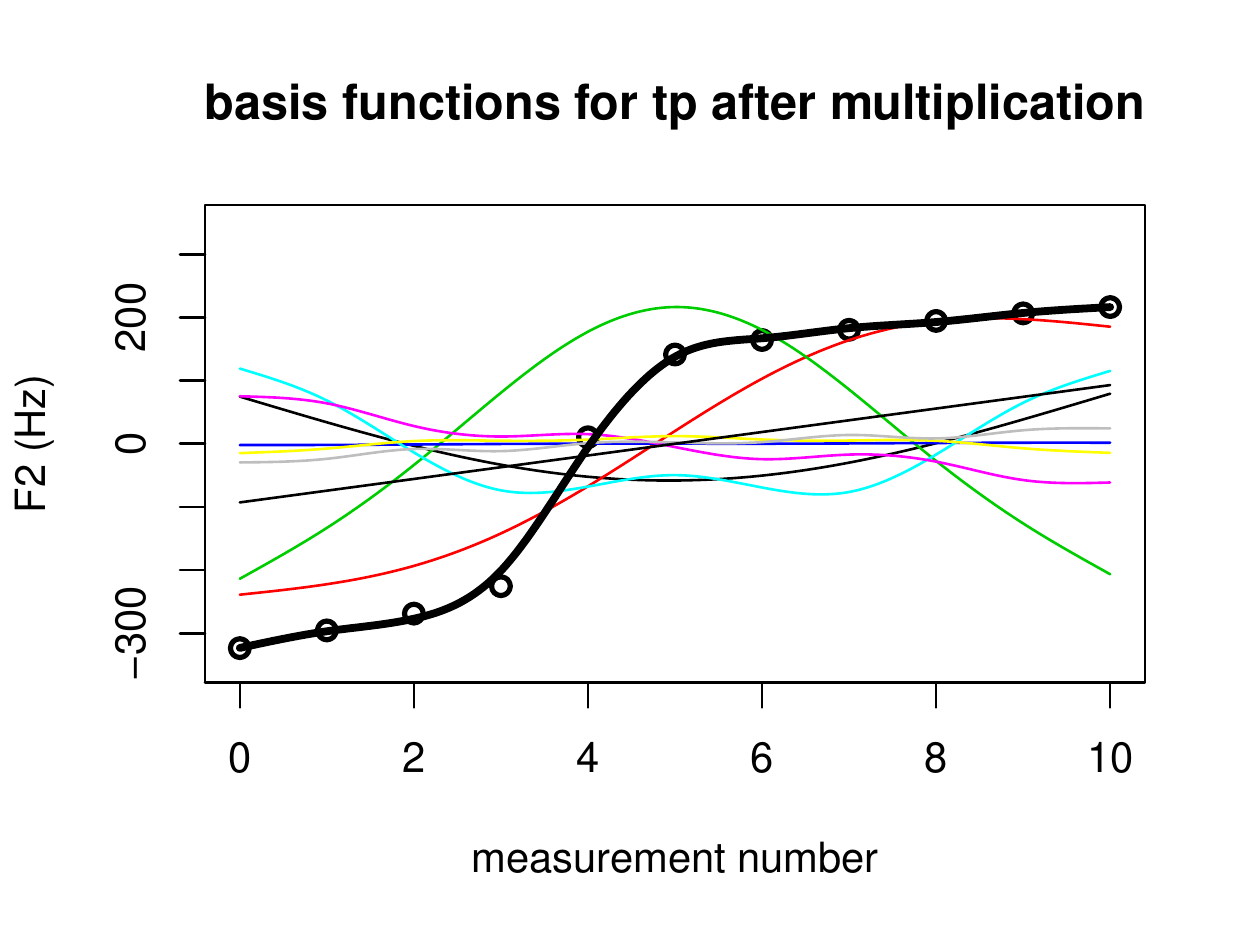} 
\includegraphics[width=0.495\textwidth]{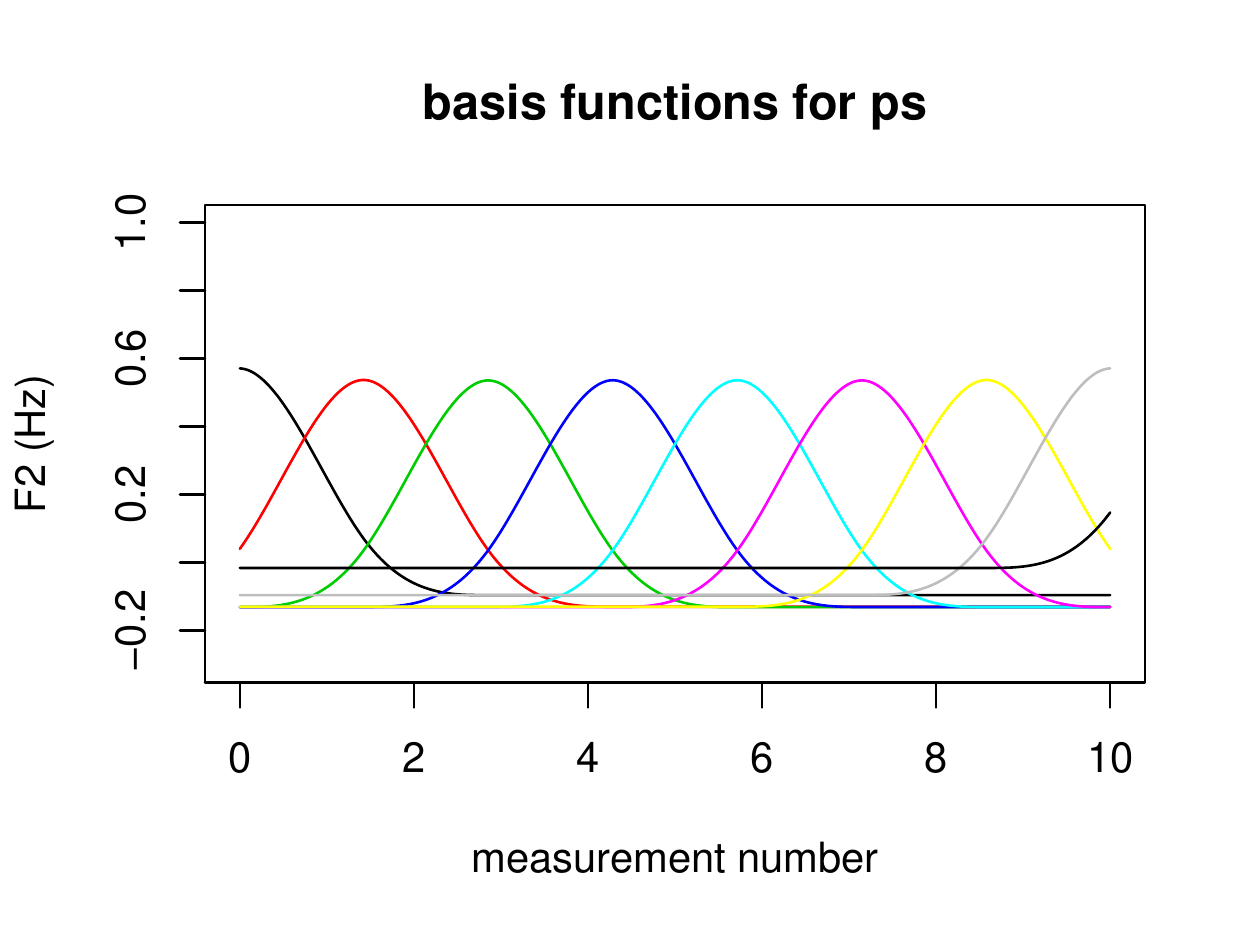} 
\includegraphics[width=0.495\textwidth]{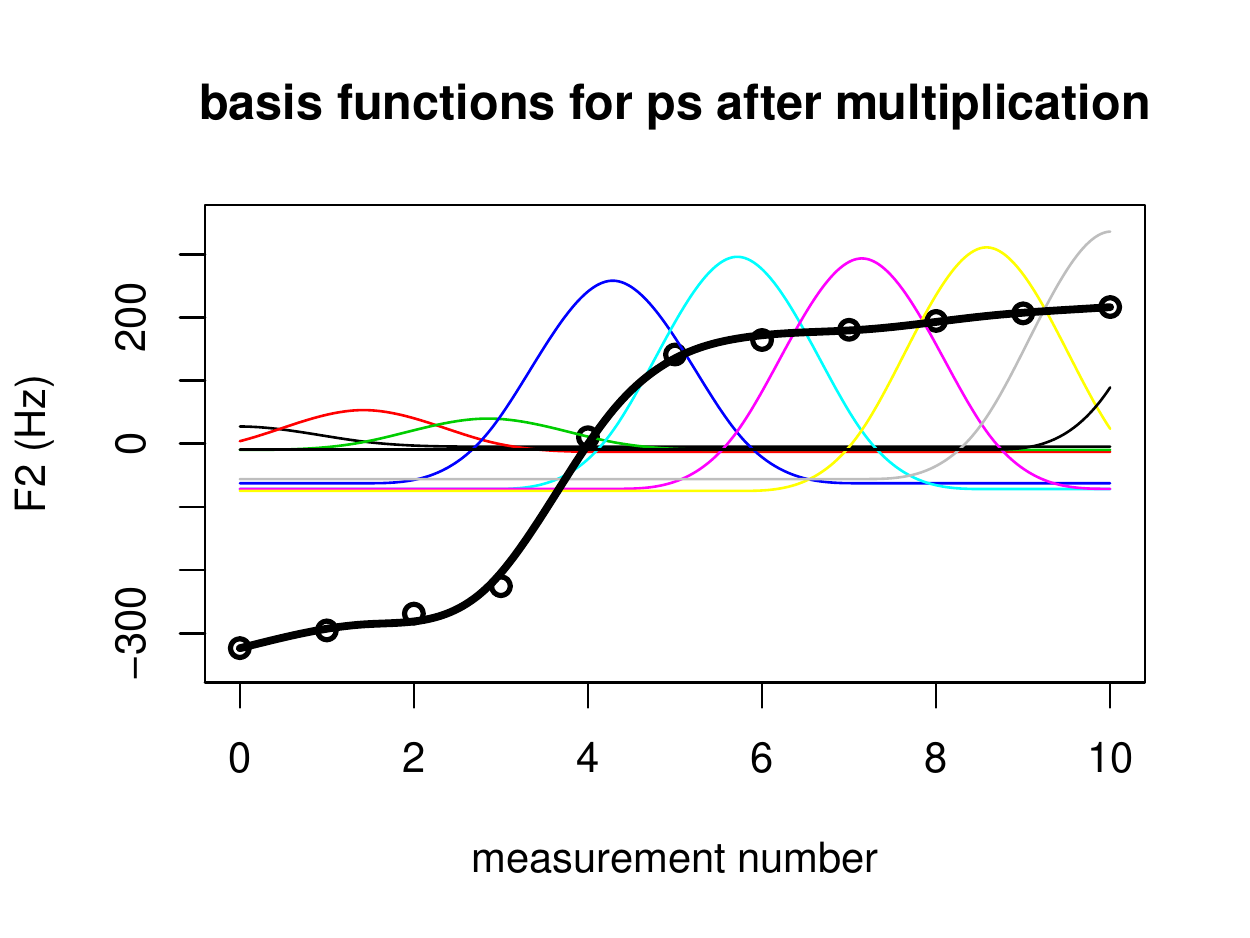} 

}

\end{knitrout}

\noindent The basis functions can be very different (compare e.g.\ \verb+tp+ vs.\ \verb+ps+), but the overall smooths are quite similar across the three models. P-splines seem to provide a slightly (though only very slightly!) worse fit in this case, while cubic and thin plate regression splines do equally well. The default smooth for the {\tt bam()} function from the {\tt mgcv} package is the thin plate regression spline.

The smooths that we've looked at so far are all univariate smooths: they fit a smooth line to the outcome variable as a function of a single predictor. However, it is possible to specify multivariate smooths as well, and certain smooth classes are capable of representing such smooths. For instance, one might want to look at how vowel duration affects the shape of F2 trajectories for a given vowel. One way to do this is to specify a bivariate smooth, where one of the predictor variables is \verb+measurement.no+ (where the measurement was taken along the trajectory), and the other one \verb+duration+ (the overall duration of the vowel in seconds). The fitted bivariate smooth will be a two-dimensional surface, which essentially consists of trajectory shapes that vary smoothly as a function of overall duration. We'll see examples of multivariate smooths in section \ref{sec:tutorial}.

One final note about smooth classes. GAMMs can use smooths to represent not only fitted curves and surfaces, but also random intercepts and slopes. For instance, the linear mixed model with intercepts and slopes on page \pageref{lmers} can be specified in two different ways: using the traditional mixed model specification from the \verb+lme4+ package \citep{lme4}, or using a GAMM model specification.\footnote{The random effect specification in the LMER model is {\tt(1 + measurement.no || vowel)} instead of the more typical {\tt(1 + measurement.no | vowel)}, which results in a model where the variance components for the random intercept and slope are estimated, but their correlation isn't. This is because linear mixed models specified using {\tt bam()} do not include correlations between random slopes and intercepts under the same grouping factor.}

\begin{knitrout}\small
\definecolor{shadecolor}{rgb}{0.969, 0.969, 0.969}\color{fgcolor}\begin{kframe}
\begin{alltt}
\hlstd{demo.slope.lmer} \hlkwb{<-} \hlkwd{lmer}\hlstd{(f2} \hlopt{~} \hlstd{measurement.no} \hlopt{+}
                             \hlstd{(}\hlnum{1} \hlopt{+} \hlstd{measurement.no} \hlopt{||} \hlstd{vowel),}
                        \hlkwc{data}\hlstd{=traj.random)}

\hlstd{demo.slope.gamm} \hlkwb{<-} \hlkwd{bam}\hlstd{(f2} \hlopt{~} \hlstd{measurement.no} \hlopt{+}
                            \hlkwd{s}\hlstd{(vowel,} \hlkwc{bs}\hlstd{=}\hlstr{"re"}\hlstd{)} \hlopt{+}
                            \hlkwd{s}\hlstd{(vowel, measurement.no,} \hlkwc{bs}\hlstd{=}\hlstr{"re"}\hlstd{),}
                       \hlkwc{data}\hlstd{=traj.random)}
\end{alltt}
\end{kframe}
\end{knitrout}

\noindent These formulae actually specify the same model. Random intercepts and slopes are represented by the random effects (\verb+bs="re"+) smooth class in GAMMs, and they behave in much the same way as random effects in linear mixed effects models. Random smooths are a different type of construct, and they do not have a straightforward equivalent in linear models. The smooth class for random smooths is called \textit{factor smooth interactions} (\verb+bs="fs"+). Its use will be illustrated later in section \ref{sec:tutorial}.

\subsection{Significance testing using GAMMs}

\subsubsection{Methods for significance testing with GAMMs}\label{sec:sigtesting_methods}

The models and data discussed in the previous section are useful in that they illustrate some of the basic concepts of GAMMs, but they are also a bit weird. There is only a single group of trajectories, and the shape of these trajectories does not vary as a function of any other predictors (although the individual F2 measurements do, of course, vary as a function of time). This type of situation does not typically arise in real examples: dynamic analyses of linguistic data are usually conducted with the goal of testing whether a given set of predictors has a significant effect on the trajectories under investigation. For instance, one might look at whether the shape of F2 trajectories is affected by vowel duration, whether pitch contours are different across questions and statements, or whether a diachronic change in spectral centre of gravity for a given fricative follows different patterns in different communities. While significance testing is relatively straightforward for linear models (though less so for linear mixed models), GAMMs offer a number of different ways of testing for significance. In this section, we briefly review these methods. It should be emphasised that they are not all equally appropriate, and some of them are, in fact, seriously flawed. The next section identifies potential issues with these methods and outlines a few recommendations about how they should be used.

We start with a simple scenario. Let's say we have two words sharing the same diphthong, and we suspect that the realisation of the diphthong differs between the two words. We'll refer to the words as \textit{A} and \textit{B}. We collect dynamic F2 measurements for 50 tokens of each word. The two sets of 50 trajectories are shown below:
\begin{knitrout}
\definecolor{shadecolor}{rgb}{0.969, 0.969, 0.969}\color{fgcolor}

{\centering \includegraphics[width=0.495\textwidth]{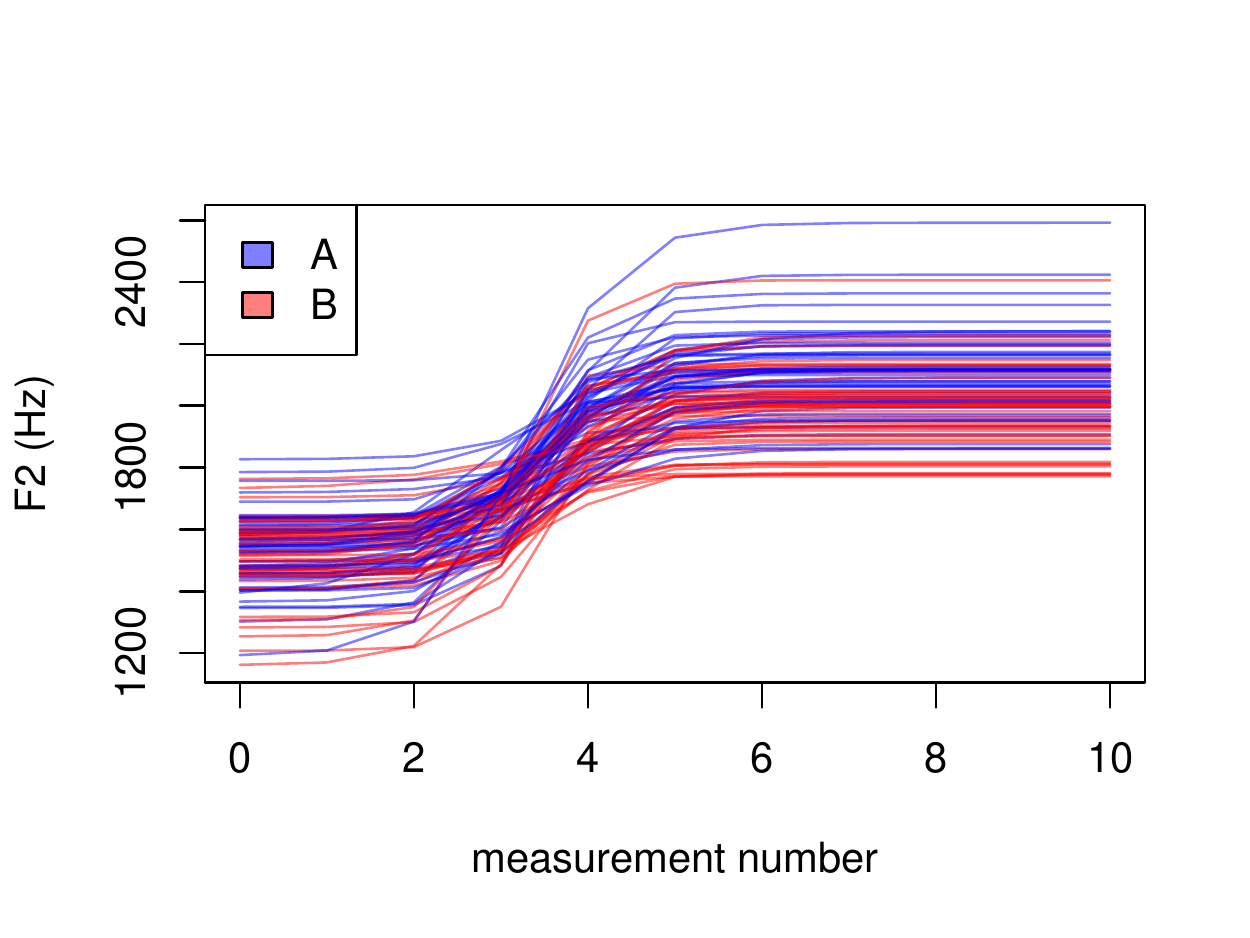} 

}

\end{knitrout}

\noindent We use a GAMM of the following structure to test for significant differences between the two words:

\begin{knitrout}\footnotesize
\definecolor{shadecolor}{rgb}{0.969, 0.969, 0.969}\color{fgcolor}\begin{kframe}
\begin{alltt}
\hlstd{demo.w.gamm} \hlkwb{<-} \hlkwd{bam}\hlstd{(f2} \hlopt{~} \hlstd{word} \hlopt{+}
                        \hlkwd{s}\hlstd{(measurement.no)} \hlopt{+}
                        \hlkwd{s}\hlstd{(measurement.no,} \hlkwc{by}\hlstd{=word)} \hlopt{+}
                        \hlkwd{s}\hlstd{(measurement.no, traj,} \hlkwc{bs}\hlstd{=}\hlstr{"fs"}\hlstd{,} \hlkwc{m}\hlstd{=}\hlnum{1}\hlstd{),}
                   \hlkwc{data}\hlstd{=dat.words,} \hlkwc{method}\hlstd{=}\hlstr{"ML"}\hlstd{)}
\end{alltt}
\end{kframe}
\end{knitrout}

\noindent Let's go through this model specification quickly. The first predictor is a parametric term that captures overall differences in the height of the trajectories as a function of the word they come from. The second predictor, \verb+s(measurement.no)+, corresponds to a single smooth fit at the reference value of the categorical predictor \verb+word+ (i.e.\ \verb+A+). The third predictor is a so-called \textit{difference smooth} that captures the difference between the trajectories for \verb+A+ and \verb+B+. We will discuss difference smooths in more detail later. The last predictor corresponds to random smooths by trajectory (i.e.\ a separate smooth for each of the 50 trajectories). Again, we'll say more about these later.

Confusingly, there are at least \textit{six} different ways of testing whether the difference between \textit{A} and \textit{B} is significant (and probably more). Four of these are available as part of standard model summaries, and the other two can be performed by plotting confidence intervals. 

Let's start with the model summary-based methods. Here is the model summary for \verb+demo.w.gamm+:\newpage

\begin{knitrout}\footnotesize
\definecolor{shadecolor}{rgb}{0.969, 0.969, 0.969}\color{fgcolor}\begin{kframe}
\begin{alltt}
\hlkwd{summary}\hlstd{(demo.w.gamm)}
\end{alltt}
\begin{verbatim}
## 
## Family: gaussian 
## Link function: identity 
## 
## Formula:
## f2 ~ word + s(measurement.no) + s(measurement.no, by = word) + 
##     s(measurement.no, traj, bs = "fs", m = 1)
## 
## Parametric coefficients:
##             Estimate Std. Error t value Pr(>|t|)    
## (Intercept)  1898.88      12.22 155.355  < 2e-16 ***
## wordB         -80.89      17.12  -4.725 3.58e-06 ***
## ---
## Signif. codes:  0 '***' 0.001 '**' 0.01 '*' 0.05 '.' 0.1 ' ' 1
## 
## Approximate significance of smooth terms:
##                             edf  Ref.df       F p-value    
## s(measurement.no)         8.936   8.953 275.314 < 2e-16 ***
## s(measurement.no):wordB   3.378   3.618   4.566 0.00272 ** 
## s(measurement.no,traj)  793.816 898.000  81.655 < 2e-16 ***
## ---
## Signif. codes:  0 '***' 0.001 '**' 0.01 '*' 0.05 '.' 0.1 ' ' 1
## 
## R-sq.(adj) =  0.997   Deviance explained = 99.9%
## -ML = 5908.6  Scale est. = 240.07    n = 1100
\end{verbatim}
\end{kframe}
\end{knitrout}

\noindent The initial section up to the first table should be familiar from other types of regression models. The \verb+Parametric coefficients+ table shows all non-smooth terms, that is, the intercept and the categorical predictor \verb+word+. The next table ({\tt Approximate significance of smooth terms}) summarises the smooth terms: the reference smooth, the difference smooth and the random smooths. Both the parametric and the smooth tables show $p$-values for the terms, although they are based on different tests: $t$-tests for parametric terms and `approximate' $F$-tests for smooth terms. \citet[191]{wood06} warns that the approximate $p$-values values for smooth terms should be taken with a pinch of salt: they can be anticonservative in certain cases.

So which $p$-values should we be looking at? First, we can look at the $p$-value for the parametric term \verb+word+ (method 1). Assuming an alpha-level of 0.05, this $p$-value is $< \alpha$, which means that there is a significant difference in the overall height of the two trajectories. We can also look at the $p$-value of the difference smooth, which, again, is $< 0.05$ -- that is, there is a significant difference between the shapes of the two trajectories (method 2). Another option is to claim a significant difference between the two trajectories if either the parametric term or the difference smooth (or both) are significant (method 3). The last non-visual option is to set up a nested model which excludes the parametric term and the difference smooth, and compare this to the original model using the \verb+compareML()+ command from the \verb+itsadug+ package, which is actually a form of \verb+anova+ (method 4):

\newpage
\begin{knitrout}\footnotesize
\definecolor{shadecolor}{rgb}{0.969, 0.969, 0.969}\color{fgcolor}\begin{kframe}
\begin{alltt}
\hlstd{demo.w.gamm2} \hlkwb{<-} \hlkwd{bam}\hlstd{(f2} \hlopt{~} \hlkwd{s}\hlstd{(measurement.no)} \hlopt{+}
                         \hlkwd{s}\hlstd{(measurement.no, traj,} \hlkwc{bs}\hlstd{=}\hlstr{"fs"}\hlstd{,} \hlkwc{m}\hlstd{=}\hlnum{1}\hlstd{),}
                    \hlkwc{data}\hlstd{=dat.words,} \hlkwc{method}\hlstd{=}\hlstr{"ML"}\hlstd{)}
\hlkwd{compareML}\hlstd{(demo.w.gamm, demo.w.gamm2,} \hlkwc{print.output}\hlstd{=F)}\hlopt{$}\hlstd{table}
\end{alltt}
\begin{verbatim}
##          Model    Score Edf  Chisq    Df   p.value Sig.
## 1 demo.w.gamm2 5923.394   5                            
## 2  demo.w.gamm 5908.614   8 14.780 3.000 1.707e-06  ***
\end{verbatim}
\end{kframe}
\end{knitrout}

\noindent According to the model comparison, the inclusion of the parametric term and the smooth difference term significantly improves the model fit. Importantly, these methods tell us little about the exact nature of the difference between the two words.

There are two visual methods for significance testing, both of which rely on confidence intervals. First, we can plot the predicted trajectories for both words with corresponding pointwise confidence intervals and check for overlap / lack of overlap at different points (method 5). Second, we can plot the difference smooth itself along with a confidence interval and check whether the confidence interval includes 0 at different points (method 6). One of the advantages of these methods is that they allow us to see where and in what way the trajectories words differ. These methods are illustrated below:

\begin{knitrout}
\definecolor{shadecolor}{rgb}{0.969, 0.969, 0.969}\color{fgcolor}

{\centering \includegraphics[width=0.495\textwidth]{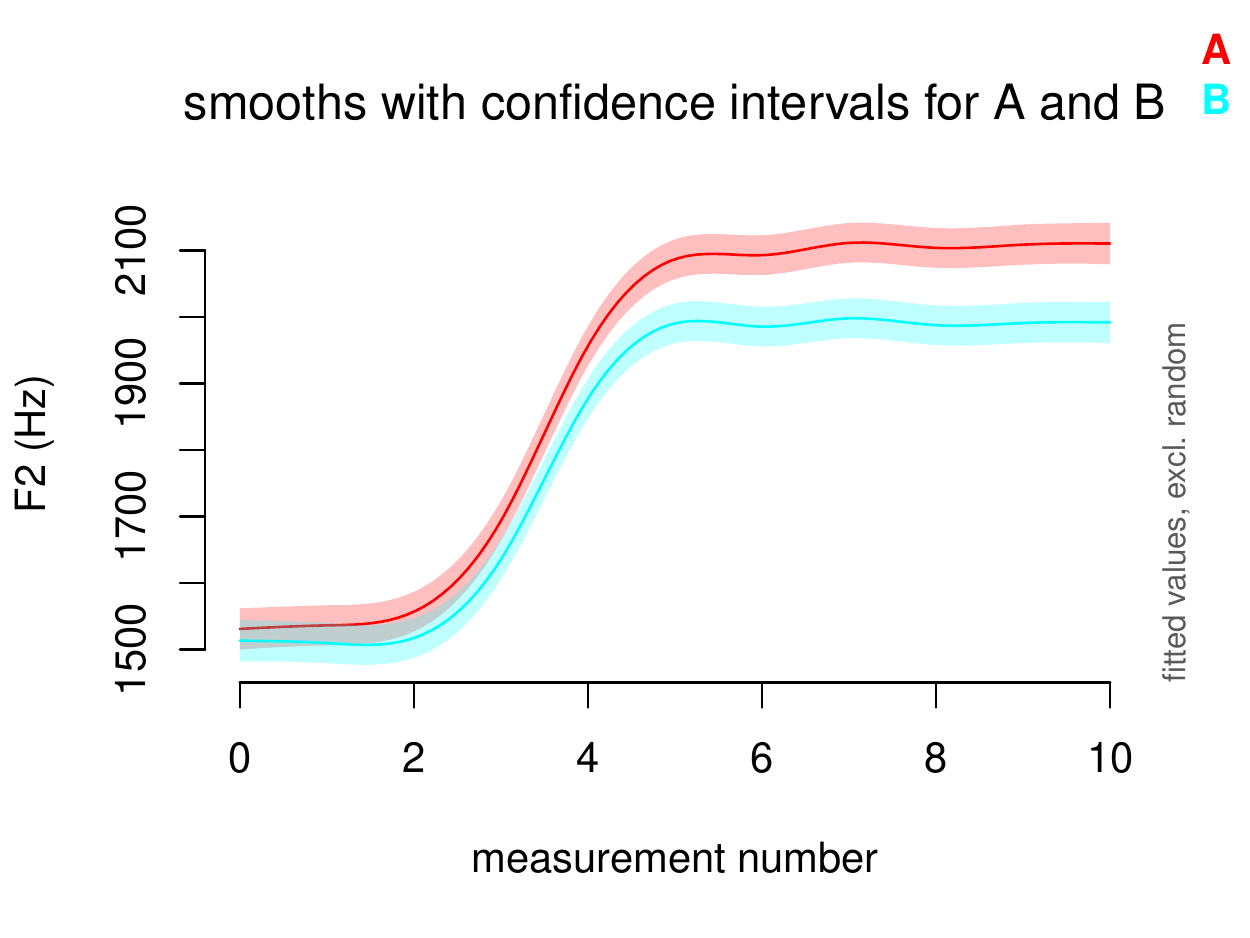} 
\includegraphics[width=0.495\textwidth]{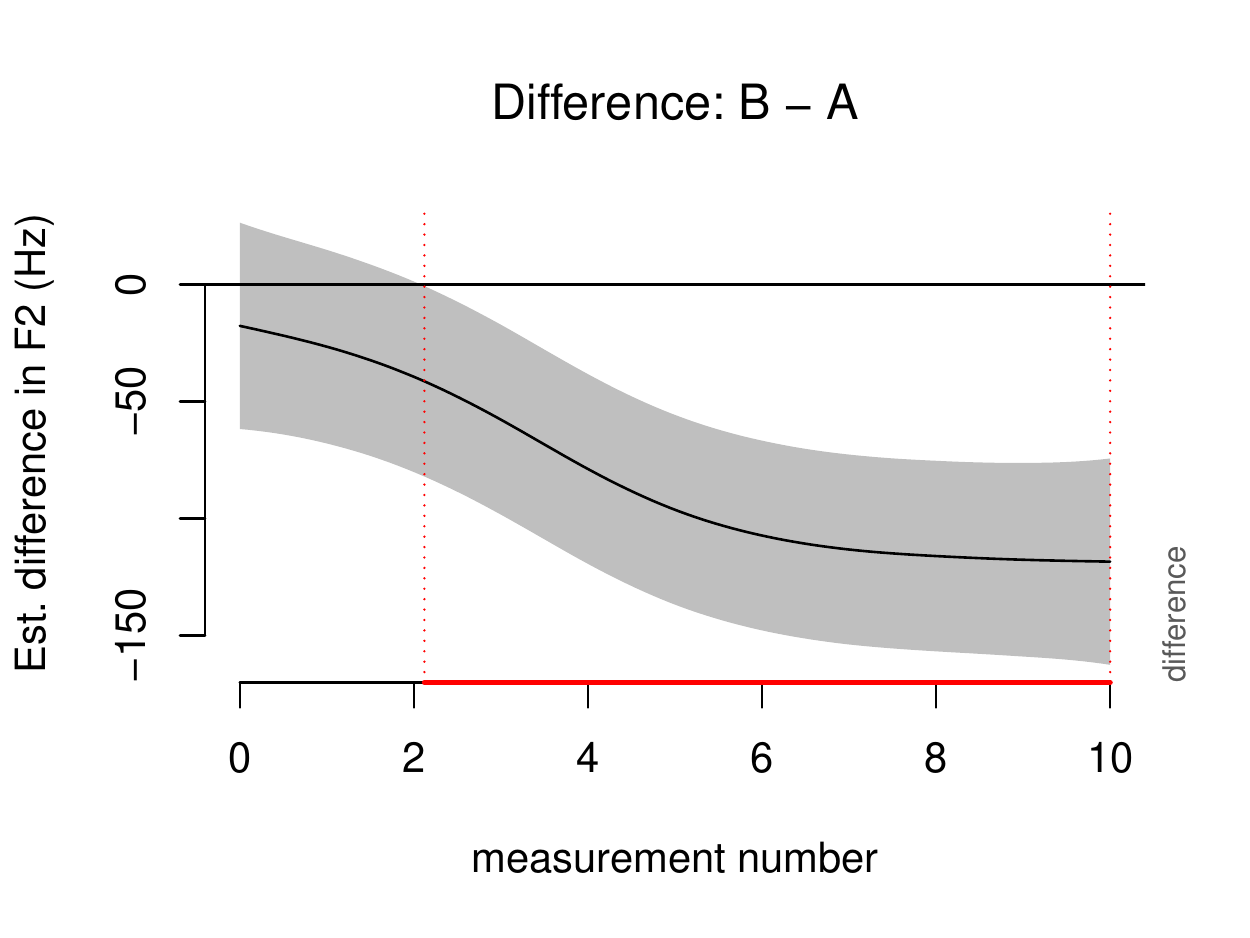} 

}

\end{knitrout}

\noindent Both methods suggest that there is a significant difference between the two sets of trajectories. Moreover, they also reveal that the trajectories only diverge after measurement number 2 (that's the 3rd measurement out of 11). In other words, the trajectories differ significantly, but only between measurement numbers 2 and 10.

As a final note, significance testing with GAMMs can be substantially more complicated when the predictor of interest is a continuous variable. Although the general principles discussed in this section apply to continuous variables as well, their implementation can be a lot trickier due to the potential complexity of smooth interactions and constraints on the software packages used to fit GAMMs. We will go through a few worked examples later in the tutorial.

\subsubsection{Recommendations}\label{sec:recommendations}

In \citet{soskuthy16labphon} and \citeauthor{soskuthy16gam} (in prep), I present simulation-based results that reveal a wide range of variation in the rate of false positives and false negatives across the different methods discussed above. A point-by-point summary of these results is presented below:
\begin{itemize}
  \item Significance testing based on the $t$/$F$-values for the parametric / smooth terms in the model summary is only justified when our predictions are directly about these individual terms. For example, if our prediction is that the average value of a trajectory will be higher in one condition than in another, but we have no predictions about trajectory shapes, we can safely rely on the $p$-value for the parametric term (method 1). Conversely, if our prediction relates to the shapes of the trajectories but does not concern their average value, we can use the $p$-value for the smooth term (method 2).
  \item Although predictions specifically about the parametric / smooth terms do arise occasionally, a lot of the time researchers are interested in overall differences between trajectories regardless of whether those differences are in their average value or their shape. In such cases, it is tempting to declare significance if either the parametric term or the smooth term is significant (method 3). However, this leads to higher-than-nominal false positive rates (just below $0.10$ at $\alpha = 0.05$). Model comparison where both the parametric \textit{and} the smooth terms are excluded in the nested model (method 4) yields close-to-nominal false positive rates. Another way to avoid this issue is through correction for multiple comparisons (two comparisons in this case), though this can lead to substantially diminished power. This introduction relies mainly on method 4, and correction for multiple comparisons won't be discussed in detail.
  \item Both visual methods (5 and 6) suffer from an anti-conservativity issue. If we report significant differences whenever there is \textit{any} point where confidence intervals are non-overlapping (method 5) / the confidence interval for the estimated difference excludes 0 (method 6), the rate of false positives is too high (around $0.12$ in the simulations at $\alpha = 0.05$). The rate of false positives decreases if we require significant differences at more than one point to report a significant overall difference, but this requires an arbitrary decision about the number of points, and can easily lead to results that are too conservative.
  \item Comparison based on whether there is overlap between two confidence intervals (method 5) is inadequate for significance testing, and graphs with separate trajectories (rather than a difference smooth) should only be used for graphical illustration. People tend to misinterpret the meaning of overlapping confidence intervals in such graphs. When the confidence intervals do not overlap, there is indeed a significant difference between the estimated quantities. However, we cannot make any conclusions about the significance of the difference when the confidence intervals do overlap: it may be significant but it may also be non-significant. Difference smooths (method 6) do not suffer from this problem, and should be the preferred option. Note that this issue is independent of the anti-conservativity problem described above.
  \item The simulations also reveal that failing to include (i) an error model to adjust the model output for residual autocorrelation within the trajectories or (ii) random smooths-per-trajectory can lead to catastrophically high false positive rates (up to 0.6 at $\alpha = 0.05$) regardless of what method is used for significance testing. These methods are discussed in more detail below.
\end{itemize}

\noindent Based on these results, the most reliable (i.e.\ least anti-conservative) option for significance testing is to first use an ANOVA (method 4) to see if there is an overall difference between groups of trajectories, and then look at difference smooths (method 6) to identify where the difference lies along the trajectory. Additionally, it is also useful to plot individual smooths (with or without confidence intervals) to provide a visual summary of the main trends in the data set, but these should not be used for significance testing.

\section{A GAMM tutorial}\label{sec:tutorial}

In this tutorial, we will work through two detailed examples. The first of these is based on simulated data, while the second one uses a real data set taken from \citet{stuartsmithetal15}.

\subsection{Analysing a simple simulated data set}

The first data set contains simulated F2 trajectories, and is very similar to the example data set introduced in section \ref{sec:sigtesting_methods}. It contains 50 F2 trajectories, each of them represented by 11 measurements taken at equal intervals (at 0\%, 10\%, 20\%, \ldots, 100\%). The observations in the data set are the individual measurements, which means that there are 550 data points altogether. The variable \verb+measurement.no+ codes the location of individual data points along the trajectory. Each of the trajectories has an ID (a number between 1--50), which is encoded in the column \verb+traj+. The trajectories represent two different words: 25 of them come from word \textit{A} and 25 of them from word \textit{B}. This is encoded by the \verb+word+ variable. The underlying curves that served as the basis of the simulated trajectories overlap at the beginning, but are different by about 100 Hz near the end. There is an additional variable termed \verb+duration+, which stands for overall vowel duration measured in seconds. The simulation was set up so that long vowels have slightly wider trajectories than short vowels. The data set is called \verb+words.50+ and can be downloaded from the github page for this introduction:\vspace{1ex}

\noindent \url{https://github.com/soskuthy/gamm_intro}.
\vspace{1ex}

\noindent Here's a small sample of the data:

\begin{knitrout}\footnotesize
\definecolor{shadecolor}{rgb}{0.969, 0.969, 0.969}\color{fgcolor}\begin{kframe}
\begin{alltt}
\hlkwd{head}\hlstd{(words.50)}
\end{alltt}
\begin{verbatim}
##     traj word measurement.no       f2  duration
## 1 traj.1    A              0 1642.761 0.1378182
## 2 traj.1    A              1 1644.162 0.1378182
## 3 traj.1    A              2 1659.948 0.1378182
## 4 traj.1    A              3 1788.793 0.1378182
## 5 traj.1    A              4 2044.977 0.1378182
## 6 traj.1    A              5 2115.984 0.1378182
\end{verbatim}
\end{kframe}
\end{knitrout}

\noindent Let's import the libraries that we'll be using.

\begin{knitrout}\footnotesize
\definecolor{shadecolor}{rgb}{0.969, 0.969, 0.969}\color{fgcolor}\begin{kframe}
\begin{alltt}
\hlkwd{library}\hlstd{(ggplot2)}
\hlkwd{library}\hlstd{(mgcv)}
\hlkwd{library}\hlstd{(itsadug)}
\hlkwd{source}\hlstd{(}\hlstr{"gamm_hacks.r"}\hlstd{)}
\end{alltt}
\end{kframe}
\end{knitrout}

\noindent First of all, let's create a simple plot to see what the raw data look like. This is good practice regardless of the type of model one is planning to fit, but it is especially useful for GAMMs, where the shape of the trajectories has implications for the model fitting procedure (e.g.\ choosing the number of basis functions). We'll use the package \verb+ggplot2+ to create the plot, as it makes it really easy to show the structure of the data set through the use of colours and other devices. The trajectories representing the two words are shown in separate panels and the overall duration of each trajectory is indicated by shading: longer trajectories are darker (the trajectories are time-normalised, so they all have the same length along the x-axis). Since this tutorial is not about \verb+ggplot2+, we won't discuss the code below.

\begin{knitrout}\footnotesize
\definecolor{shadecolor}{rgb}{0.969, 0.969, 0.969}\color{fgcolor}\begin{kframe}
\begin{alltt}
\hlkwd{ggplot}\hlstd{(words.50,} \hlkwd{aes}\hlstd{(}\hlkwc{x}\hlstd{=measurement.no,} \hlkwc{y}\hlstd{=f2,} \hlkwc{group}\hlstd{=traj,}
                     \hlkwc{alpha}\hlstd{=duration))} \hlopt{+}
  \hlkwd{facet_grid}\hlstd{(}\hlopt{~}\hlstd{word)} \hlopt{+} \hlkwd{geom_line}\hlstd{()}
\end{alltt}
\end{kframe}

{\centering \includegraphics[width=0.990\textwidth]{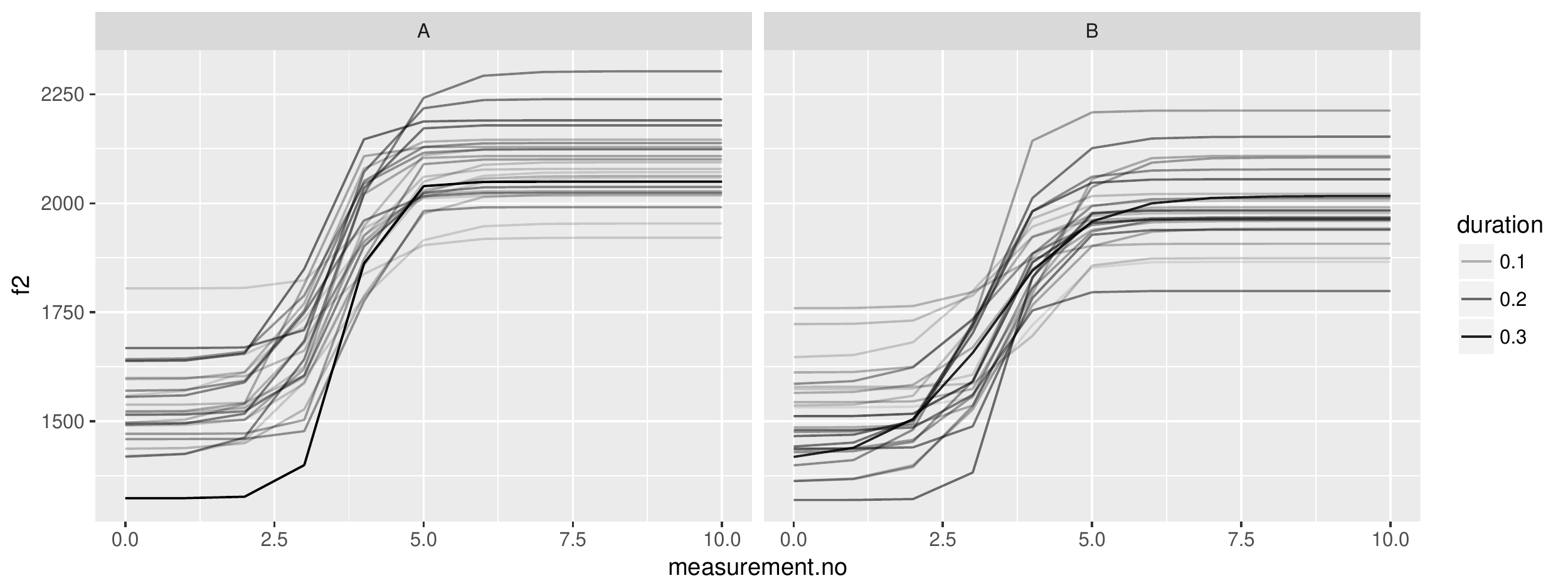} 

}

\end{knitrout}

\noindent So what do we want to capture in our model? First, we want to fit separate smooths to the two trajectories, and we'll want to use model comparison and difference smooths to see whether they are different. We also want to include some type of interaction between duration and the shape of the trajectories. Finally, we want to include (i) random smooths by trajectory and (ii) a residual error model in order to avoid false positives and obtain more accurate estimates of the underlying curves.

Let's start with a very simple model that fits separate smooths to the two words. There are a number of different ways of doing this; we'll look at two of these. First, we can specify a model that simply includes two smooths: one for word \textit{A} and another one for word \textit{B}. Though this is probably the simplest way of fitting this model, it's not necessarily the most useful one. The second method is to fit one smooth to word \textit{A} and then another smooth that represents the \textit{difference} between \textit{A} and \textit{B}. This is identical to the first model in terms of the model fit, but it is easier to interpret the model output when a difference smooth is included, as it tells us directly whether there is a significant difference between the shapes of the two trajectories. Both types of models are shown below along with relevant bits of the model summary. The models use cubic regression splines (\verb+bs="cr"+) with the default number of knots (\verb+k=10+). Using different basis functions does not substantially alter the results. The models are fitted using maximum likelihood estimation (\verb+method="ML"+), which is necessary since we want to perform model comparison between models with different fixed effects. The default method for \verb+bam()+ is fREML or fast restricted maximum likelihood estimation, but this cannot be used for comparing models with different fixed effects.

\begin{knitrout}\footnotesize
\definecolor{shadecolor}{rgb}{0.969, 0.969, 0.969}\color{fgcolor}\begin{kframe}
\begin{alltt}
\hlcom{# model with separate smooths}

\hlstd{words.50}\hlopt{$}\hlstd{word} \hlkwb{<-} \hlkwd{as.factor}\hlstd{(words.50}\hlopt{$}\hlstd{word)}
\hlstd{words.50.gam.sep} \hlkwb{<-} \hlkwd{bam}\hlstd{(f2} \hlopt{~} \hlstd{word} \hlopt{+} \hlkwd{s}\hlstd{(measurement.no,} \hlkwc{by}\hlstd{=word,} \hlkwc{bs}\hlstd{=}\hlstr{"cr"}\hlstd{),}
                        \hlkwc{data}\hlstd{=words.50,} \hlkwc{method}\hlstd{=}\hlstr{"ML"}\hlstd{)}
\hlkwd{summary.coefs}\hlstd{(words.50.gam.sep)}
\end{alltt}
\begin{verbatim}
## Parametric coefficients:
##             Estimate Std. Error t value Pr(>|t|)    
## (Intercept) 1885.149      5.755 327.588  < 2e-16 ***
## wordB        -67.372      8.138  -8.278 1.01e-15 ***
## 
## Approximate significance of smooth terms:
##                           edf Ref.df     F p-value    
## s(measurement.no):wordA 7.311  8.258 213.0  <2e-16 ***
## s(measurement.no):wordB 6.832  7.871 169.4  <2e-16 ***
\end{verbatim}
\begin{alltt}
\hlcom{# model with smooth for A & difference smooth}

\hlstd{words.50}\hlopt{$}\hlstd{word.ord} \hlkwb{<-} \hlkwd{as.ordered}\hlstd{(words.50}\hlopt{$}\hlstd{word)}
\hlkwd{contrasts}\hlstd{(words.50}\hlopt{$}\hlstd{word.ord)} \hlkwb{<-} \hlstr{"contr.treatment"}
\hlstd{words.50.gam.diff} \hlkwb{<-} \hlkwd{bam}\hlstd{(f2} \hlopt{~} \hlstd{word.ord} \hlopt{+} \hlkwd{s}\hlstd{(measurement.no,} \hlkwc{bs}\hlstd{=}\hlstr{"cr"}\hlstd{)} \hlopt{+}
                              \hlkwd{s}\hlstd{(measurement.no,} \hlkwc{by}\hlstd{=word.ord,} \hlkwc{bs}\hlstd{=}\hlstr{"cr"}\hlstd{),}
                         \hlkwc{data}\hlstd{=words.50,} \hlkwc{method}\hlstd{=}\hlstr{"ML"}\hlstd{)}
\hlkwd{summary.coefs}\hlstd{(words.50.gam.diff)}
\end{alltt}
\begin{verbatim}
## Parametric coefficients:
##             Estimate Std. Error t value Pr(>|t|)    
## (Intercept) 1885.149      5.735 328.713  < 2e-16 ***
## word.ordB    -67.372      8.110  -8.307 7.99e-16 ***
## 
## Approximate significance of smooth terms:
##                               edf Ref.df      F p-value    
## s(measurement.no)           7.805  8.594 234.40 < 2e-16 ***
## s(measurement.no):word.ordB 1.040  1.079  10.84 0.00098 ***
\end{verbatim}
\end{kframe}
\end{knitrout}

\noindent \verb+summary.coefs()+ (a function from \verb+gamm_hacks.r+) is used to save space by excluding parts of the model summary. Readers are encouraged to use the standard \verb+summary()+ function with GAMMs, as it includes some additional useful information about the model fit.

Fitting the model with two separate smooths is easy: we need to include a parametric term that captures overall differences between the trajectories,%
\footnote{This parametric term needs to be included in both models. Otherwise, the models could not capture overall differences in F2 and would (incorrectly) force both fitted smooths to have the same average value over the trajectory.} and add the option \verb+by=word+ to the smooth term (which asks for separate smooths to be fit at each level of the factor \verb+word+).

Models with difference smooths require a different approach. First, the categorical grouping variable for the words needs to be converted to a so-called `ordered factor' using \verb+as.ordered()+, and the contrasts for this ordered factor need to be set to \verb+"contr.treatment"+ (this latter step is important, or otherwise the model estimates will be off). Second, the model formula needs to include (i) a parametric term for \verb+word.ord+, (ii) a smooth over \verb+measurement.no+ without any grouping specification and (iii) a smooth over \verb+measurement.no+ with the grouping specification \verb+by=word.ord+. 

In the first model summary, the separate smooths simply represent the two different words. The significance values in the model summary refer to the individual terms: they suggest that both curves are significantly different from 0 (i.e.\ not simply flat lines). However, they do not tell us anything about the difference between the two terms. They could both be significant even if the two underlying curves were exactly the same. In the second model summary, the term \verb+s(measurement.no)+ represents the \textit{reference smooth}, that is, a curve fit to trajectories at the reference level of the ordered factor \verb+word.ord+ (i.e.\ \textit{A}). The term \verb+s(measurement.no):word.ordB+ represents the difference smooth, that is, the difference between the trajectories for \textit{A} and \textit{B}. 

The fact that the difference smooth and the parametric term for \verb+word.ord+ are both significant suggests that the trajectories for the two words are indeed different. In order to confirm this, we can perform model comparison using the function \verb+compareML()+. Following the recommendations in section \ref{sec:recommendations}, the nested model excludes both the parametric term and the smooth difference term (i.e.\ all terms that relate to the effect of \verb+word.ord+).

\begin{knitrout}\footnotesize
\definecolor{shadecolor}{rgb}{0.969, 0.969, 0.969}\color{fgcolor}\begin{kframe}
\begin{alltt}
\hlcom{# fitting a nested model without the difference smooth}
\hlstd{words.50.gam.diff.0} \hlkwb{<-} \hlkwd{bam}\hlstd{(f2} \hlopt{~} \hlkwd{s}\hlstd{(measurement.no,} \hlkwc{bs}\hlstd{=}\hlstr{"cr"}\hlstd{),}
                           \hlkwc{data}\hlstd{=words.50,} \hlkwc{method}\hlstd{=}\hlstr{"ML"}\hlstd{)}

\hlcom{# model comparison using the compareML() function from itsadug}
\hlcom{# this is very similar to the anova() function, but better suited}
\hlcom{# to models fitted using bam(); some parts of the output are suppressed}
\hlkwd{compareML}\hlstd{(words.50.gam.diff, words.50.gam.diff.0,} \hlkwc{print.output}\hlstd{=F)}\hlopt{$}\hlstd{table}
\end{alltt}
\begin{verbatim}
##                 Model    Score Edf  Chisq    Df   p.value Sig.
## 1 words.50.gam.diff.0 3334.300   3                            
## 2   words.50.gam.diff 3296.538   6 37.762 3.000 2.798e-16  ***
\end{verbatim}
\end{kframe}
\end{knitrout}

\noindent The model comparison suggests that the inclusion of the difference smooth improves the model fit significantly.

Let's create a plot of the model predictions and the difference smooth. These can be generated using the \verb+plot_smooth+ and the \verb+plot_diff+ functions from \verb+itsadug+.

\begin{knitrout}\footnotesize
\definecolor{shadecolor}{rgb}{0.969, 0.969, 0.969}\color{fgcolor}\begin{kframe}
\begin{alltt}
\hlkwd{plot_smooth}\hlstd{(words.50.gam.diff,} \hlkwc{view}\hlstd{=}\hlstr{"measurement.no"}\hlstd{,}
            \hlkwc{plot_all}\hlstd{=}\hlstr{"word.ord"}\hlstd{,} \hlkwc{rug}\hlstd{=F)}
\hlkwd{plot_diff}\hlstd{(words.50.gam.diff,} \hlkwc{view}\hlstd{=}\hlstr{"measurement.no"}\hlstd{,}
          \hlkwc{comp}\hlstd{=}\hlkwd{list}\hlstd{(}\hlkwc{word.ord}\hlstd{=}\hlkwd{c}\hlstd{(}\hlstr{"B"}\hlstd{,}\hlstr{"A"}\hlstd{)))}
\end{alltt}
\end{kframe}

{\centering \includegraphics[width=0.495\textwidth]{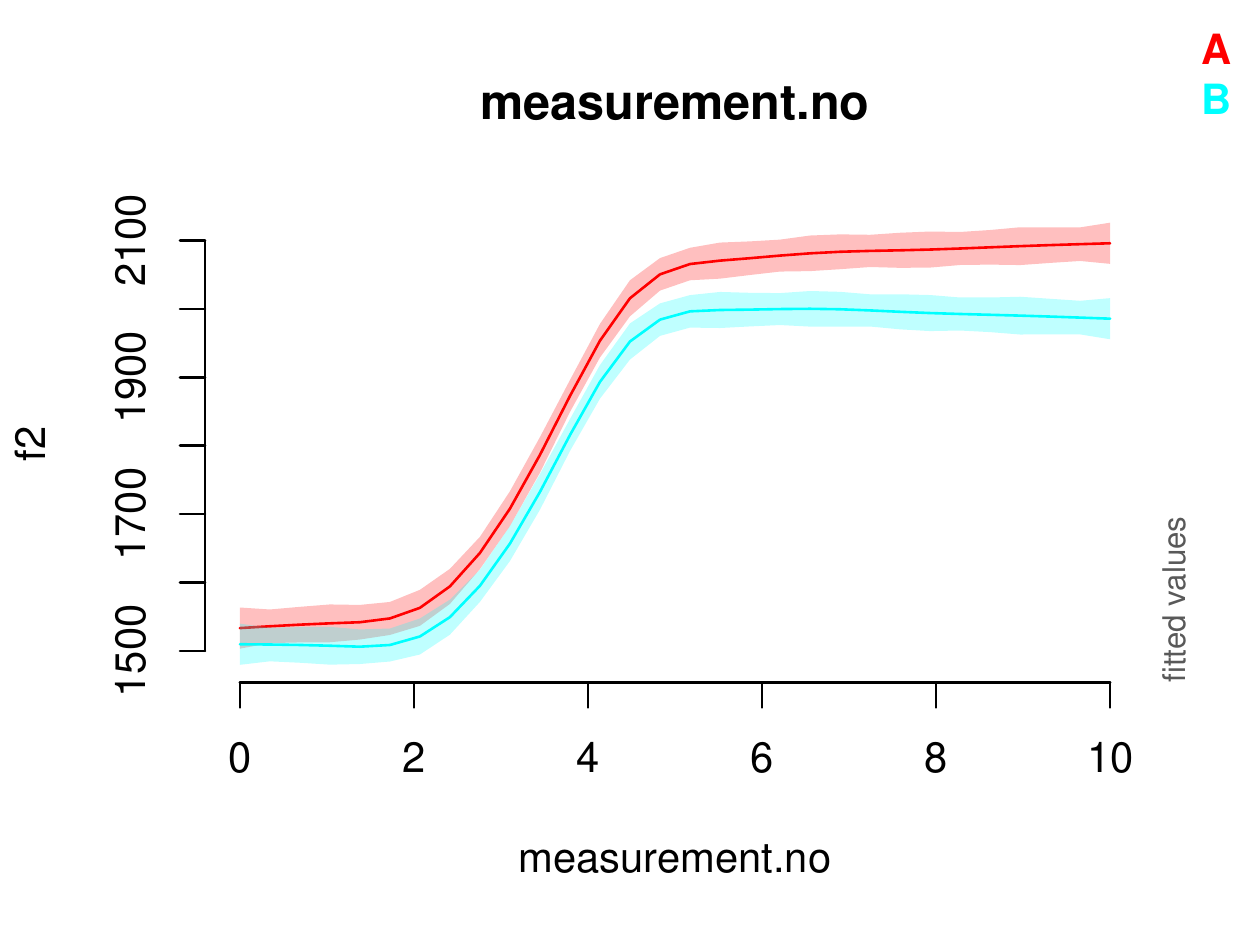} 
\includegraphics[width=0.495\textwidth]{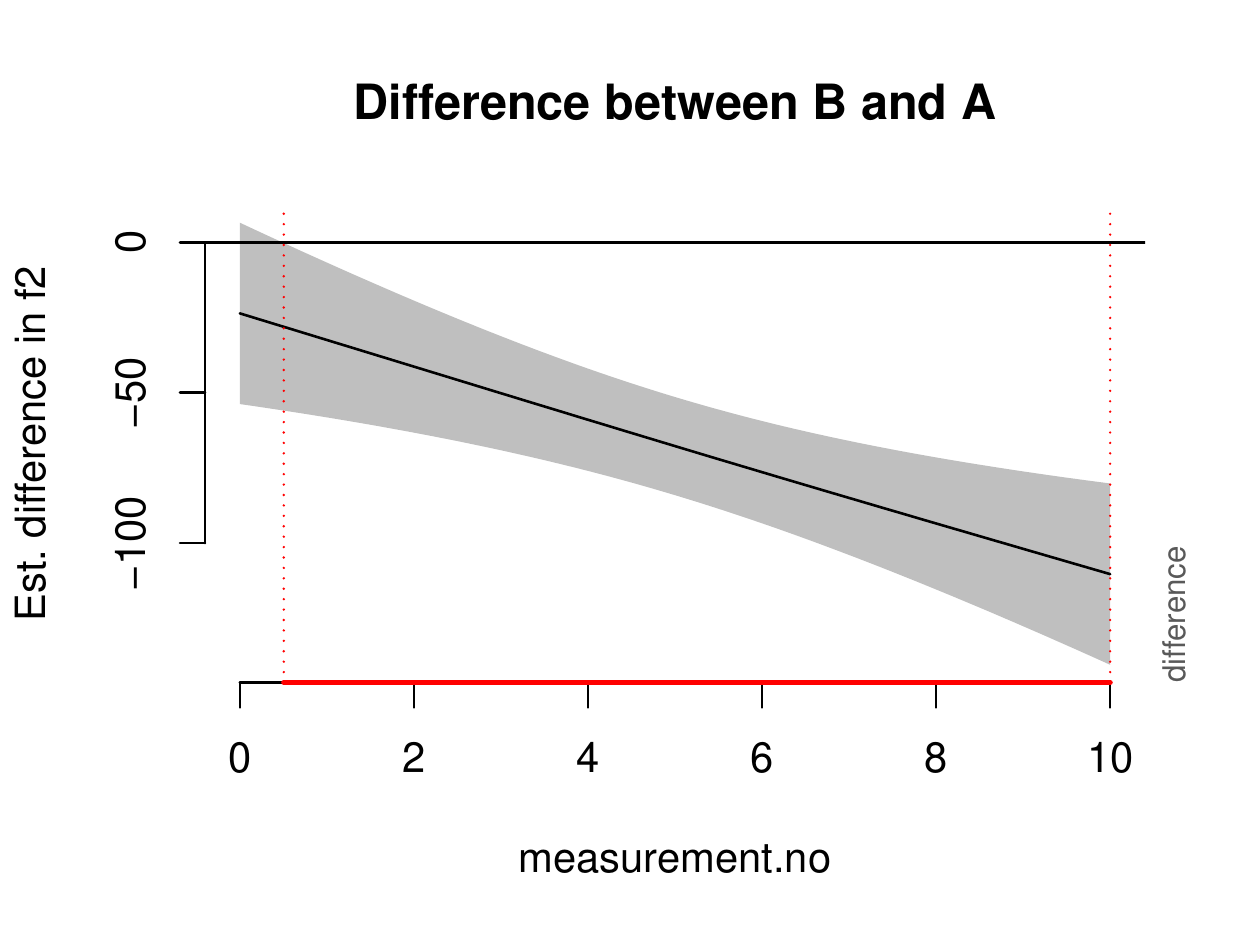} 

}

\end{knitrout}
\noindent The \verb+view+ option determines what variable to show along the $x$-axis; the {\tt plot\_all = "word.ord"} option tells R to plot predictions separately for each value of the factor \verb+word.ord+; the \verb+rug=F+ option tells R not to include ticks for each data point along the \textit{x}-axis (these can make plotting very slow and plot files very large when there are many thousands of data points); and the \verb+comp+ option specifies the levels of \verb+word.ord+ that the difference smooth is based on. In this case, the difference smooth shows $B - A$. Note that the confidence interval for the difference smooth seems a bit too narrow: the underlying curves were specified in such a way that there is no actual difference between the curves until about \verb+measurement.no+ 2, but the difference smooth shows a significant difference along almost the entire trajectory.

As a second step, let's try to account for the influence of \verb+duration+ on the trajectories. Simply including it in the model as a parametric term or even as a smooth on its own won't work: its effect is not on average F2 values, but on the shapes of the trajectories. So what we really want is a non-linear interaction between \verb+duration+ and the smooths for \verb+measurement.no+. Moreover, it would be useful if we could separate this interaction term from the main effects of \verb+duration+ and \verb+measurement.no+. The solution is to use so-called `tensor product interactions'. Although the name is quite intimidating, these are conceptually very similar to interactions in linear models. All we need to do is include three terms: \verb+s(measurement.no)+ (a smooth for the main effect of \verb+measurement.no+), \verb+s(duration)+ (main effect of \verb+duration+) and \verb+ti(measurement.no, duration)+ (the interaction between the two variables).

\begin{knitrout}\footnotesize
\definecolor{shadecolor}{rgb}{0.969, 0.969, 0.969}\color{fgcolor}\begin{kframe}
\begin{alltt}
\hlstd{words.50.gam.dur} \hlkwb{<-} \hlkwd{bam}\hlstd{(f2} \hlopt{~} \hlstd{word.ord} \hlopt{+} \hlkwd{s}\hlstd{(measurement.no,} \hlkwc{bs}\hlstd{=}\hlstr{"cr"}\hlstd{)} \hlopt{+}
                              \hlkwd{s}\hlstd{(duration,} \hlkwc{bs}\hlstd{=}\hlstr{"cr"}\hlstd{)} \hlopt{+}
                              \hlkwd{ti}\hlstd{(measurement.no, duration)} \hlopt{+}
                              \hlkwd{s}\hlstd{(measurement.no,} \hlkwc{by}\hlstd{=word.ord,} \hlkwc{bs}\hlstd{=}\hlstr{"cr"}\hlstd{),}
                         \hlkwc{data}\hlstd{=words.50,} \hlkwc{method}\hlstd{=}\hlstr{"ML"}\hlstd{)}
\hlkwd{summary.coefs}\hlstd{(words.50.gam.dur)}
\end{alltt}
\begin{verbatim}
## Parametric coefficients:
##             Estimate Std. Error t value Pr(>|t|)    
## (Intercept) 1887.721      5.315 355.170   <2e-16 ***
## word.ordB    -72.516      7.595  -9.548   <2e-16 ***
## 
## Approximate significance of smooth terms:
##                               edf Ref.df       F  p-value    
## s(measurement.no)           7.958  8.681 267.519  < 2e-16 ***
## s(duration)                 3.339  4.014   7.821 3.78e-06 ***
## ti(measurement.no,duration) 6.409  8.551   8.222 6.08e-11 ***
## s(measurement.no):word.ordB 1.662  2.062  10.137 3.91e-05 ***
\end{verbatim}
\end{kframe}
\end{knitrout}

\noindent The interaction term is significant according to the model summary. If we wanted to run a more rigorous test, we could fit a nested model where both smooths that include \verb+duration+ are dropped, and then compare it to the full model using \verb+compareML()+.

There are two ways to plot this interaction. First, we can plot smooths at a few different values of \verb+duration+. Or, alternatively, we can plot the surface that represents the interaction between \verb+duration+ and \verb+measurement.no+ using colours (with warmer colours representing higher values). Both options are shown below:
\newpage
\begin{knitrout}\footnotesize
\definecolor{shadecolor}{rgb}{0.969, 0.969, 0.969}\color{fgcolor}\begin{kframe}
\begin{alltt}
\hlkwd{plot_smooth}\hlstd{(words.50.gam.dur,} \hlkwc{view}\hlstd{=}\hlstr{"measurement.no"}\hlstd{,} \hlkwc{cond}\hlstd{=}\hlkwd{list}\hlstd{(}\hlkwc{duration}\hlstd{=}\hlnum{0.16}\hlstd{),}
            \hlkwc{rug}\hlstd{=F,} \hlkwc{col}\hlstd{=}\hlstr{"red"}\hlstd{)}
\hlkwd{plot_smooth}\hlstd{(words.50.gam.dur,} \hlkwc{view}\hlstd{=}\hlstr{"measurement.no"}\hlstd{,} \hlkwc{cond}\hlstd{=}\hlkwd{list}\hlstd{(}\hlkwc{duration}\hlstd{=}\hlnum{0.08}\hlstd{),}
            \hlkwc{rug}\hlstd{=F,} \hlkwc{col}\hlstd{=}\hlstr{"blue"}\hlstd{,} \hlkwc{add}\hlstd{=T)}
\hlkwd{fvisgam}\hlstd{(words.50.gam.dur,} \hlkwc{view}\hlstd{=}\hlkwd{c}\hlstd{(}\hlstr{"measurement.no"}\hlstd{,}\hlstr{"duration"}\hlstd{),}
        \hlkwc{ylim}\hlstd{=}\hlkwd{quantile}\hlstd{(words.50}\hlopt{$}\hlstd{duration,} \hlkwd{c}\hlstd{(}\hlnum{0.1}\hlstd{,} \hlnum{0.9}\hlstd{)))}
\end{alltt}
\end{kframe}

{\centering \includegraphics[width=0.495\textwidth]{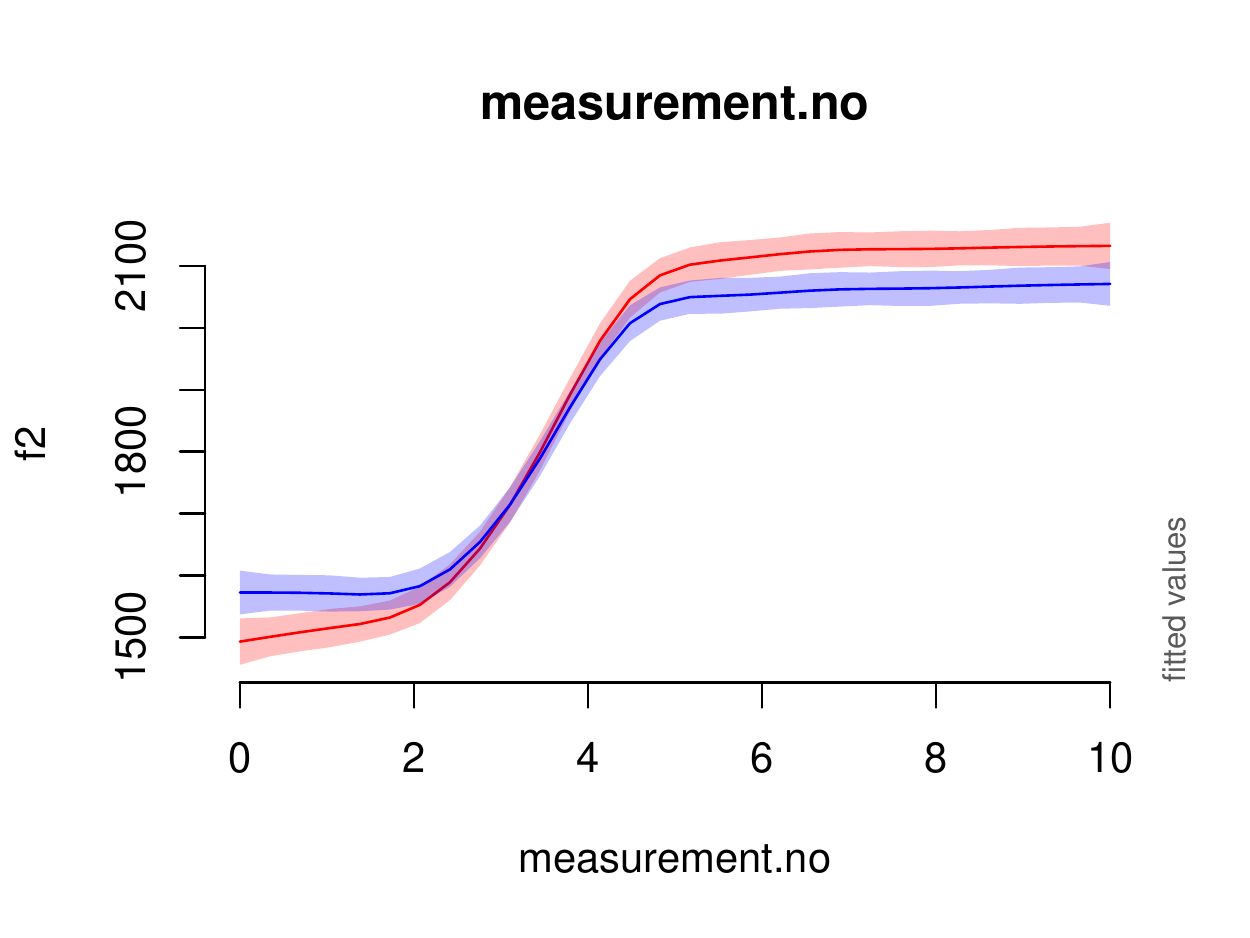} 
\includegraphics[width=0.495\textwidth]{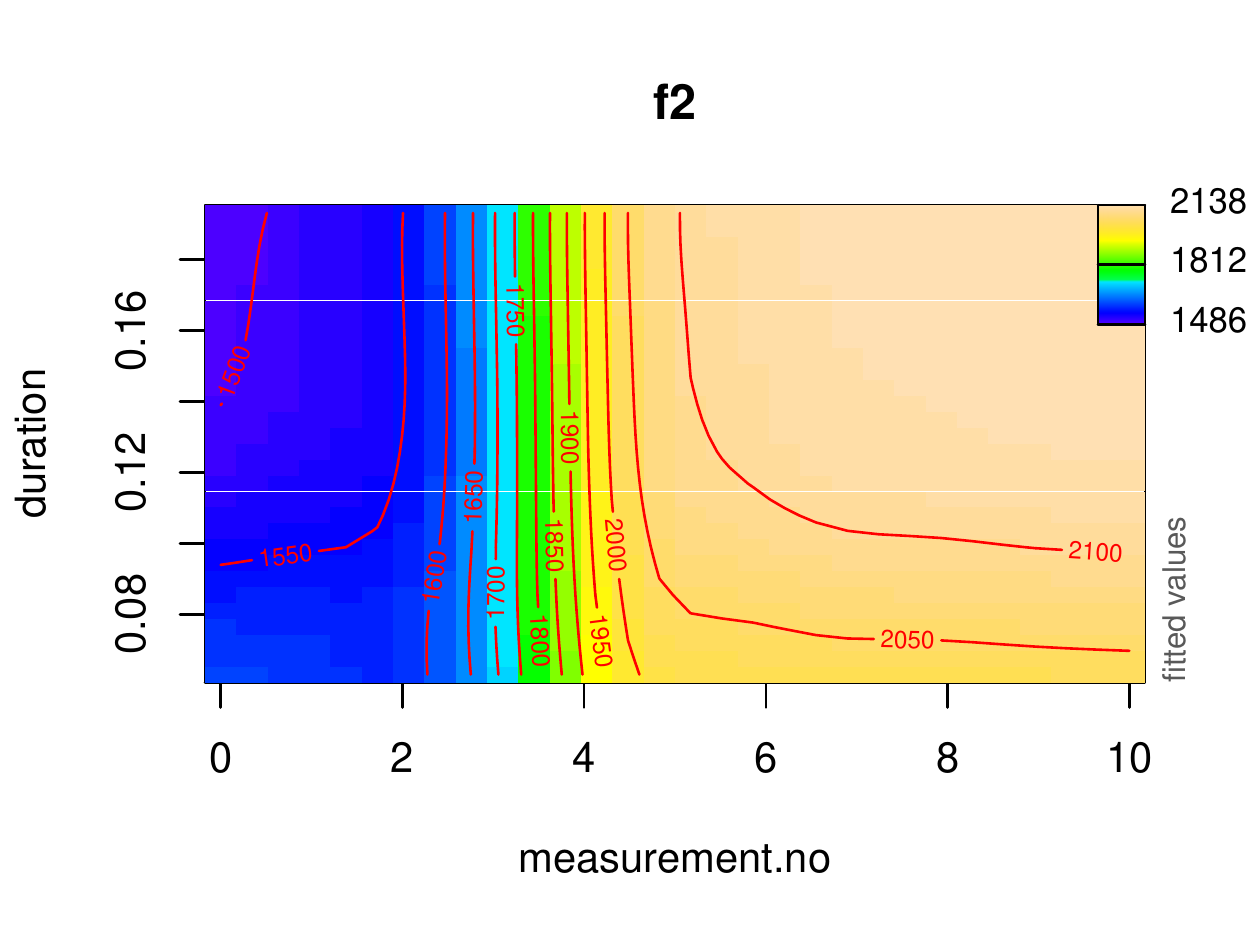} 

}

\end{knitrout}

\noindent The function \verb+plot_smooth+ needs to be run twice to generate the separate smooths: once to create the plotting area and a smooth for longer trajectories (with a duration of 0.16 s, set using the \verb+cond+ option; this smooth is shown in red) and once to add another smooth for shorter trajectories (with a duration of 0.08 s; shown in blue). Adding a smooth to an existing plot is done by including the \verb+add=T+ option. The result is the plot on the left, which suggests that shorter vowels have a flatter f2 trajectory.

The plot on the right is not easy to interpret, but let's give it a go. First, let's pick a specific point along the \textit{y}-axis, say, 0.1 s (the second notch from the bottom). This will allow us to look at a predicted trajectory for a given duration value. Now imagine a horizontal line crossing this point, and inspect the colours along this line: we start with dark blue and move towards warmer colours, finally arriving at light ochre. This indicates a rise in values, which is also shown by the labelled contour lines (which should be read similarly to contour lines in topographic maps). The colour doesn't change much between measurement points 0--2 and between 5--10, indicating that the predicted trajectory is stable in these ranges, and that most of the change takes place between 2--5. Now let's zoom out a bit and try to look at a range of different duration values at the same time. What we see is that the trajectory starts relatively high (over 1550 Hz) at short durations, and a bit lower (close to 1500 Hz) at long durations. Conversely, the trajectory ends high for long durations (over 2100 Hz), but ends low for short durations (lower than 2050 Hz). In the current case, this heatmap is not particularly useful, as the difference between trajectories with different durations is relatively small. However, it is often worth inspecting both types of graphs to get a better sense of what the model fit actually looks like.

Though our model already includes a lot of nuance, it is not yet complete: in its current form it does not recognise the fact that the data consists of a set of 50 trajectories rather than 550 completely independent measurements, and is therefore likely overconfident in its estimates. In other words, the model is not capturing dependencies within individual trajectories, which likely leave patterns in the residuals. This should show up in a residual autocorrelation plot, so let's create one. So far, we have only looked at autocorrelation within a single residual series. In the current case, there are 50 separate trajectories with 50 separate residual series (though the model is not aware that the residuals come from different trajectories), so it is impractical to create separate autocorrelation plots for each of them. Instead, we will look at average autocorrelation values across the 50 residual series. The plot is shown below.

\begin{knitrout}\footnotesize
\definecolor{shadecolor}{rgb}{0.969, 0.969, 0.969}\color{fgcolor}\begin{kframe}
\begin{alltt}
\hlkwd{acf_plot}\hlstd{(}\hlkwd{resid}\hlstd{(words.50.gam.dur),} \hlkwc{split_by}\hlstd{=}\hlkwd{list}\hlstd{(words.50}\hlopt{$}\hlstd{traj))}
\end{alltt}
\end{kframe}

{\centering \includegraphics[width=0.495\textwidth]{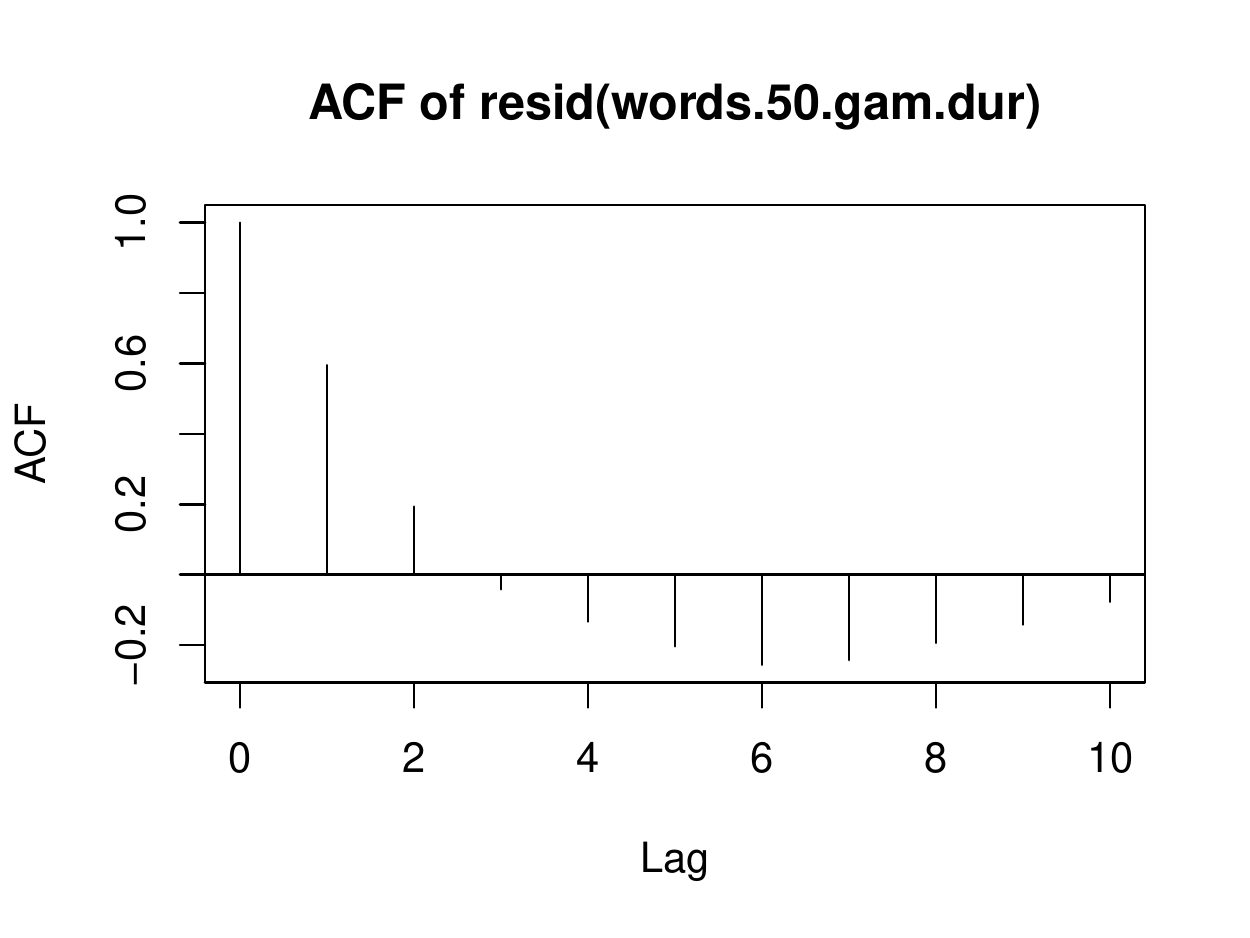} 

}

\end{knitrout}

\noindent The function \verb+resid(words.50.gam.dur)+ extracts the residuals from the model. These are the main argument to the function \verb+acf_plot()+, which creates an averaged residual plot for multiple trajectories. Since the residuals are just a series of numbers without any structure, the function needs to be told how to chop them up into separate trajectories: this is done using the option \verb+split_by=list(words.50$traj)+, which tells \verb+acf_plot()+ to calculate the autocorrelations separately for each trajectory and then average over them. Note also that \verb+acf_plot()+ (like most other autocorrelation plotting functions) assumes that the order of the observations in your data frame (\verb+words.50+) is the same as their order in the time series over which the autocorrelations are calculated (so e.g.\ an observation at \verb+measurement.no = 2+ comes before an observation from the same trajectory at \verb+measurement.no = 5+).

The residual autocorrelation plot suggests that there is a fair bit of positive autocorrelation at lag 1. This confirms our suspicion that there are patterns in the residuals of the model, which are likely affecting the model output as well. The remaining autocorrelation values are somewhat less worrying with low values mainly between $0.2$ and $-0.2$.

We'll look at two solutions for addressing this issue: (i) using random structures and (ii) using an autoregressive error model within trajectories. Both of these are effective at keeping type I error rates close to nominal values in type I error simulations using data sets with a structure very similar to the current one (\citeauthor{soskuthy16gam}, in prep).

Let's start with the random structures: we'll try out a few different options and perform model comparisons to see what type of random structure is best suited to our data. We'll explore three different options: (i) random intercepts only, (ii) random intercepts plus slopes and (iii) random smooths. The code for fitting these models is shown below. The last model may take a while to fit, so be prepared to wait for a few minutes.
\newpage
\begin{knitrout}\footnotesize
\definecolor{shadecolor}{rgb}{0.969, 0.969, 0.969}\color{fgcolor}\begin{kframe}
\begin{alltt}
\hlcom{# random intercepts only}
\hlstd{words.50.gam.int} \hlkwb{<-} \hlkwd{bam}\hlstd{(f2} \hlopt{~} \hlstd{word.ord} \hlopt{+} \hlkwd{s}\hlstd{(measurement.no,} \hlkwc{bs}\hlstd{=}\hlstr{"cr"}\hlstd{)} \hlopt{+}
                             \hlkwd{s}\hlstd{(duration,} \hlkwc{bs}\hlstd{=}\hlstr{"cr"}\hlstd{)} \hlopt{+}
                             \hlkwd{ti}\hlstd{(measurement.no, duration)} \hlopt{+}
                             \hlkwd{s}\hlstd{(measurement.no,} \hlkwc{by}\hlstd{=word.ord,} \hlkwc{bs}\hlstd{=}\hlstr{"cr"}\hlstd{)} \hlopt{+}
                             \hlkwd{s}\hlstd{(traj,} \hlkwc{bs}\hlstd{=}\hlstr{"re"}\hlstd{),}
                         \hlkwc{data}\hlstd{=words.50,} \hlkwc{method}\hlstd{=}\hlstr{"fREML"}\hlstd{)}
\hlcom{# random intercepts + slopes}
\hlstd{words.50.gam.slope} \hlkwb{<-} \hlkwd{bam}\hlstd{(f2} \hlopt{~} \hlstd{word.ord} \hlopt{+} \hlkwd{s}\hlstd{(measurement.no,} \hlkwc{bs}\hlstd{=}\hlstr{"cr"}\hlstd{)} \hlopt{+}
                               \hlkwd{s}\hlstd{(duration,} \hlkwc{bs}\hlstd{=}\hlstr{"cr"}\hlstd{)} \hlopt{+}
                               \hlkwd{ti}\hlstd{(measurement.no, duration)} \hlopt{+}
                               \hlkwd{s}\hlstd{(measurement.no,} \hlkwc{by}\hlstd{=word.ord,} \hlkwc{bs}\hlstd{=}\hlstr{"cr"}\hlstd{)} \hlopt{+}
                               \hlkwd{s}\hlstd{(traj,} \hlkwc{bs}\hlstd{=}\hlstr{"re"}\hlstd{)} \hlopt{+}
                               \hlkwd{s}\hlstd{(traj, measurement.no,} \hlkwc{bs}\hlstd{=}\hlstr{"re"}\hlstd{),}
                         \hlkwc{data}\hlstd{=words.50,} \hlkwc{method}\hlstd{=}\hlstr{"fREML"}\hlstd{)}
\hlcom{# random smooths}
\hlstd{words.50.gam.smooth} \hlkwb{<-} \hlkwd{bam}\hlstd{(f2} \hlopt{~} \hlstd{word.ord} \hlopt{+} \hlkwd{s}\hlstd{(measurement.no,} \hlkwc{bs}\hlstd{=}\hlstr{"cr"}\hlstd{)} \hlopt{+}
                           \hlkwd{s}\hlstd{(duration,} \hlkwc{bs}\hlstd{=}\hlstr{"cr"}\hlstd{)} \hlopt{+}
                           \hlkwd{ti}\hlstd{(measurement.no, duration)} \hlopt{+}
                           \hlkwd{s}\hlstd{(measurement.no,} \hlkwc{by}\hlstd{=word.ord,} \hlkwc{bs}\hlstd{=}\hlstr{"cr"}\hlstd{)} \hlopt{+}
                           \hlkwd{s}\hlstd{(measurement.no, traj,} \hlkwc{bs}\hlstd{=}\hlstr{"fs"}\hlstd{,} \hlkwc{xt}\hlstd{=}\hlstr{"cr"}\hlstd{,} \hlkwc{m}\hlstd{=}\hlnum{1}\hlstd{,} \hlkwc{k}\hlstd{=}\hlnum{5}\hlstd{),}
                         \hlkwc{data}\hlstd{=words.50,} \hlkwc{method}\hlstd{=}\hlstr{"fREML"}\hlstd{)}
\end{alltt}
\end{kframe}
\end{knitrout}

\noindent Random intercepts are coded by including a smooth over the grouping variable with the smoothing class specified as \verb+bs="re"+. This is really a technicality: \verb+bam()+ is able to use the mathematics of smooths to estimate various random structures, and this is reflected in its syntax as well. Random slopes are coded by adding the slope variable (\verb+duration+) after the grouping variable (\verb+traj+) inside the smooth, and keeping the smoothing class as \verb+bs="re"+. This is a bit confusing, since random smooths have the opposite syntax: here, the continuous variable comes first, followed by the grouping variable. Let's repeat this below just to be on the safe side:
\begin{itemize}
  \item random slopes: grouping factor, continuous variable
  \item random smooths: continuous variable, grouping factor
\end{itemize}
For random smooths, we also have to change the smoothing class to \verb+bs="fs"+, which is short for `factor smooth interactions'. The \verb+m=1+ specification is recommended in several papers (e.g.\ \citealt{baayenetal16}) for random smooths; what it does is slightly change the way the smoothing penalty is estimated (the default value is 2). The \verb+xt="cr"+ option sets the smooth class for the individual random smooths to cubic regression splines -- this may not be necessary, but my impression was that cubic regression splines did a better job at capturing the current vowel trajectories. Finally, \verb+k=5+ sets a relatively low upper limit on the wiggliness of the individual random smooths. Although a higher number might result in a marginally better fit, this model already takes a long time to fit, and increasing {\tt k} for random smooths can drastically increase the amount of resources (memory and time) needed to fit GAMMs.

The estimation method for the three models above is set to \verb+method="fREML"+. We need restricted maximum likelihood estimation in this case since we want to compare models with the same fixed effects, but different random effects. In principle, we could also use \verb+method="REML"+, but fREML (`fast REML') is much faster and tends to yield essentially the same results. If the comparison was between models with different fixed effects but the same random effects (the typical case), the models would have to be estimated using \verb+method="ML"+ (see also \citealt{zuuretal09}; unfortunately, there is no `fast ML', so models fitted with ML can take a long time to converge).

In order to find out which model fits the data best, we'll use a statistic called AIC (Akaike Information Criterion). This is necessary since the models are not all properly nested within each other. AIC is a combination of two quantities: how surprising the data are given our fitted model (the lower this number, the better the fit) and how many parameters are used in the model. That is, AIC penalises both bad model fits and unnecessary model complexity. When comparing two models, the one with a lower AIC should be preferred (AIC comparisons are slightly more complicated, but we will go with this simple heuristic for the current case). Here are the AIC values for the three models above:

\begin{knitrout}\footnotesize
\definecolor{shadecolor}{rgb}{0.969, 0.969, 0.969}\color{fgcolor}\begin{kframe}
\begin{alltt}
\hlkwd{AIC}\hlstd{(words.50.gam.int, words.50.gam.slope, words.50.gam.smooth)}
\end{alltt}
\begin{verbatim}
##                            df      AIC
## words.50.gam.int     70.02953 6173.767
## words.50.gam.slope  117.16199 5634.792
## words.50.gam.smooth 212.68868 5269.876
\end{verbatim}
\end{kframe}
\end{knitrout}

\noindent Based on these values, the model with random smooths is a clear winner: the added model complexity is more than compensated for by the improvement in model fit.

Let's recreate the model prediction plots that we previously generated for the simple model without random smooths. The plots below show predicted trajectories for words $A$ and $B$ (left), and a difference smooth for $B - A$.

\begin{knitrout}\footnotesize
\definecolor{shadecolor}{rgb}{0.969, 0.969, 0.969}\color{fgcolor}\begin{kframe}
\begin{alltt}
\hlkwd{plot_smooth}\hlstd{(words.50.gam.smooth,} \hlkwc{view}\hlstd{=}\hlstr{"measurement.no"}\hlstd{,} \hlkwc{plot_all}\hlstd{=}\hlstr{"word.ord"}\hlstd{,}
            \hlkwc{rug}\hlstd{=F,} \hlkwc{rm.ranef}\hlstd{=T)}
\hlkwd{plot_diff}\hlstd{(words.50.gam.smooth,} \hlkwc{view}\hlstd{=}\hlstr{"measurement.no"}\hlstd{,}
          \hlkwc{comp}\hlstd{=}\hlkwd{list}\hlstd{(}\hlkwc{word.ord}\hlstd{=}\hlkwd{c}\hlstd{(}\hlstr{"B"}\hlstd{,}\hlstr{"A"}\hlstd{)),} \hlkwc{rm.ranef}\hlstd{=T)}
\end{alltt}
\end{kframe}

{\centering \includegraphics[width=0.495\textwidth]{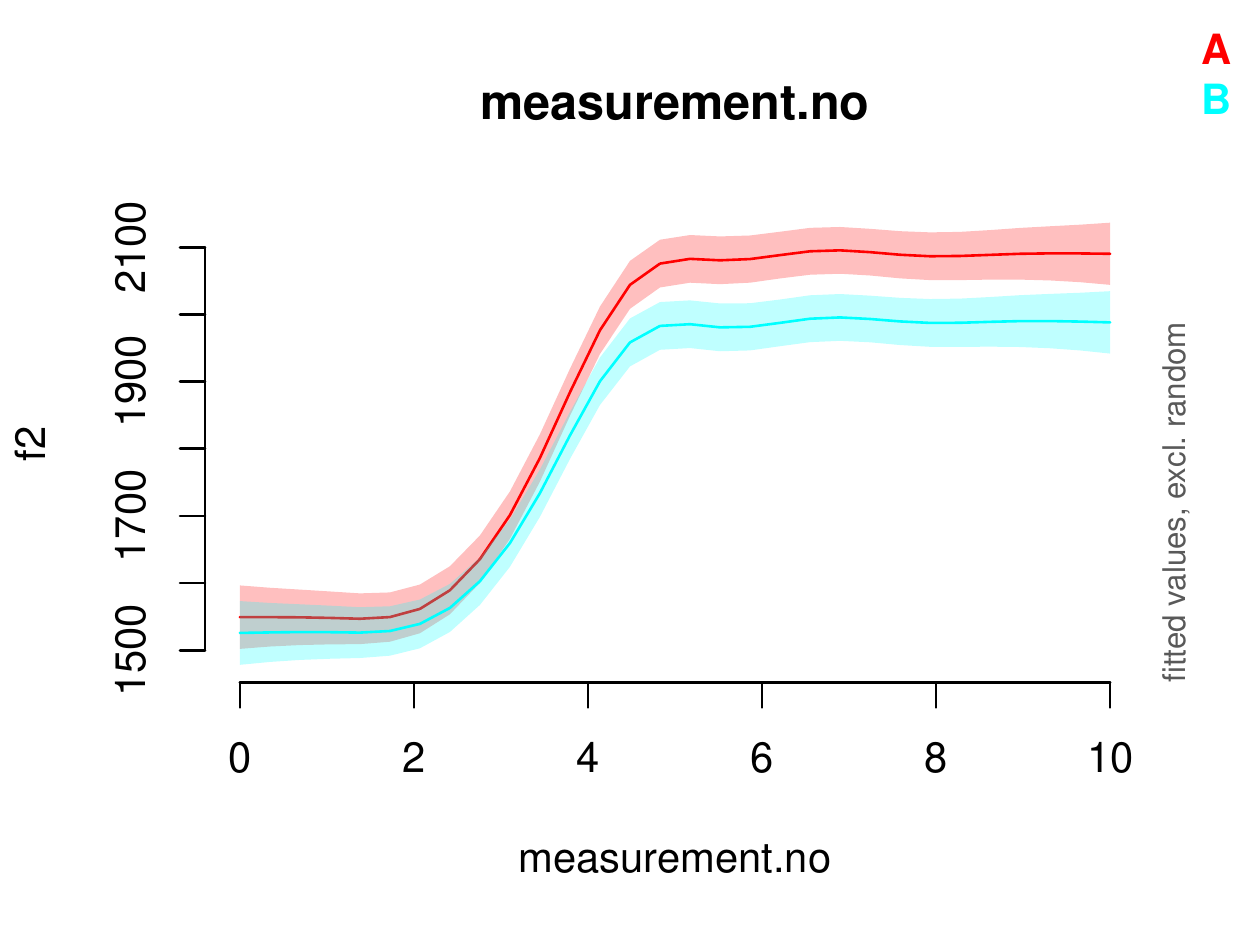} 
\includegraphics[width=0.495\textwidth]{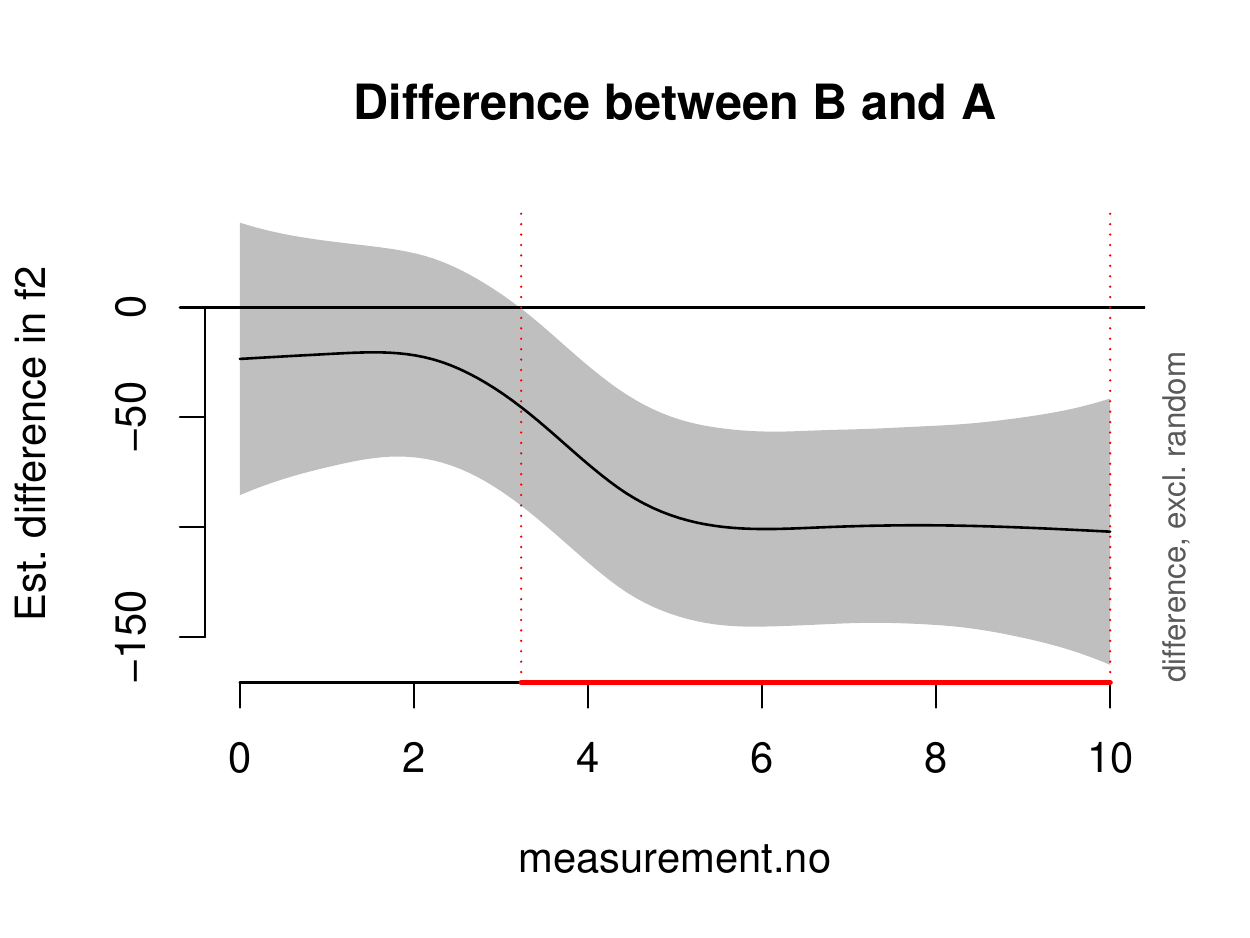} 

}

\end{knitrout}

\noindent This is very similar to what we did for the simple model, but both functions include an additional option: \verb+rm.ranef=T+. This option ensures that the predictions are shown for words $A$ vs.\ $B$ in general. If this option was left out, the predictions would relate to a given value of the grouping factor for the random effect, that is, \verb+traj+. In other words, the plot would show predictions for a specific trajectory. But instead of just plotting that single trajectory, R would attempt to figure out what that trajectory would look like if it belonged to word $A$ and what it would look like if it belonged to word $B$. Since a given trajectory can only belong to one word, this can lead to unreliable confidence intervals and results that are difficult to interpret.

As expected, the confidence intervals for the model with random smooths are much wider than they are for the simple model (cf.\ the earlier graphs). Moreover, the shape of the difference smooth changes considerably between the two models: it looks more non-linear for the current model, and the confidence interval includes 0 until about \verb+measurement.no+ 3. Since the initial sections of the underlying curves for the two words are identical, this is a clear improvement on the previous model.

Can the added by-trajectory random smooths deal with the autocorrelation issue in the model? Below is the residual autocorrelation plot for the updated model.

\begin{knitrout}\footnotesize
\definecolor{shadecolor}{rgb}{0.969, 0.969, 0.969}\color{fgcolor}\begin{kframe}
\begin{alltt}
\hlkwd{acf_resid}\hlstd{(words.50.gam.smooth,} \hlkwc{split_pred}\hlstd{=}\hlkwd{c}\hlstd{(}\hlstr{"traj"}\hlstd{))}
\end{alltt}
\end{kframe}

{\centering \includegraphics[width=0.495\textwidth]{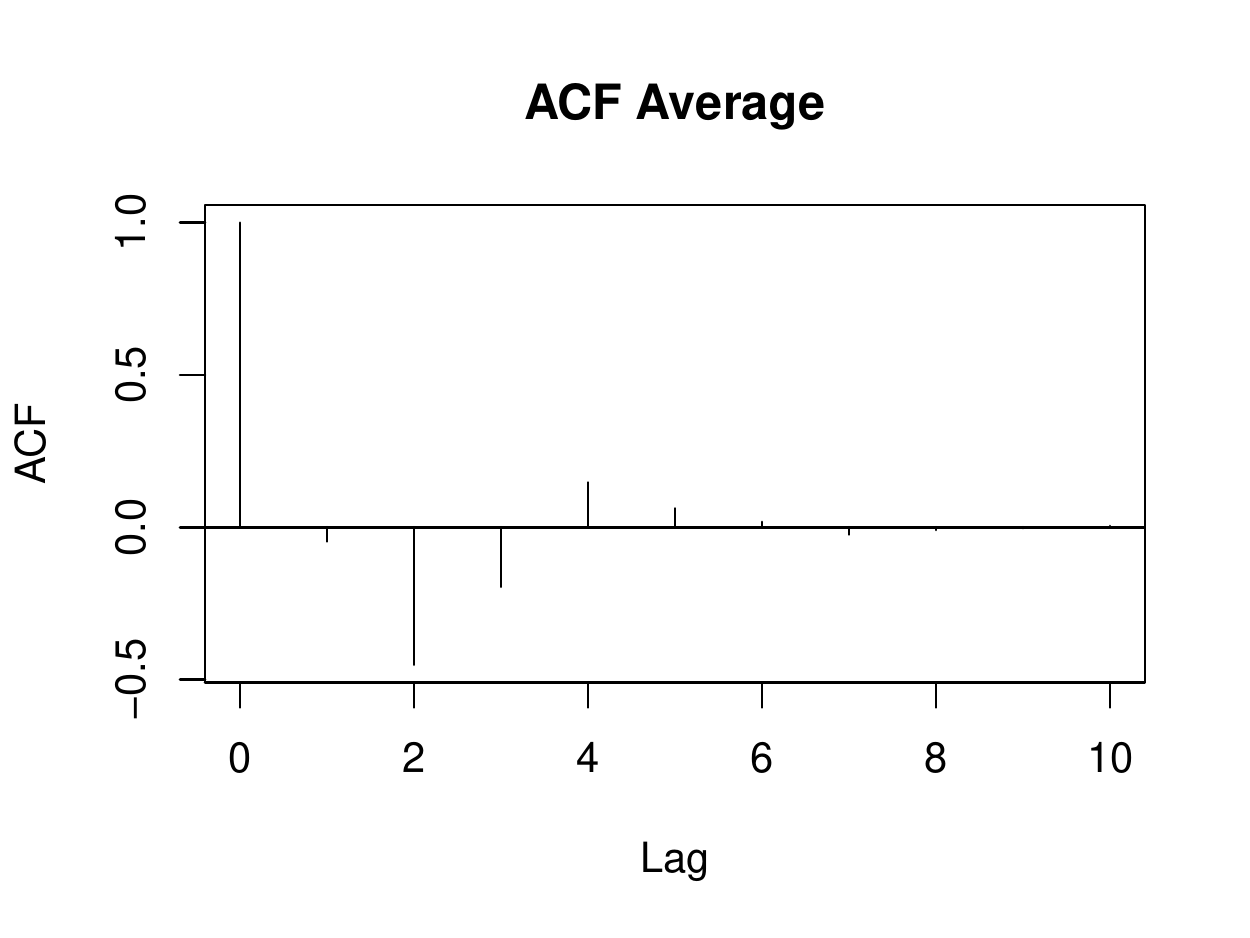} 

}

\end{knitrout}

\noindent Since this model includes \verb+traj+ as a predictor, a different function can be used (though \verb+acf_plot()+ would also work fine): \verb+acf_resid()+ from \verb+itsadug+. This function is a bit simpler in that it automatically extracts the residuals of the model and it's sufficient to specify the name of the predictor identifying the trajectories (using the \verb+split_pred+ option). The plot suggests that the autocorrelation at lag 1 is completely gone, although a certain amount of negative autocorrelation is introduced at lag 2. This is likely because of the fact that the initial and final sections of the raw trajectories are completely straight, but the fitted trajectories are actually a bit wavy. However, this negative autocorrelation in the residuals does not seem to lead to increased false positive rates in type I error simulations (at least not for this type of trajectory). This may be a consequence of the fact that there is very little variance left in the data after the addition of the by-trajectory random smooths.

The second option for reducing autocorrelation in the residuals is to use an autoregressive error model. There are many different kinds of autoregressive models, and \verb+bam()+ only supports the simplest of these: the so-called AR1 error model. An AR1 error model estimates the model parameters under the assumption that the errors%
\footnote{
The notion of `error' is closely related to that of `residual': errors are the deviations of the actual observations from the true underlying quantitites, while residuals are the deviations of the actual observations from the model predictions. Modelling correlations between errors has the practical effect of removing correlations in the residuals (assuming that we choose the right error model).
} 
for neighbouring observations in a time series are correlated: for instance, the error at \verb+measurement.no+ 4 is partly determined by the error at \verb+measurement.no+ 3. The AR1 model assumes that such correlations only exist between immediately adjacent points in the time series. An AR2 model would assume that the error at \verb+measurement.no+ 4 depends both on \verb+measurement.no+'s 2 and 3.

There are three things that need to be done before adding an AR1 model to a GAMM. First, the data set needs to be set up so that the order of observations in the data frame reflects their order in the time series (this is already how the current data frame is formatted). Second, we need to mark the starting point of each time series in the data set (i.e.\ \verb+measurement.no+ 0 for each trajectory) using a separate column -- otherwise \verb+bam()+ wouldn't be able to determine which adjacent points in the data frame actually belong to the same time series, and which of them span two time series that just happen to be next to each other in the data set (e.g.\ the last \verb+measurement.no+ from one trajectory and the first \verb+measurement.no+ from the next trajectory). Third, \verb+bam()+ cannot estimate the degree of correlation between the errors, so that has to be specified manually. We'll use a rough estimate from the original model. Here's the code for fitting the model:

\begin{knitrout}\footnotesize
\definecolor{shadecolor}{rgb}{0.969, 0.969, 0.969}\color{fgcolor}\begin{kframe}
\begin{alltt}
\hlcom{# 1) data frame is already ordered correctly}
\hlcom{# 2) marking starting points of each trajectory}

\hlstd{words.50}\hlopt{$}\hlstd{start.event} \hlkwb{<-} \hlstd{words.50}\hlopt{$}\hlstd{measurement.no} \hlopt{==} \hlnum{0}

\hlcom{# 3) getting a rough estimate of the correlation between adjacent errors}

\hlstd{r1} \hlkwb{<-} \hlkwd{start_value_rho}\hlstd{(words.50.gam.dur)}

\hlcom{# fitting the model}

\hlstd{words.50.gam.AR} \hlkwb{<-} \hlkwd{bam}\hlstd{(f2} \hlopt{~} \hlstd{word.ord} \hlopt{+} \hlkwd{s}\hlstd{(measurement.no,} \hlkwc{bs}\hlstd{=}\hlstr{"cr"}\hlstd{)} \hlopt{+}
                            \hlkwd{s}\hlstd{(duration,} \hlkwc{bs}\hlstd{=}\hlstr{"cr"}\hlstd{)} \hlopt{+}
                            \hlkwd{ti}\hlstd{(measurement.no, duration)} \hlopt{+}
                            \hlkwd{s}\hlstd{(measurement.no,} \hlkwc{by}\hlstd{=word.ord,} \hlkwc{bs}\hlstd{=}\hlstr{"cr"}\hlstd{),}
                       \hlkwc{data}\hlstd{=words.50,} \hlkwc{method}\hlstd{=}\hlstr{"fREML"}\hlstd{,}
                       \hlkwc{rho}\hlstd{=r1,} \hlkwc{AR.start}\hlstd{=words.50}\hlopt{$}\hlstd{start.event)}
\end{alltt}
\end{kframe}
\end{knitrout}

\noindent The first line of code adds an indicator column to the data frame that has the value \verb+TRUE+ marking the beginning of each trajectory. The function \verb+start_value_rho()+ is from the package \verb+itsadug+. It takes a GAMM without an autoregressive error model and returns the residual autocorrelation at lag 1. Note that this value should be treated as a rough estimate, and other values may, in fact, do a better job at reducing the autocorrelation in the residuals (see \citealt{baayenetal16} for further discussion). The autoregressive model is added to the GAMM by setting the \verb+rho+ argument to the estimated correlation parameter and specifying the starting points of the time series using the \verb+AR.start+ argument.

Two plots are shown below, which are both generated using residual autocorrelation plotting functions.

\begin{knitrout}\footnotesize
\definecolor{shadecolor}{rgb}{0.969, 0.969, 0.969}\color{fgcolor}\begin{kframe}
\begin{alltt}
\hlkwd{acf_plot}\hlstd{(}\hlkwd{resid}\hlstd{(words.50.gam.AR),} \hlkwc{split_by}\hlstd{=}\hlkwd{list}\hlstd{(words.50}\hlopt{$}\hlstd{traj))}
\hlkwd{acf_resid}\hlstd{(words.50.gam.AR,} \hlkwc{split_pred}\hlstd{=}\hlstr{"AR.start"}\hlstd{)}
\end{alltt}
\end{kframe}

{\centering \includegraphics[width=0.495\textwidth]{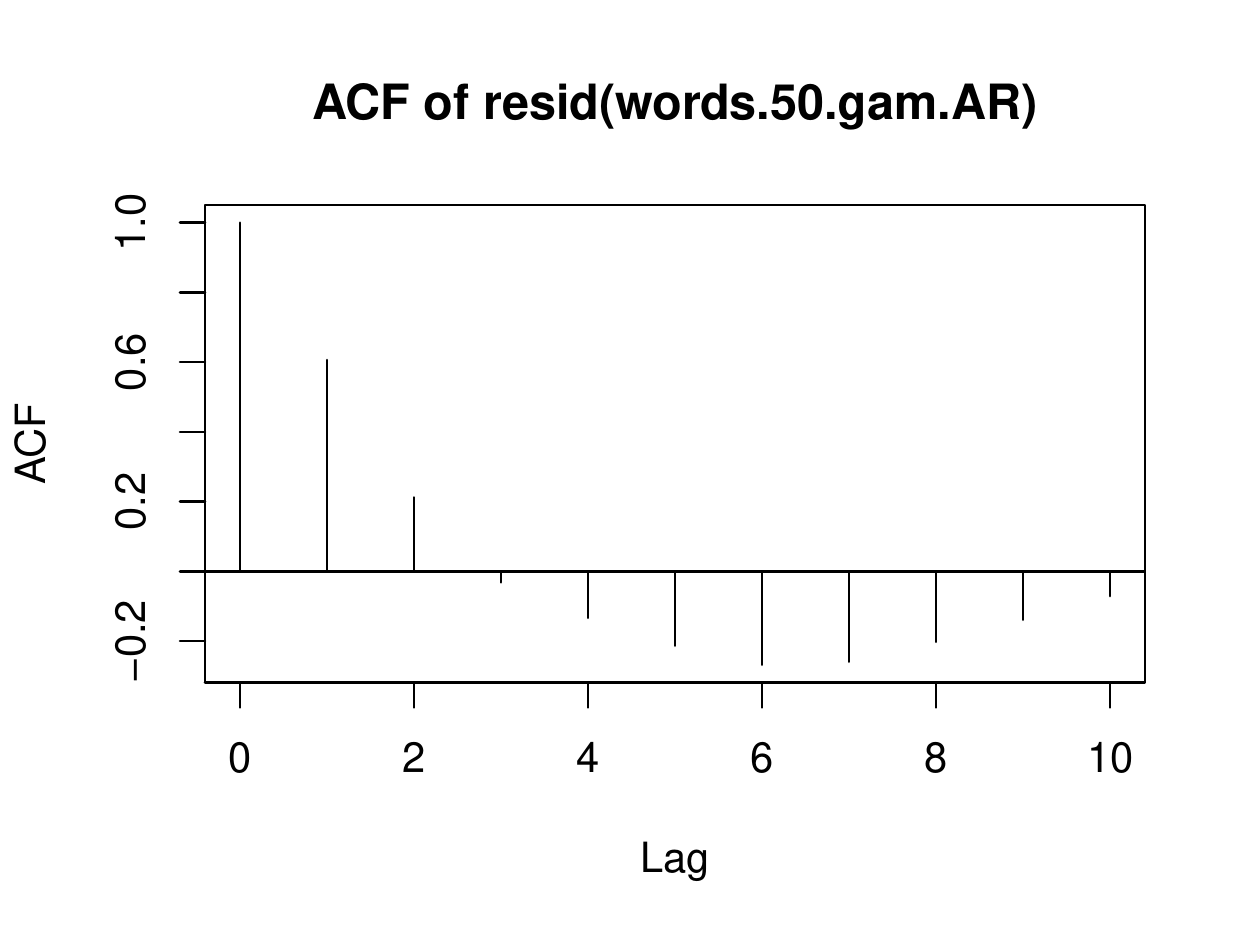} 
\includegraphics[width=0.495\textwidth]{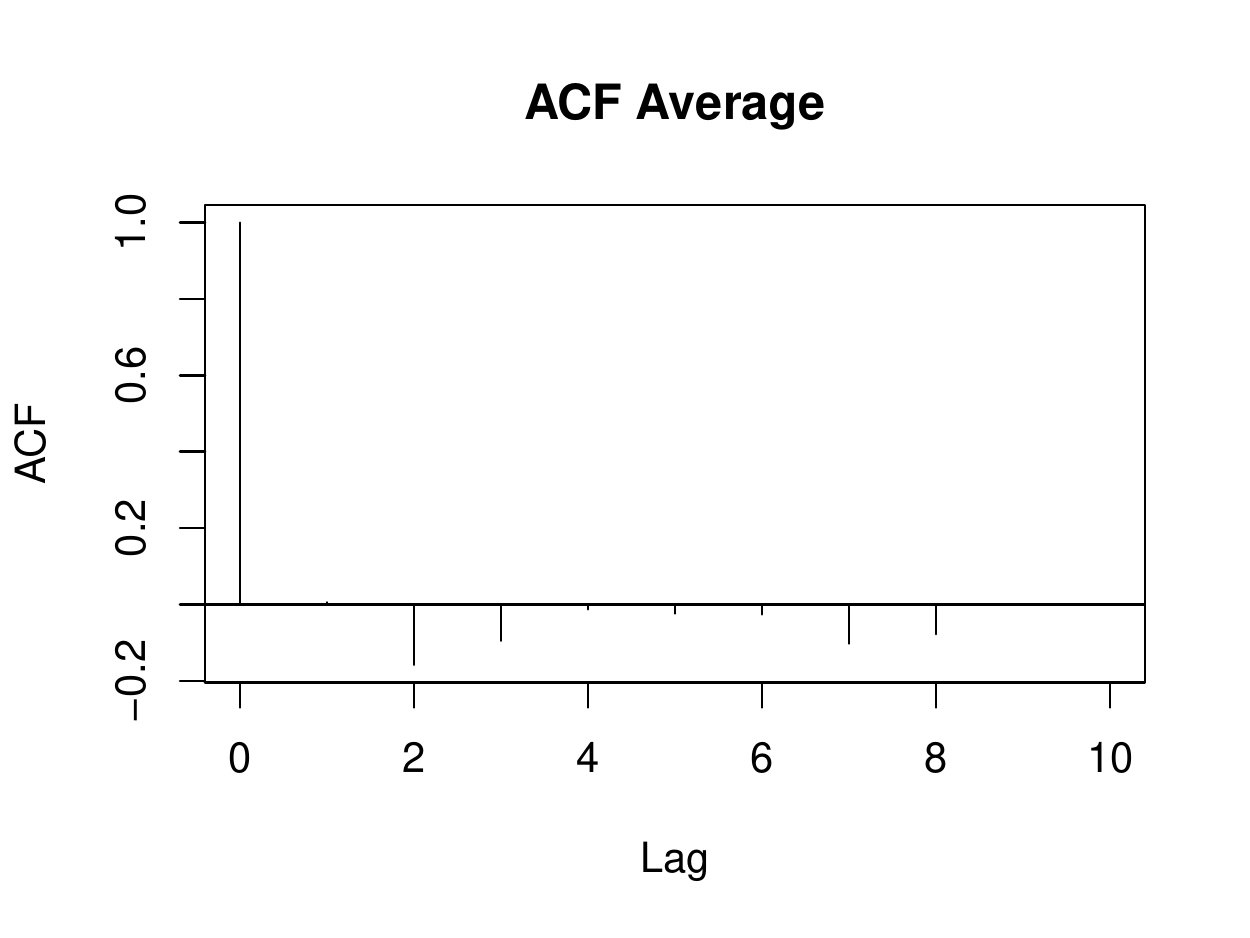} 

}

\end{knitrout}

\noindent As should be clear from the syntax, both of these plots actually show the residuals from our updated model. However, the one on the left seems to suggest that the residual autocorrelation has not been affected by the AR1 model, while the one on the right suggests the opposite, that is, the AR1 model has been successful at reducing autocorrelation. This is because the plot on the left shows the raw residuals (which don't take the fitted AR model into account), while the model on the right shows the normalised residuals (which do). When an AR model is used, the plot generated using the raw residuals is misleading, so the plot on the right should be used. Note that this is created by the \verb+acf_resid()+ function, which works with the normalised residuals by default (but needs to be told that the data was split into separate time series, which were identified by the \verb+AR.start+ argument of the \verb+bam()+ function).

The plots below show the model predictions for words \textit{A} and \textit{B}, and the difference smooth.

\begin{knitrout}\footnotesize
\definecolor{shadecolor}{rgb}{0.969, 0.969, 0.969}\color{fgcolor}\begin{kframe}
\begin{alltt}
\hlkwd{plot_smooth}\hlstd{(words.50.gam.AR,} \hlkwc{view}\hlstd{=}\hlstr{"measurement.no"}\hlstd{,} \hlkwc{plot_all}\hlstd{=}\hlstr{"word.ord"}\hlstd{,} \hlkwc{rug}\hlstd{=F)}
\hlkwd{plot_diff}\hlstd{(words.50.gam.AR,} \hlkwc{view}\hlstd{=}\hlstr{"measurement.no"}\hlstd{,}
          \hlkwc{comp}\hlstd{=}\hlkwd{list}\hlstd{(}\hlkwc{word.ord}\hlstd{=}\hlkwd{c}\hlstd{(}\hlstr{"B"}\hlstd{,}\hlstr{"A"}\hlstd{)))}
\end{alltt}
\end{kframe}

{\centering \includegraphics[width=0.495\textwidth]{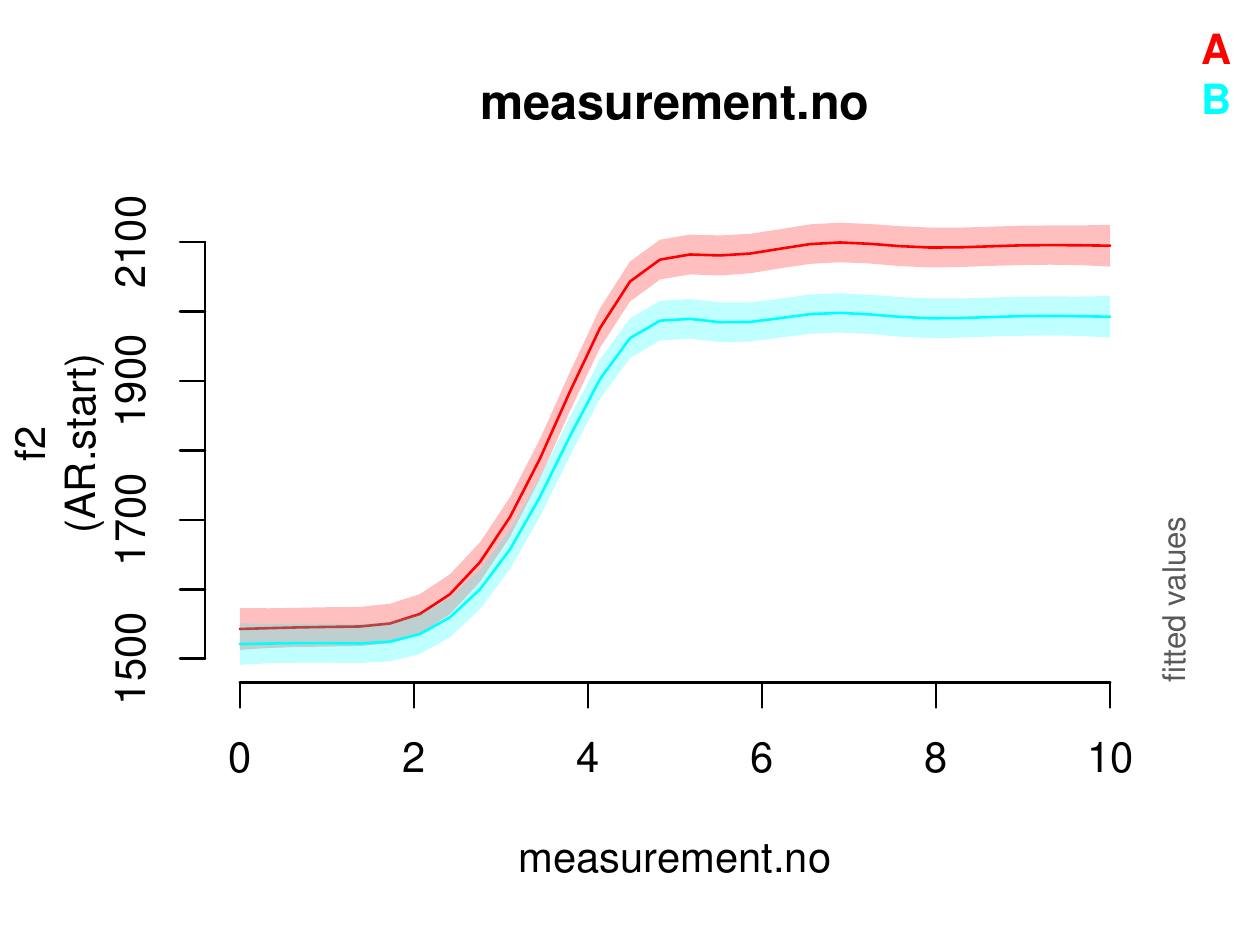} 
\includegraphics[width=0.495\textwidth]{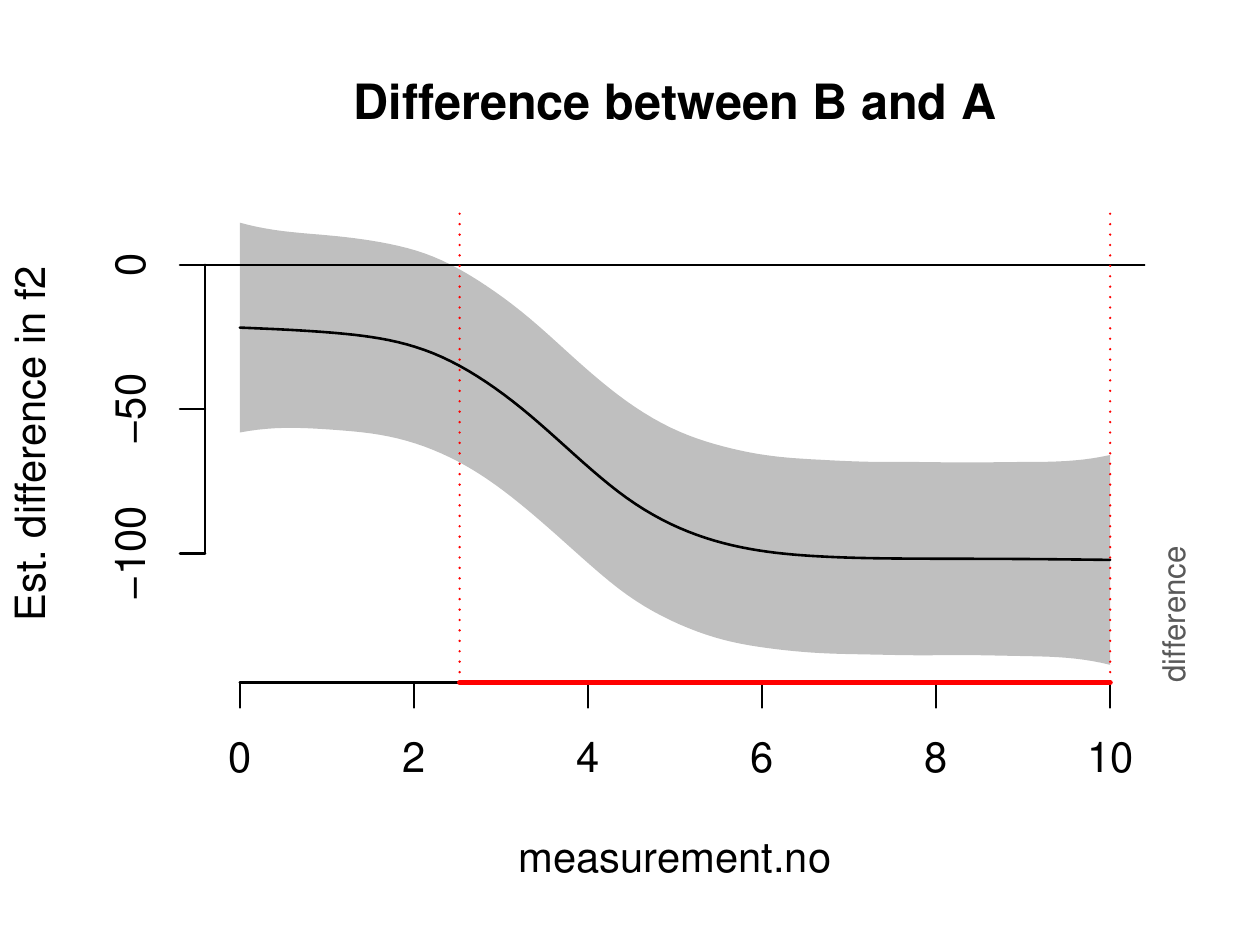} 

}

\end{knitrout}

\noindent These plots look quite similar to the ones from the model including random smooths, though the confidence intervals are slightly narrower and the shapes of the fitted trajectories are subtly different.

So which method should be used: random smooths by trajectory or an AR1 model? In the current case, both of them seem to work fine and lead to very similar conclusions. However, that may not be the case for all data sets. For instance, \citet{baayenetal16} discuss a case where both random smooths and an AR model seem to be necessary in order to appropriately account for all dependencies within a data set. In deciding between approaches, it is important to look at residual autocorrelation plots and also to check how much variance is actually left in the residuals.

There is, however, one practical consideration that makes the AR1 approach slightly more attractive. Adding an AR1 model to a GAMM is computationally inexpensive, while adding large numbers of random smooths can be very time- and memory-consuming. This may not be a problem for small data sets, but the computational cost of random smooths can be prohibitive for models with large numbers of trajectories.

What if we still want to fit random smooths to a large data set? The function \verb+bam()+ has an optional argument \verb+discrete+, which speeds up computation substantially when set to \verb+TRUE+. Moreover, when \verb+discrete=TRUE+, a GAMM can be fitted using multiple processor cores. The number of processor cores used in the computation is specified by the \verb+nthreads+ argument. Note that this technique only works with the fREML fitting method, so evaluating significance through model comparison may not be possible.

\section{Analysing data from \citet{stuartsmithetal15}}

We've applied GAMMs to various simulated data sets, but we haven't yet looked at any real data. In this section, we will use the methods introduced above to analyse a subset of the data presented in \citet{stuartsmithetal15}. This data set contains dynamic F3 measurements of word-final /r/ in spontaneous recordings of Glaswegian. The speakers are older males recorded at four different time points between 1970 and 2000. The data set is a fairly typical example of dynamic speech data. It is thoroughly unbalanced with varying numbers of tokens across speakers, decades, words and environments. Moreover, it likely shows the influence of a large number of different variables, though only a few of these are of interest for our present purposes. It is also larger than the toy data sets that we've worked with so far: the subset that we will look at contains 420 individual trajectories.

The data set consists of F3 trajectories measured at 11 evenly spaced points. The trajectories include both /r/ and the preceding vowel. The data come from four sets of three speakers recorded in the 1970s, the 1980s, the 1990s and the 2000s (i.e.\ 12 speakers altogether). The following variables will be used in the analysis:
\begin{itemize}
  \item \verb+measurement.no+: see above
  \item \verb+duration+: overall duration of the vowel + /r/ sequence
  \item \verb+decade+: decade of recording coded as a continuous variable
  \item \verb+stress+: whether the previous vowel is an unstressed schwa or a full vowel
  \item \verb+traj+: a grouping variable by trajectory
  \item \verb+speaker+: a grouping variable by speaker
\end{itemize}
The main question that we'll try to answer is whether the acoustic characteristics of final /r/ have changed over time. That is, we'll be looking for an effect of {\tt decade} on the F3 trajectories. However, there are a few complicating factors. First, the trajectories vary quite substantially in terms of their duration, and this is likely to affect their shapes: we might expect to see flatter trajectories for shorter vowel + /r/ sequences. Second, the vowels in the vowel + /r/ sequences are not always the same, but their distribution is unbalanced: about 65\% of all the vowels are schwas. Since the quality of the vowel may affect the F3 trajectory (though F3 is not expected to vary as much as F1 or F2 would), it is important to bring this factor into the analysis. To keep things simple, the quality of the vowel is encoded by the {\tt stress} variable, which splits the data set according to whether the vowel is an unstressed schwa or a full vowel. We may also expect that stressed vs.\ unstressed vowel + /r/ sequences will change differently, though we have no clear prediction about what such a difference would look like.

Before we start analysing the data, there is one further important point that should be discussed. Although the analysis presented here will be shown in a relatively streamlined form, the model fitting procedure was actually preceded by a detailed exploration of the data through various plots. These plots won't be shown here, but many of the analytical decisions below are actually based on observations that emerged from this exploratory analysis. This type of data exploration is absolutely crucial, and there is almost no point in starting to fit GAMMs until we get a sense of the range of variation in the trajectories that we are trying to model.

To avoid starting with a very complex model, let's first simply try to capture the effect of {\tt decade} on the trajectories. Since {\tt decade} only across but not within speakers, it's essential to include {\tt speaker} as a random smooth in the model. Otherwise the estimated effect of {\tt decade} will be based on the assumption that there are 420 independent data points where, in reality, there are really only 12.  We'll also record and display the amount of time it takes to fit the model using the \verb+system.time()+ function. This function displays three timing values. The last one of these shows how long it takes to fit the model overall; the other two are less relevant.

\begin{knitrout}\footnotesize
\definecolor{shadecolor}{rgb}{0.969, 0.969, 0.969}\color{fgcolor}\begin{kframe}
\begin{alltt}
\hlstd{gl.r} \hlkwb{<-} \hlkwd{read.csv}\hlstd{(}\hlstr{"glasgow_r.csv"}\hlstd{)}
\hlstd{gl.r.gamm.simple.t} \hlkwb{<-} \hlkwd{system.time}\hlstd{(}
  \hlstd{gl.r.gamm.simple} \hlkwb{<-} \hlkwd{bam}\hlstd{(f3} \hlopt{~} \hlkwd{s}\hlstd{(measurement.no)} \hlopt{+}
                               \hlkwd{s}\hlstd{(decade,} \hlkwc{k}\hlstd{=}\hlnum{4}\hlstd{)} \hlopt{+}
                               \hlkwd{ti}\hlstd{(measurement.no, decade,} \hlkwc{k}\hlstd{=}\hlkwd{c}\hlstd{(}\hlnum{10}\hlstd{,}\hlnum{4}\hlstd{))} \hlopt{+}
                               \hlkwd{s}\hlstd{(measurement.no, speaker,} \hlkwc{bs}\hlstd{=}\hlstr{"fs"}\hlstd{,} \hlkwc{m}\hlstd{=}\hlnum{1}\hlstd{,} \hlkwc{k}\hlstd{=}\hlnum{4}\hlstd{),}
                          \hlkwc{dat}\hlstd{=gl.r,} \hlkwc{method}\hlstd{=}\hlstr{"ML"}\hlstd{)}
\hlstd{)}
\hlstd{gl.r.gamm.simple.t}
\end{alltt}
\begin{verbatim}
##    user  system elapsed 
##   2.508   0.136   2.678
\end{verbatim}
\begin{alltt}
\hlkwd{summary.coefs}\hlstd{(gl.r.gamm.simple)}
\end{alltt}
\begin{verbatim}
## Parametric coefficients:
##             Estimate Std. Error t value Pr(>|t|)    
## (Intercept)  2327.37      27.48   84.71   <2e-16 ***
## 
## Approximate significance of smooth terms:
##                              edf Ref.df      F  p-value    
## s(measurement.no)          1.012  1.018 43.760 3.23e-11 ***
## s(decade)                  1.242  1.246 11.620 0.000181 ***
## ti(measurement.no,decade)  1.673  1.750  3.125 0.090800 .  
## s(measurement.no,speaker) 30.990 46.000 19.176  < 2e-16 ***
\end{verbatim}
\end{kframe}
\end{knitrout}

\noindent There are a few things to note here. \verb+k+ cannot be set higher than 4 for \verb+s(decade)+, since \verb+decade+ only has 4 values. This also holds for the \verb+ti()+ interaction term, where \verb+k+ needs to be set separately for the two main terms. \verb+k+ is also set to 4 for the by-speaker random smooths, partly because the raw data don't show that much wiggliness and partly because we will eventually want to include some further random smooths, so it's worth keeping things reasonably simple.

Here's a plot illustrating the effect of {\tt decade} on the trajectories:\footnote{A slightly modified version of {\tt plot\_smooth()} is used to keep things simple. This function can be loaded by sourcing the file {\tt gamm\_hacks.r}.}

\begin{knitrout}\footnotesize
\definecolor{shadecolor}{rgb}{0.969, 0.969, 0.969}\color{fgcolor}\begin{kframe}
\begin{alltt}
\hlkwd{source}\hlstd{(}\hlstr{"gamm_hacks.r"}\hlstd{)}
\hlkwd{plot_smooth.cont}\hlstd{(gl.r.gamm.simple,} \hlkwc{view}\hlstd{=}\hlstr{"measurement.no"}\hlstd{,} \hlkwc{plot_all.c}\hlstd{=}\hlstr{"decade"}\hlstd{,}
                 \hlkwc{rug}\hlstd{=F,} \hlkwc{rm.ranef}\hlstd{=T)}
\end{alltt}
\end{kframe}

{\centering \includegraphics[width=0.495\textwidth]{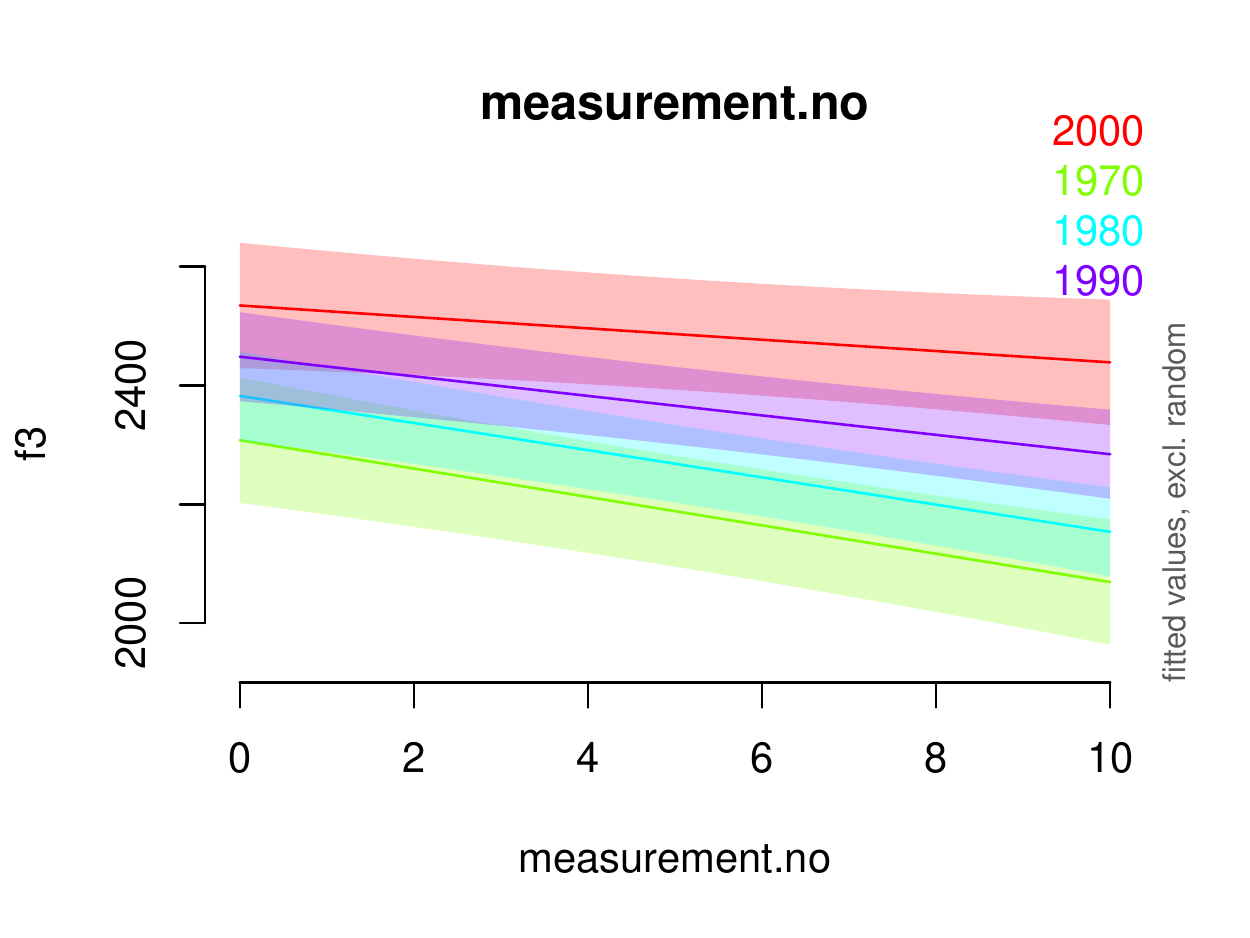} 

}

\end{knitrout}

\noindent We won't run model comparisons for this model, as a few things are still missing, but {\tt decade} does seem to have a significant effect on average F3 and possibly even the slope of the trajectories. Interestingly, the predicted trajectories come out as completely straight (this can also be read off the model summary, where the EDF for all three fixed smooth terms is close to 1). This is an example of oversmoothing, which often happens when residual patterns aren't appropriately accounted for.

Next, we'll add two further variables: {\tt duration} and {\tt stress}. These two variables actually show a slight correlation, as unstressed schwas are shorter on average than full vowels, but there is sufficient overlap between the {\tt duration} values for the two {\tt stress} groups to estimate these effects separately. We'll add a main smooth term for {\tt duration} as well as a \verb+ti()+ interaction with \verb+measurement.no+. The predictor {\tt stress} is included as a parametric term and a difference smooth (using \verb+by=stress+).

\begin{knitrout}\footnotesize
\definecolor{shadecolor}{rgb}{0.969, 0.969, 0.969}\color{fgcolor}\begin{kframe}
\begin{alltt}
\hlstd{gl.r}\hlopt{$}\hlstd{stress} \hlkwb{<-} \hlkwd{as.ordered}\hlstd{(gl.r}\hlopt{$}\hlstd{stress)}
\hlkwd{contrasts}\hlstd{(gl.r}\hlopt{$}\hlstd{stress)} \hlkwb{<-} \hlstr{"contr.treatment"}
\hlstd{gl.r.gamm.covs.t} \hlkwb{<-} \hlkwd{system.time}\hlstd{(}
  \hlstd{gl.r.gamm.covs} \hlkwb{<-} \hlkwd{bam}\hlstd{(f3} \hlopt{~} \hlstd{stress} \hlopt{+}
                          \hlkwd{s}\hlstd{(measurement.no)} \hlopt{+} \hlkwd{s}\hlstd{(measurement.no,} \hlkwc{by}\hlstd{=stress)} \hlopt{+}
                          \hlkwd{s}\hlstd{(duration)} \hlopt{+} \hlkwd{ti}\hlstd{(measurement.no, duration)} \hlopt{+}
                          \hlkwd{s}\hlstd{(decade,} \hlkwc{k}\hlstd{=}\hlnum{4}\hlstd{)} \hlopt{+}
                          \hlkwd{ti}\hlstd{(measurement.no, decade,} \hlkwc{k}\hlstd{=}\hlkwd{c}\hlstd{(}\hlnum{10}\hlstd{,}\hlnum{4}\hlstd{))} \hlopt{+}
                          \hlkwd{s}\hlstd{(measurement.no, speaker,} \hlkwc{bs}\hlstd{=}\hlstr{"fs"}\hlstd{,} \hlkwc{m}\hlstd{=}\hlnum{1}\hlstd{,} \hlkwc{k}\hlstd{=}\hlnum{4}\hlstd{),}
                        \hlkwc{dat}\hlstd{=gl.r,} \hlkwc{method}\hlstd{=}\hlstr{"ML"}\hlstd{)}
\hlstd{)}
\hlstd{gl.r.gamm.covs.t}
\end{alltt}
\begin{verbatim}
##    user  system elapsed 
##   2.728   0.105   2.851
\end{verbatim}
\end{kframe}
\end{knitrout}

\noindent The inclusion of \verb+stress+ raises an additional problem: it is possible that the effect of stress varies across subjects. In a linear mixed effects model, this issue would be dealt with through random slopes (e.g.\ one could include by-subject random slopes for the interaction between \verb+measurement.no+ and \verb+stress+ and the corresponding main terms). Although such solutions could also be explored for the current case, we'll follow a random smooth-based technique from \citet{wielingetal16} instead. We will fit separate smooths for each speaker at each level of \verb+stress+, which will allow us to capture speaker-specific trends in the \verb+stress+ effect. This can be achieved by using a combined \verb+speaker+ $\times$ \verb+stress+ variable as the grouping factor for the random smooths. The model is shown below.

\begin{knitrout}\footnotesize
\definecolor{shadecolor}{rgb}{0.969, 0.969, 0.969}\color{fgcolor}\begin{kframe}
\begin{alltt}
\hlcom{# creating combined grouping variable}
\hlstd{gl.r}\hlopt{$}\hlstd{speakerStress} \hlkwb{<-} \hlkwd{interaction}\hlstd{(gl.r}\hlopt{$}\hlstd{speaker, gl.r}\hlopt{$}\hlstd{stress)}

\hlstd{gl.r.gamm.covs.2.t} \hlkwb{<-} \hlkwd{system.time}\hlstd{(}
  \hlstd{gl.r.gamm.covs.2} \hlkwb{<-} \hlkwd{bam}\hlstd{(f3} \hlopt{~} \hlstd{stress} \hlopt{+}
                             \hlkwd{s}\hlstd{(measurement.no)} \hlopt{+} \hlkwd{s}\hlstd{(measurement.no,} \hlkwc{by}\hlstd{=stress)} \hlopt{+}
                             \hlkwd{s}\hlstd{(duration)} \hlopt{+} \hlkwd{ti}\hlstd{(measurement.no, duration)} \hlopt{+}
                             \hlkwd{s}\hlstd{(decade,} \hlkwc{k}\hlstd{=}\hlnum{4}\hlstd{)} \hlopt{+}
                             \hlkwd{ti}\hlstd{(measurement.no, decade,} \hlkwc{k}\hlstd{=}\hlkwd{c}\hlstd{(}\hlnum{10}\hlstd{,}\hlnum{4}\hlstd{))} \hlopt{+}
                             \hlkwd{s}\hlstd{(measurement.no, speakerStress,} \hlkwc{bs}\hlstd{=}\hlstr{"fs"}\hlstd{,} \hlkwc{m}\hlstd{=}\hlnum{1}\hlstd{,} \hlkwc{k}\hlstd{=}\hlnum{4}\hlstd{),}
                        \hlkwc{dat}\hlstd{=gl.r,} \hlkwc{method}\hlstd{=}\hlstr{"ML"}\hlstd{)}
\hlstd{)}
\hlstd{gl.r.gamm.covs.2.t}
\end{alltt}
\begin{verbatim}
##    user  system elapsed 
##   6.802   0.346   7.226
\end{verbatim}
\end{kframe}
\end{knitrout}

\noindent Note that the function \verb+interaction()+ simply combines \verb+speaker+ and \verb+stress+ in a single variable. To save space, the model summary is not shown, but here's a graphical summary of the effects of {\tt duration} and {\tt stress}.

\begin{knitrout}\footnotesize
\definecolor{shadecolor}{rgb}{0.969, 0.969, 0.969}\color{fgcolor}\begin{kframe}
\begin{alltt}
\hlkwd{plot_smooth}\hlstd{(gl.r.gamm.covs.2,} \hlkwc{view}\hlstd{=}\hlstr{"measurement.no"}\hlstd{,} \hlkwc{cond}\hlstd{=}\hlkwd{list}\hlstd{(}\hlkwc{duration}\hlstd{=}\hlnum{0.3}\hlstd{),}
            \hlkwc{rug}\hlstd{=F,} \hlkwc{rm.ranef}\hlstd{=T,} \hlkwc{col}\hlstd{=}\hlstr{"blue"}\hlstd{,} \hlkwc{main}\hlstd{=}\hlstr{"duration"}\hlstd{)}
\hlkwd{plot_smooth}\hlstd{(gl.r.gamm.covs.2,} \hlkwc{view}\hlstd{=}\hlstr{"measurement.no"}\hlstd{,} \hlkwc{cond}\hlstd{=}\hlkwd{list}\hlstd{(}\hlkwc{duration}\hlstd{=}\hlnum{0.1}\hlstd{),}
            \hlkwc{rug}\hlstd{=F,} \hlkwc{rm.ranef}\hlstd{=T,} \hlkwc{col}\hlstd{=}\hlstr{"red"}\hlstd{,} \hlkwc{add}\hlstd{=T)}
\hlkwd{plot_smooth}\hlstd{(gl.r.gamm.covs.2,} \hlkwc{view}\hlstd{=}\hlstr{"measurement.no"}\hlstd{,} \hlkwc{plot_all}\hlstd{=}\hlstr{"stress"}\hlstd{,}
            \hlkwc{rug}\hlstd{=F,} \hlkwc{rm.ranef}\hlstd{=T,} \hlkwc{main}\hlstd{=}\hlstr{"stress"}\hlstd{)}
\end{alltt}
\end{kframe}

{\centering \includegraphics[width=0.495\textwidth]{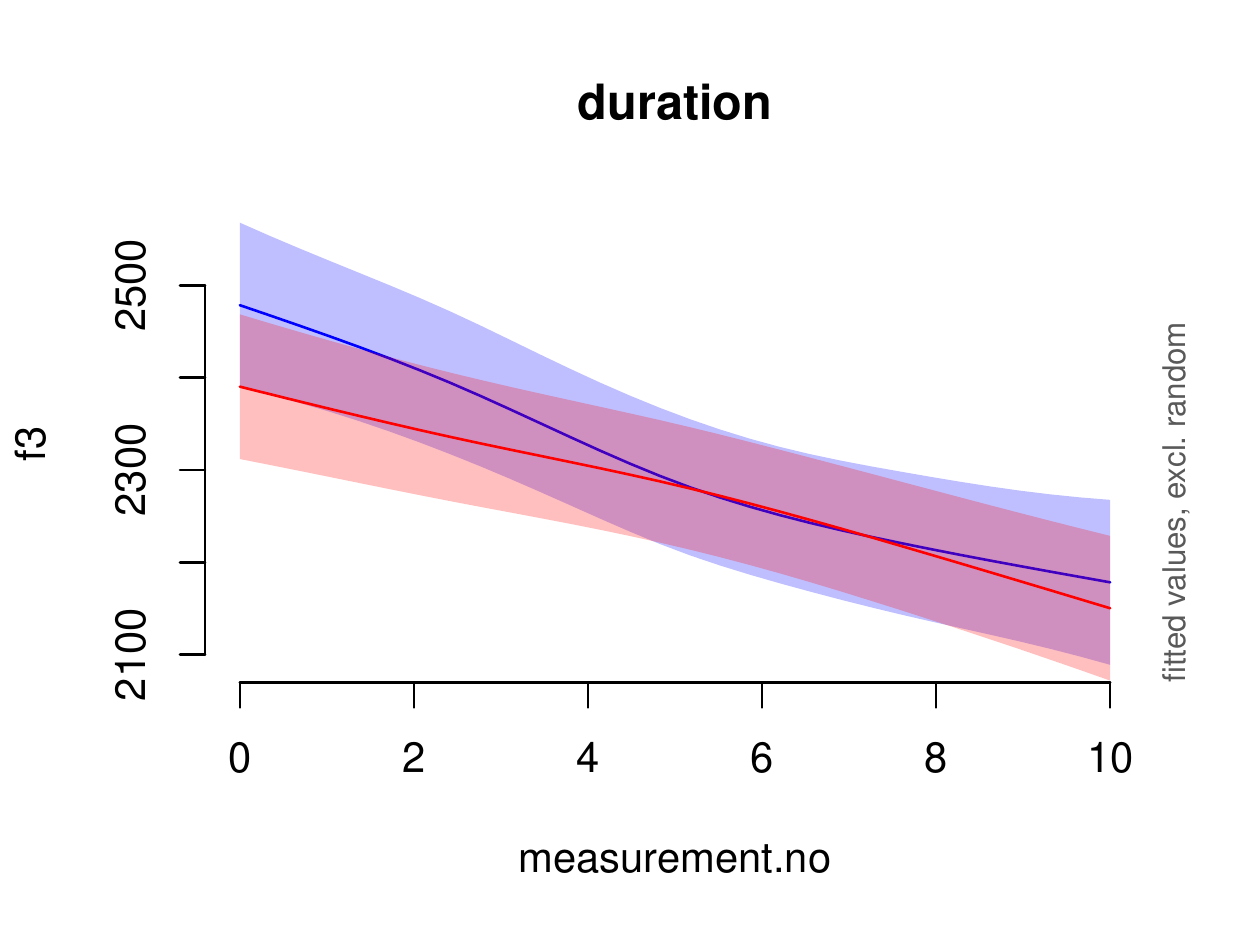} 
\includegraphics[width=0.495\textwidth]{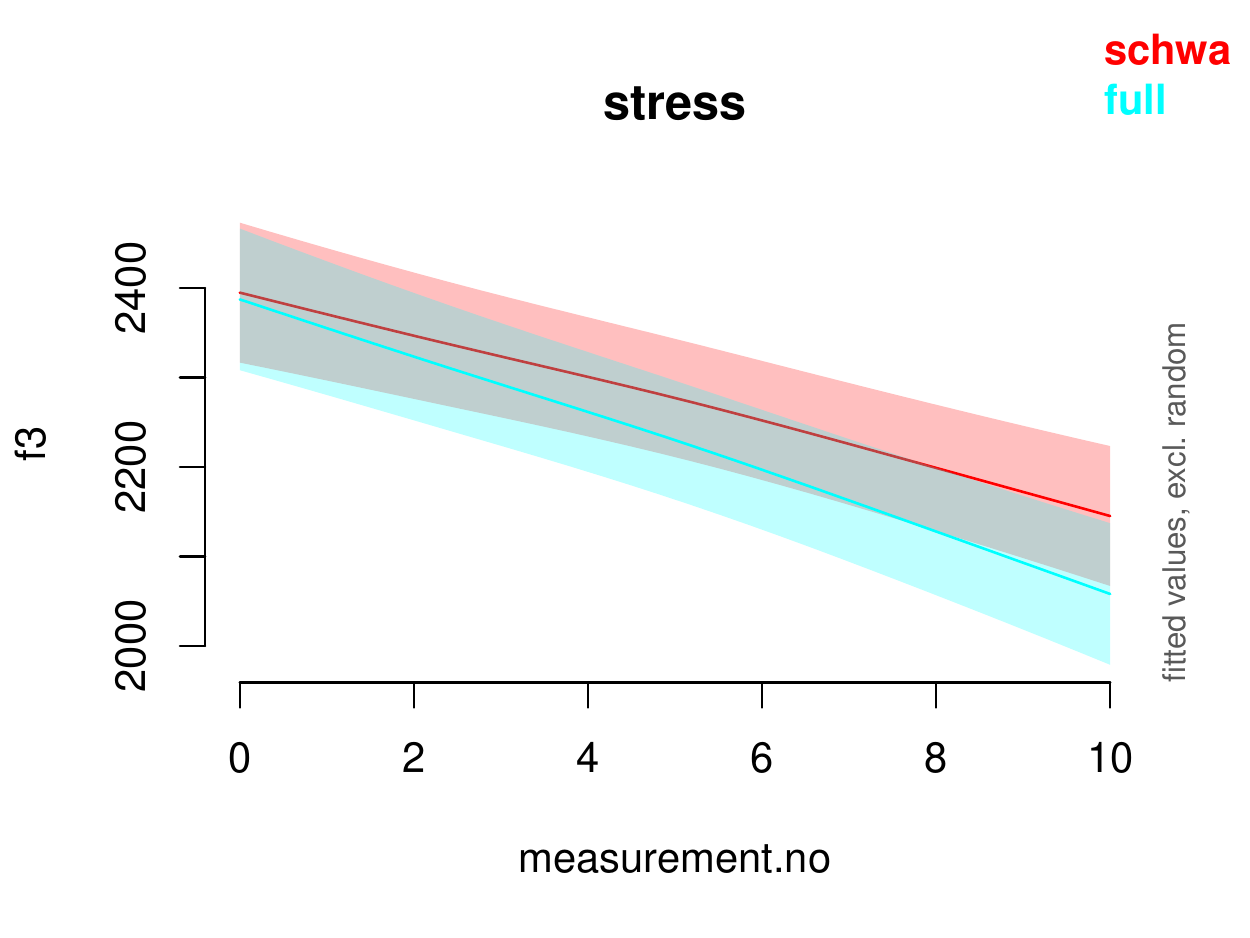} 

}

\end{knitrout}

\noindent The F3 trajectories start lower for shorter /Vr/ sequences and are a bit flatter (the red line represents a short trajectory). This suggests that short vowels are more strongly coarticulated with the following /r/. Moreover, there is a more pronounced dip near the /r/ part of the sequence for sequences with full vowels. Again, the unstressed sequence shows a flatter trajectory. One possible interpretation is that /r/ is more likely to be weakened or deleted after unstressed vowels.

Finally, let's see whether {\tt stress} interacts with the effect of {\tt decade}. Two further \verb+by+ terms need to be added: one for the \verb+decade+ main term and another one for the \verb+ti()+ interaction between \verb+measurement.no+ and \verb+decade+. Only the added terms are shown in the code chunk below.

\begin{knitrout}\footnotesize
\definecolor{shadecolor}{rgb}{0.969, 0.969, 0.969}\color{fgcolor}\begin{kframe}
\begin{alltt}
gl.r.gamm.intr <- \hlkwd{bam}(f3 ~ stress +
                           ...
                           \hlkwd{s}(decade, k=4, by=stress) + 
                           \hlkwd{ti}(measurement.no, decade, k=\hlkwd{c}(10,4), by=stress),
                        dat=gl.r, method=\hlstr{"ML"})

\hlcom{##   user  system elapsed }
\hlcom{## 12.390   0.411  12.818 }
\end{alltt}
\end{kframe}
\end{knitrout}

\noindent The summary for this model (not shown above) suggests that {\tt stress} and {\tt decade} do not interact significantly.

How does this model fare with respect to residual autocorrelation? Since we haven't yet included by-trajectory random smooths and/or an autoregressive error model, there will likely be some autocorrelation in the residuals. This is confirmed by the residual autocorrelation plot below:

\begin{knitrout}\footnotesize
\definecolor{shadecolor}{rgb}{0.969, 0.969, 0.969}\color{fgcolor}\begin{kframe}
\begin{alltt}
\hlkwd{acf_plot}\hlstd{(}\hlkwd{resid}\hlstd{(gl.r.gamm.intr),} \hlkwc{split_by}\hlstd{=}\hlkwd{list}\hlstd{(gl.r}\hlopt{$}\hlstd{traj))}
\end{alltt}
\end{kframe}

{\centering \includegraphics[width=0.495\textwidth]{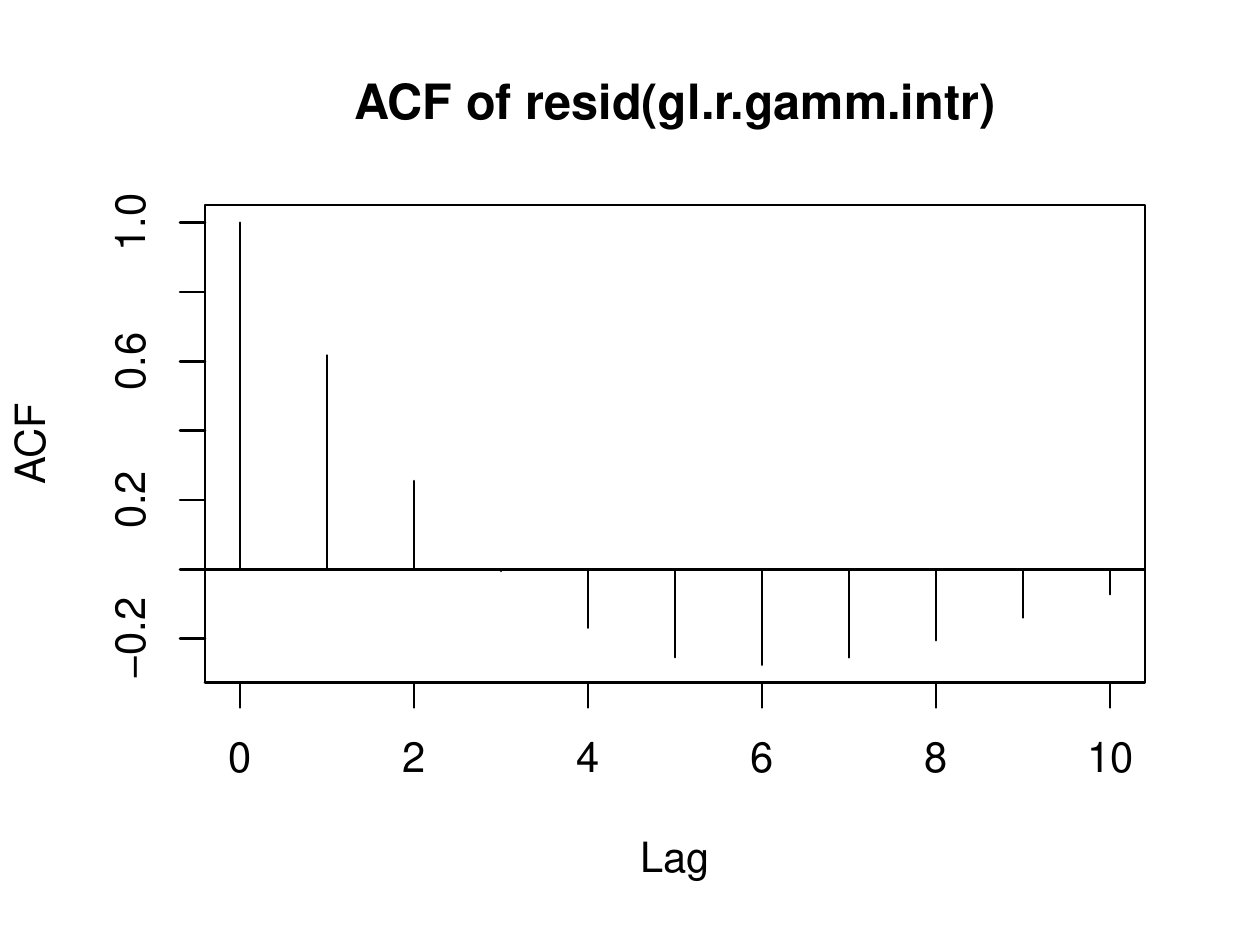} 

}

\end{knitrout}

\noindent The residual patterns look very similar to those from the simpler models of the previous section, and are due to the fact that the grouping structure in the data is ignored. This likely has an influence on the estimated effects of \verb+stress+ and \verb+duration+, which should really be estimated from differences across trajectories, not individual data points. Since the model does not include by-trajectory random smooths or an AR error model, it is not aware of dependencies among data points from the same trajectories. 

Let's first try fitting the model using by-trajectory random smooths. We'll change the estimation method to fREML as ML is simply too slow and memory-intensive. This, of course, means that we cannot check for the significance of fixed terms using model comparison. Note also that the value of \verb+discrete+ is set to \verb+TRUE+ to speed up computation. If your machine has more than one processor core, you can try setting \verb+nthreads+ to a value higher than one, which may speed up processing even further.

\begin{knitrout}\footnotesize
\definecolor{shadecolor}{rgb}{0.969, 0.969, 0.969}\color{fgcolor}\begin{kframe}
\begin{alltt}
\hlstd{gl.r.gamm.traj.t} \hlkwb{<-} \hlkwd{system.time}\hlstd{(}
  \hlstd{gl.r.gamm.traj} \hlkwb{<-} \hlkwd{bam}\hlstd{(f3} \hlopt{~} \hlstd{stress} \hlopt{+}
                         \hlkwd{s}\hlstd{(measurement.no)} \hlopt{+} \hlkwd{s}\hlstd{(measurement.no,} \hlkwc{by}\hlstd{=stress)} \hlopt{+}
                         \hlkwd{s}\hlstd{(duration)} \hlopt{+} \hlkwd{ti}\hlstd{(measurement.no, duration)} \hlopt{+}
                         \hlkwd{s}\hlstd{(decade,} \hlkwc{k}\hlstd{=}\hlnum{4}\hlstd{)} \hlopt{+} \hlkwd{s}\hlstd{(decade,} \hlkwc{k}\hlstd{=}\hlnum{4}\hlstd{,} \hlkwc{by}\hlstd{=stress)} \hlopt{+}
                         \hlkwd{ti}\hlstd{(measurement.no, decade,} \hlkwc{k}\hlstd{=}\hlkwd{c}\hlstd{(}\hlnum{10}\hlstd{,}\hlnum{4}\hlstd{))} \hlopt{+}
                         \hlkwd{ti}\hlstd{(measurement.no, decade,} \hlkwc{k}\hlstd{=}\hlkwd{c}\hlstd{(}\hlnum{10}\hlstd{,}\hlnum{4}\hlstd{),} \hlkwc{by}\hlstd{=stress)} \hlopt{+}
                         \hlkwd{s}\hlstd{(measurement.no, speakerStress,} \hlkwc{bs}\hlstd{=}\hlstr{"fs"}\hlstd{,} \hlkwc{m}\hlstd{=}\hlnum{1}\hlstd{,} \hlkwc{k}\hlstd{=}\hlnum{4}\hlstd{)} \hlopt{+}
                         \hlkwd{s}\hlstd{(measurement.no, traj,} \hlkwc{bs}\hlstd{=}\hlstr{"fs"}\hlstd{,} \hlkwc{m}\hlstd{=}\hlnum{1}\hlstd{,} \hlkwc{k}\hlstd{=}\hlnum{4}\hlstd{),}
                        \hlkwc{dat}\hlstd{=gl.r,} \hlkwc{method}\hlstd{=}\hlstr{"fREML"}\hlstd{,} \hlkwc{discrete}\hlstd{=T)}
  \hlstd{)}
\hlstd{gl.r.gamm.traj.t}
\end{alltt}
\begin{verbatim}
##    user  system elapsed 
## 220.948  19.134 243.193
\end{verbatim}
\begin{alltt}
\hlkwd{summary.coefs}\hlstd{(gl.r.gamm.traj,} \hlkwc{digits}\hlstd{=}\hlnum{3}\hlstd{)}
\end{alltt}
\begin{verbatim}
## Parametric coefficients:
##             Estimate Std. Error t value Pr(>|t|)    
## (Intercept)   2303.8       31.7   72.72   <2e-16 ***
## stressschwa     42.2       43.6    0.97     0.33    
## 
## Approximate significance of smooth terms:
##                                           edf  Ref.df      F p-value    
## s(measurement.no)                        6.43    7.35   9.65 2.7e-12 ***
## s(measurement.no):stressschwa            5.24    6.19   2.74 0.01106 *  
## s(duration)                              1.87    1.87   0.94 0.29569    
## ti(measurement.no,duration)              5.59    6.35   3.60 0.00120 ** 
## s(decade)                                1.93    1.93  10.44 0.00022 ***
## s(decade):stressschwa                    1.00    1.00   0.24 0.62743    
## ti(decade,measurement.no)               11.57   14.12   2.01 0.01198 *  
## ti(decade,measurement.no):stressschwa    6.40    8.34   0.34 0.95583    
## s(measurement.no,speakerStress)         45.31   94.00   1.54 < 2e-16 ***
## s(measurement.no,traj)                1536.71 1676.00 101.81 < 2e-16 ***
\end{verbatim}
\end{kframe}
\end{knitrout}

\noindent The same model fit with an AR1 error term is shown on the next page.
\newpage
\begin{knitrout}\footnotesize
\definecolor{shadecolor}{rgb}{0.969, 0.969, 0.969}\color{fgcolor}\begin{kframe}
\begin{alltt}
\hlcom{# marking start of trajectories}
\hlstd{gl.r}\hlopt{$}\hlstd{start.event} \hlkwb{<-} \hlstd{gl.r}\hlopt{$}\hlstd{measurement.no} \hlopt{==} \hlnum{0}
\hlcom{# getting rough estimate of autocorrelation parameter}
\hlstd{gl.autocorr} \hlkwb{<-} \hlkwd{start_value_rho}\hlstd{(gl.r.gamm.intr)}

\hlcom{# fit model}
\hlstd{gl.r.gamm.AR.t} \hlkwb{<-} \hlkwd{system.time}\hlstd{(}
  \hlstd{gl.r.gamm.AR} \hlkwb{<-} \hlkwd{bam}\hlstd{(f3} \hlopt{~} \hlstd{stress} \hlopt{+}
                         \hlkwd{s}\hlstd{(measurement.no)} \hlopt{+} \hlkwd{s}\hlstd{(measurement.no,} \hlkwc{by}\hlstd{=stress)} \hlopt{+}
                         \hlkwd{s}\hlstd{(duration)} \hlopt{+} \hlkwd{ti}\hlstd{(measurement.no, duration)} \hlopt{+}
                         \hlkwd{s}\hlstd{(decade,} \hlkwc{k}\hlstd{=}\hlnum{4}\hlstd{)} \hlopt{+} \hlkwd{s}\hlstd{(decade,} \hlkwc{k}\hlstd{=}\hlnum{4}\hlstd{,} \hlkwc{by}\hlstd{=stress)} \hlopt{+}
                         \hlkwd{ti}\hlstd{(measurement.no, decade,} \hlkwc{k}\hlstd{=}\hlkwd{c}\hlstd{(}\hlnum{10}\hlstd{,}\hlnum{4}\hlstd{))} \hlopt{+}
                         \hlkwd{ti}\hlstd{(measurement.no, decade,} \hlkwc{k}\hlstd{=}\hlkwd{c}\hlstd{(}\hlnum{10}\hlstd{,}\hlnum{4}\hlstd{),} \hlkwc{by}\hlstd{=stress)} \hlopt{+}
                         \hlkwd{s}\hlstd{(measurement.no, speakerStress,} \hlkwc{bs}\hlstd{=}\hlstr{"fs"}\hlstd{,} \hlkwc{m}\hlstd{=}\hlnum{1}\hlstd{,} \hlkwc{k}\hlstd{=}\hlnum{4}\hlstd{),}
                        \hlkwc{dat}\hlstd{=gl.r,} \hlkwc{method}\hlstd{=}\hlstr{"ML"}\hlstd{,}
                        \hlkwc{AR.start}\hlstd{=gl.r}\hlopt{$}\hlstd{start.event,} \hlkwc{rho}\hlstd{=gl.autocorr)}
  \hlstd{)}
\hlstd{gl.r.gamm.AR.t}
\end{alltt}
\begin{verbatim}
##    user  system elapsed 
##  19.309   0.868  20.683
\end{verbatim}
\begin{alltt}
\hlkwd{summary.coefs}\hlstd{(gl.r.gamm.AR,} \hlkwc{digits}\hlstd{=}\hlnum{3}\hlstd{)}
\end{alltt}
\begin{verbatim}
## Parametric coefficients:
##             Estimate Std. Error t value Pr(>|t|)    
## (Intercept)   2295.4       28.6   80.16   <2e-16 ***
## stressschwa     47.3       39.9    1.19     0.24    
## 
## Approximate significance of smooth terms:
##                                         edf Ref.df     F p-value    
## s(measurement.no)                      4.78   6.07  9.84 6.9e-11 ***
## s(measurement.no):stressschwa          1.01   1.01  1.38   0.241    
## s(duration)                            2.35   2.99  2.94   0.035 *  
## ti(measurement.no,duration)            6.26   8.10  4.61 1.2e-05 ***
## s(decade)                              1.50   1.52 11.49 6.7e-05 ***
## s(decade):stressschwa                  1.00   1.00  0.37   0.542    
## ti(measurement.no,decade)              2.45   3.13  1.61   0.190    
## ti(measurement.no,decade):stressschwa  1.01   1.01  0.04   0.846    
## s(measurement.no,speakerStress)       67.45  92.00  6.39 < 2e-16 ***
\end{verbatim}
\end{kframe}
\end{knitrout}

\noindent First of all, note that the AR model converges about 10 times faster than the model with random smooths, even though it is fitted with ML, which is generally a bit slower (ML was chosen so that we can perform model comparison later). There are also differences in significance across the two models. For some variables, the model with random smooths is more conservative (\verb+duration+); for others, it is less conservative (\verb+stress+). Such differences only really occur in cases where the \textit{p}-values are relatively close to 0.05. The fixed effects that are significant in both models are \verb+s(decade)+ and \verb+ti(measurement.no,duration)+.

The residual autocorrelation plots for the two models are shown below. The results are very similar to those from the previous sections: the model with random smooths by trajectory introduces an artefactual negative autocorrelation at lag 2 (though, as before, there is little residual variance left in the data, so this may be less problematic than it appears from the autocorrelation plot), while the model with an autoregressive error model gets rid of most of the autocorrelation without introducing any artefacts.

\begin{knitrout}\footnotesize
\definecolor{shadecolor}{rgb}{0.969, 0.969, 0.969}\color{fgcolor}\begin{kframe}
\begin{alltt}
\hlcom{# autocorrelation plots}
\hlkwd{acf_resid}\hlstd{(gl.r.gamm.traj,} \hlkwc{split_pred}\hlstd{=}\hlstr{"traj"}\hlstd{)}
\hlkwd{acf_resid}\hlstd{(gl.r.gamm.AR,} \hlkwc{split_pred}\hlstd{=}\hlstr{"AR.start"}\hlstd{)}
\end{alltt}
\end{kframe}

{\centering \includegraphics[width=0.495\textwidth]{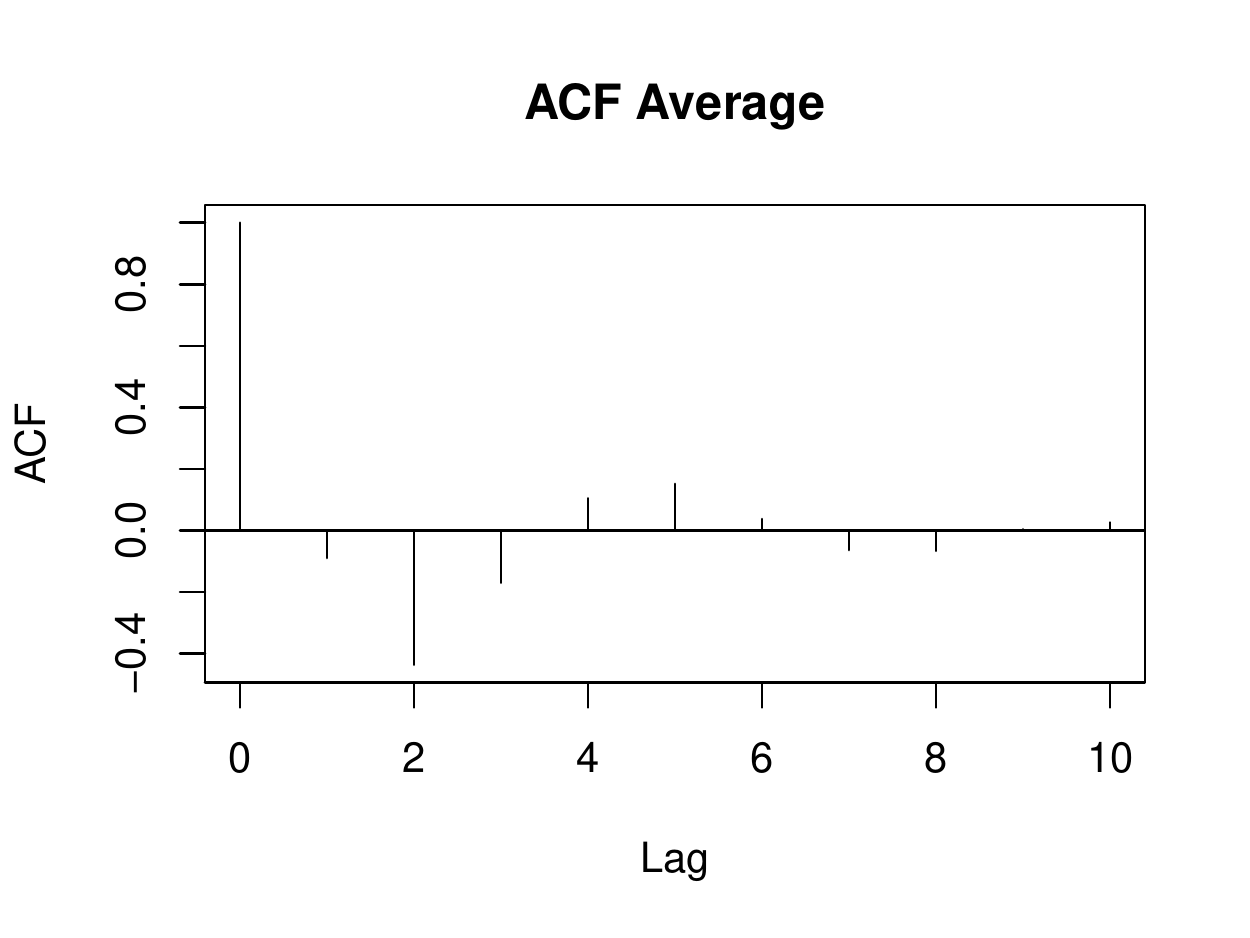} 
\includegraphics[width=0.495\textwidth]{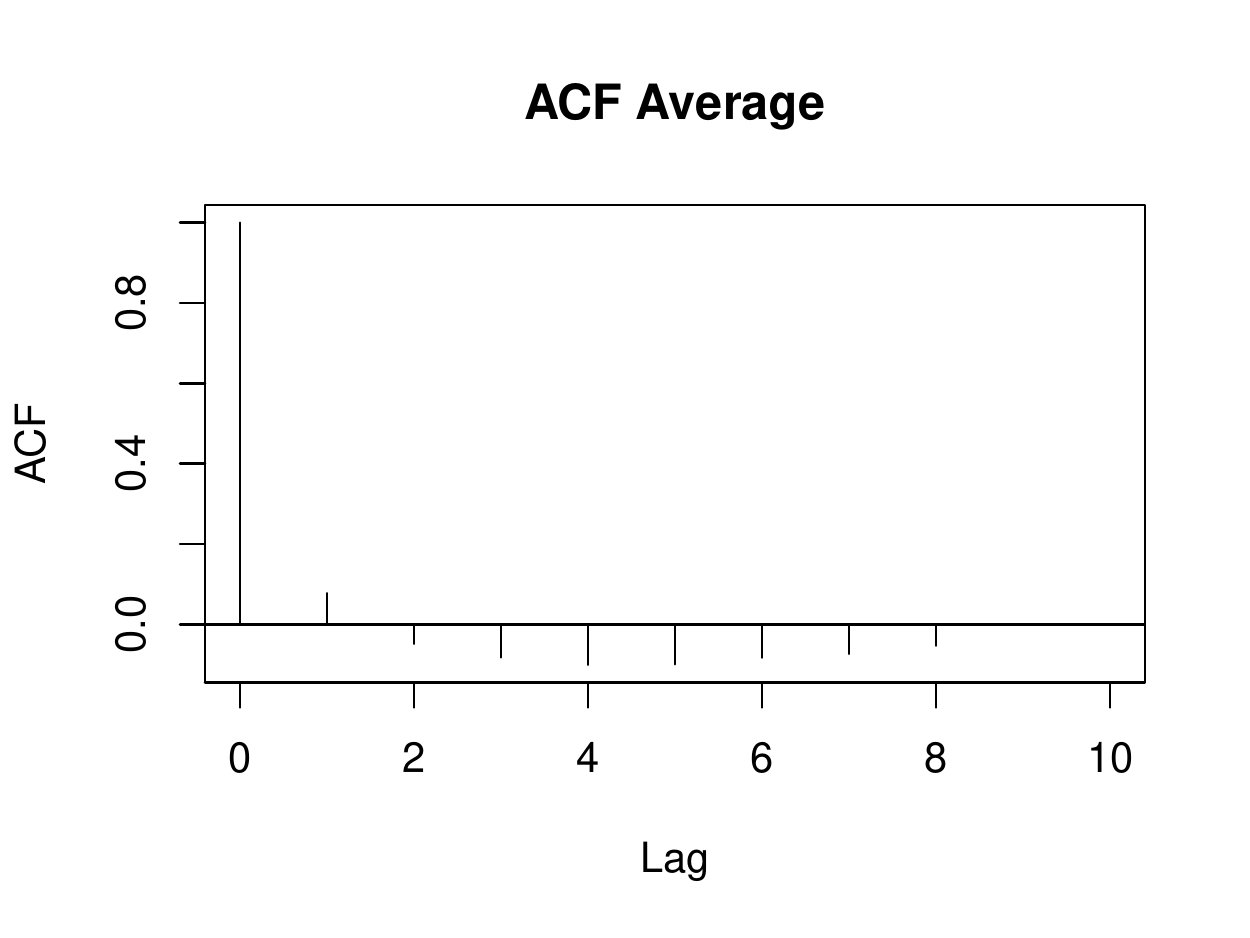} 

}

\end{knitrout}

\noindent Let's compare the predictions of the models. The plots below show the estimated {\tt decade} effect (for schwa) from the two models. 

\begin{knitrout}\footnotesize
\definecolor{shadecolor}{rgb}{0.969, 0.969, 0.969}\color{fgcolor}\begin{kframe}
\begin{alltt}
\hlkwd{plot_smooth.cont}\hlstd{(gl.r.gamm.traj,} \hlkwc{view}\hlstd{=}\hlstr{"measurement.no"}\hlstd{,} \hlkwc{plot_all.c}\hlstd{=}\hlstr{"decade"}\hlstd{,}
                 \hlkwc{cond}\hlstd{=}\hlkwd{list}\hlstd{(}\hlkwc{stress}\hlstd{=}\hlstr{"schwa"}\hlstd{),} \hlkwc{rug}\hlstd{=F,} \hlkwc{rm.ranef}\hlstd{=T,}
                 \hlkwc{main}\hlstd{=}\hlstr{"decade - random smooths"}\hlstd{,} \hlkwc{ylim}\hlstd{=}\hlkwd{c}\hlstd{(}\hlnum{1900}\hlstd{,}\hlnum{2700}\hlstd{))}
\hlkwd{plot_smooth.cont}\hlstd{(gl.r.gamm.AR,} \hlkwc{view}\hlstd{=}\hlstr{"measurement.no"}\hlstd{,} \hlkwc{plot_all.c}\hlstd{=}\hlstr{"decade"}\hlstd{,}
                 \hlkwc{cond}\hlstd{=}\hlkwd{list}\hlstd{(}\hlkwc{stress}\hlstd{=}\hlstr{"schwa"}\hlstd{),} \hlkwc{rug}\hlstd{=F,} \hlkwc{rm.ranef}\hlstd{=T,}
                 \hlkwc{main}\hlstd{=}\hlstr{"decade - AR"}\hlstd{,} \hlkwc{ylim}\hlstd{=}\hlkwd{c}\hlstd{(}\hlnum{1900}\hlstd{,}\hlnum{2700}\hlstd{))}
\end{alltt}
\end{kframe}

{\centering \includegraphics[width=0.495\textwidth]{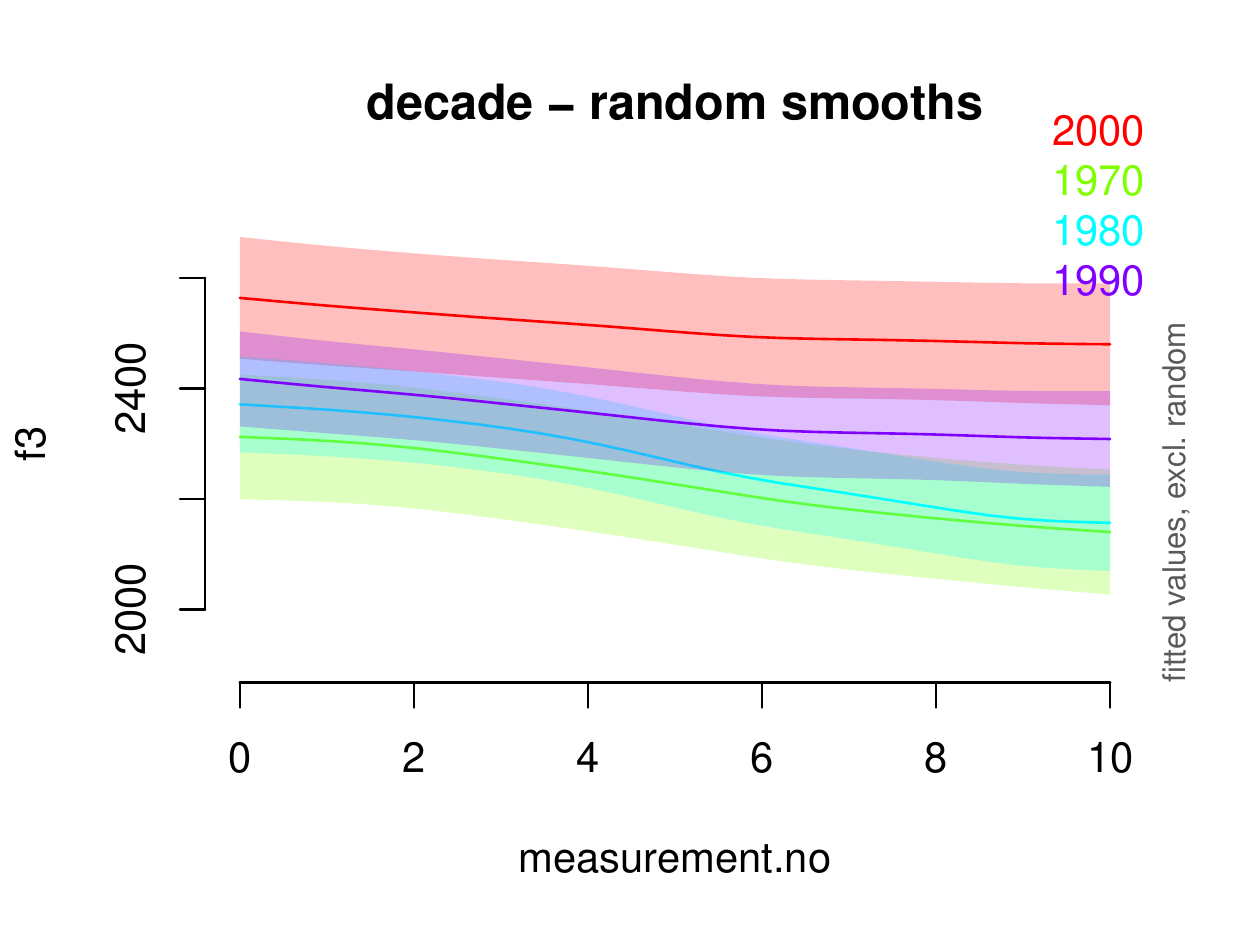} 
\includegraphics[width=0.495\textwidth]{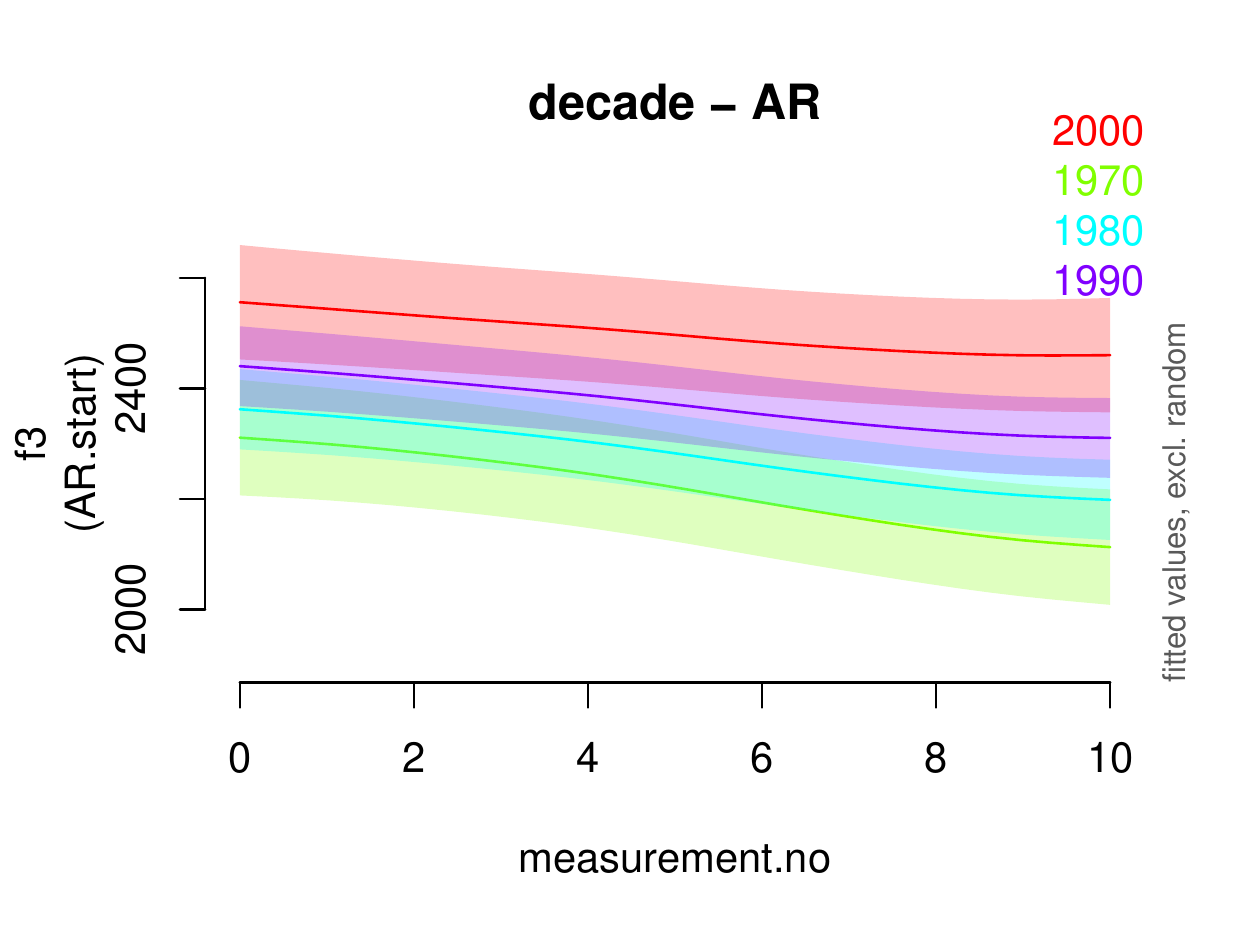} 

}

\end{knitrout}

\noindent Though the two plots are very similar, the estimated trajectory shapes are slightly different, especially for the 1980s speakers (this is likely a reflection of the fact that the smooth interaction between \verb+decade+ and \verb+measurement.no+ is significant in the model with random smooths, but not in the one with an AR1 model). The trajectories estimated by the GAMM with the AR model are also generally a bit smoother.

Here are the predictions for stressed vowel + /r/ \textit{vs.} unstressed vowel + /r/ sequences.

\begin{knitrout}\footnotesize
\definecolor{shadecolor}{rgb}{0.969, 0.969, 0.969}\color{fgcolor}\begin{kframe}
\begin{alltt}
\hlkwd{plot_smooth}\hlstd{(gl.r.gamm.traj,} \hlkwc{view}\hlstd{=}\hlstr{"measurement.no"}\hlstd{,} \hlkwc{plot_all}\hlstd{=}\hlstr{"stress"}\hlstd{,}
            \hlkwc{rug}\hlstd{=F,} \hlkwc{rm.ranef}\hlstd{=T,} \hlkwc{main}\hlstd{=}\hlstr{"stress - random smooths"}\hlstd{,} \hlkwc{ylim}\hlstd{=}\hlkwd{c}\hlstd{(}\hlnum{2000}\hlstd{,}\hlnum{2450}\hlstd{))}
\hlkwd{plot_smooth}\hlstd{(gl.r.gamm.AR,} \hlkwc{view}\hlstd{=}\hlstr{"measurement.no"}\hlstd{,} \hlkwc{plot_all}\hlstd{=}\hlstr{"stress"}\hlstd{,}
            \hlkwc{rug}\hlstd{=F,} \hlkwc{rm.ranef}\hlstd{=T,} \hlkwc{main}\hlstd{=}\hlstr{"stress - AR"}\hlstd{,} \hlkwc{ylim}\hlstd{=}\hlkwd{c}\hlstd{(}\hlnum{2000}\hlstd{,}\hlnum{2450}\hlstd{))}
\end{alltt}
\end{kframe}

{\centering \includegraphics[width=0.495\textwidth]{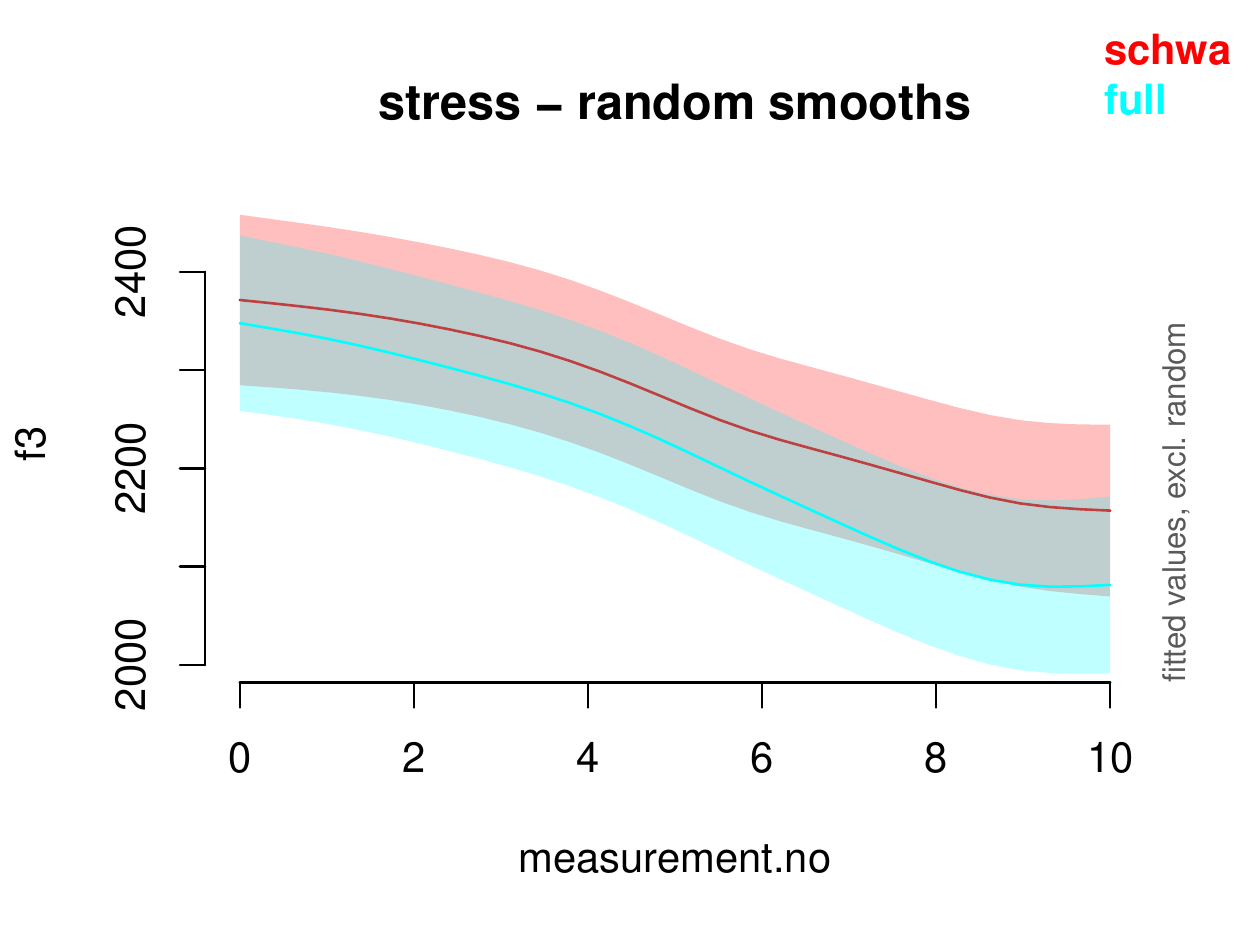} 
\includegraphics[width=0.495\textwidth]{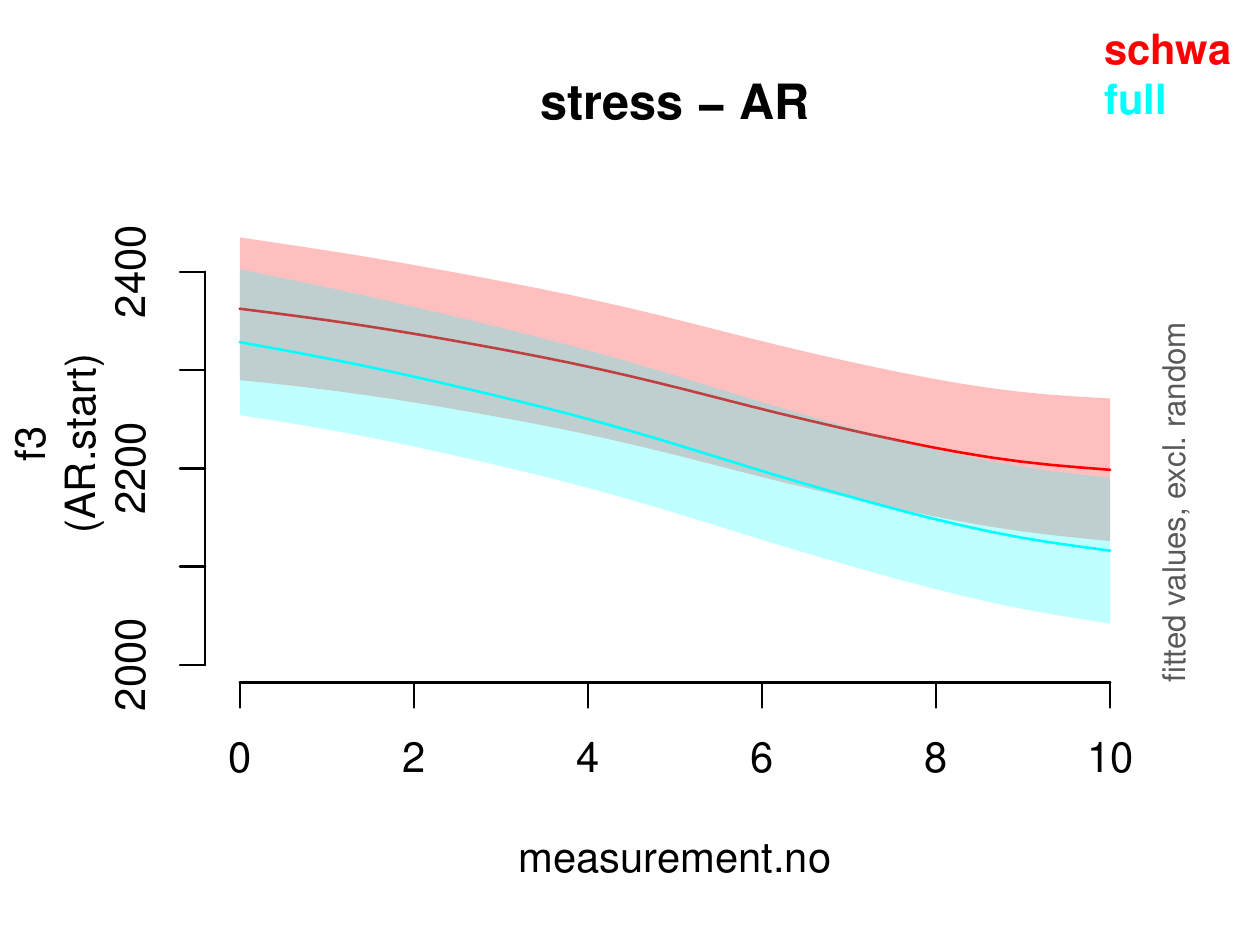} 

}

\end{knitrout}

\noindent Again, the estimated trajectories look similar, but the AR ones are somewhat smoother (and since they run almost in parallel, the shape difference that is significant for the random smooth model is not significant here -- see the model summaries).

The predictions for V + /r/ sequences with short \textit{vs.} long durations:

\begin{knitrout}\footnotesize
\definecolor{shadecolor}{rgb}{0.969, 0.969, 0.969}\color{fgcolor}\begin{kframe}
\begin{alltt}
\hlkwd{plot_smooth}\hlstd{(gl.r.gamm.traj,} \hlkwc{view}\hlstd{=}\hlstr{"measurement.no"}\hlstd{,} \hlkwc{cond}\hlstd{=}\hlkwd{list}\hlstd{(}\hlkwc{duration}\hlstd{=}\hlnum{0.3}\hlstd{),}
            \hlkwc{rug}\hlstd{=F,} \hlkwc{rm.ranef}\hlstd{=T,} \hlkwc{col}\hlstd{=}\hlstr{"blue"}\hlstd{,} \hlkwc{main}\hlstd{=}\hlstr{"duration - random smooths"}\hlstd{,}
            \hlkwc{ylim}\hlstd{=}\hlkwd{c}\hlstd{(}\hlnum{2050}\hlstd{,}\hlnum{2550}\hlstd{))}
\hlkwd{plot_smooth}\hlstd{(gl.r.gamm.traj,} \hlkwc{view}\hlstd{=}\hlstr{"measurement.no"}\hlstd{,} \hlkwc{cond}\hlstd{=}\hlkwd{list}\hlstd{(}\hlkwc{duration}\hlstd{=}\hlnum{0.1}\hlstd{),}
            \hlkwc{rug}\hlstd{=F,} \hlkwc{rm.ranef}\hlstd{=T,} \hlkwc{col}\hlstd{=}\hlstr{"red"}\hlstd{,} \hlkwc{add}\hlstd{=T)}
\hlkwd{plot_smooth}\hlstd{(gl.r.gamm.AR,} \hlkwc{view}\hlstd{=}\hlstr{"measurement.no"}\hlstd{,} \hlkwc{cond}\hlstd{=}\hlkwd{list}\hlstd{(}\hlkwc{duration}\hlstd{=}\hlnum{0.3}\hlstd{),}
            \hlkwc{rug}\hlstd{=F,} \hlkwc{rm.ranef}\hlstd{=T,} \hlkwc{col}\hlstd{=}\hlstr{"blue"}\hlstd{,} \hlkwc{main}\hlstd{=}\hlstr{"duration - AR"}\hlstd{,}
            \hlkwc{ylim}\hlstd{=}\hlkwd{c}\hlstd{(}\hlnum{2050}\hlstd{,}\hlnum{2550}\hlstd{))}
\hlkwd{plot_smooth}\hlstd{(gl.r.gamm.AR,} \hlkwc{view}\hlstd{=}\hlstr{"measurement.no"}\hlstd{,} \hlkwc{cond}\hlstd{=}\hlkwd{list}\hlstd{(}\hlkwc{duration}\hlstd{=}\hlnum{0.1}\hlstd{),}
            \hlkwc{rug}\hlstd{=F,} \hlkwc{rm.ranef}\hlstd{=T,} \hlkwc{col}\hlstd{=}\hlstr{"red"}\hlstd{,} \hlkwc{add}\hlstd{=T)}
\end{alltt}
\end{kframe}

{\centering \includegraphics[width=0.495\textwidth]{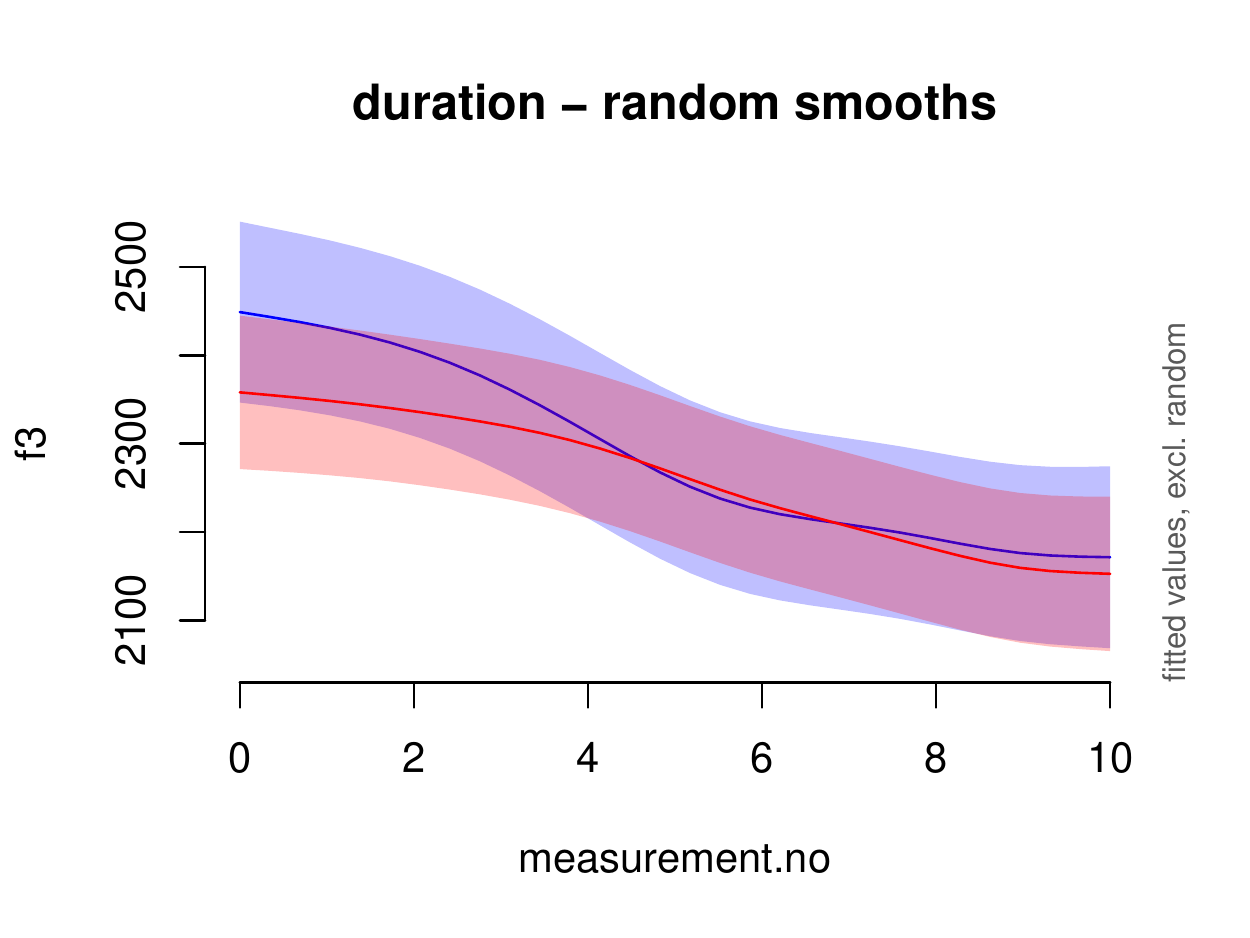} 
\includegraphics[width=0.495\textwidth]{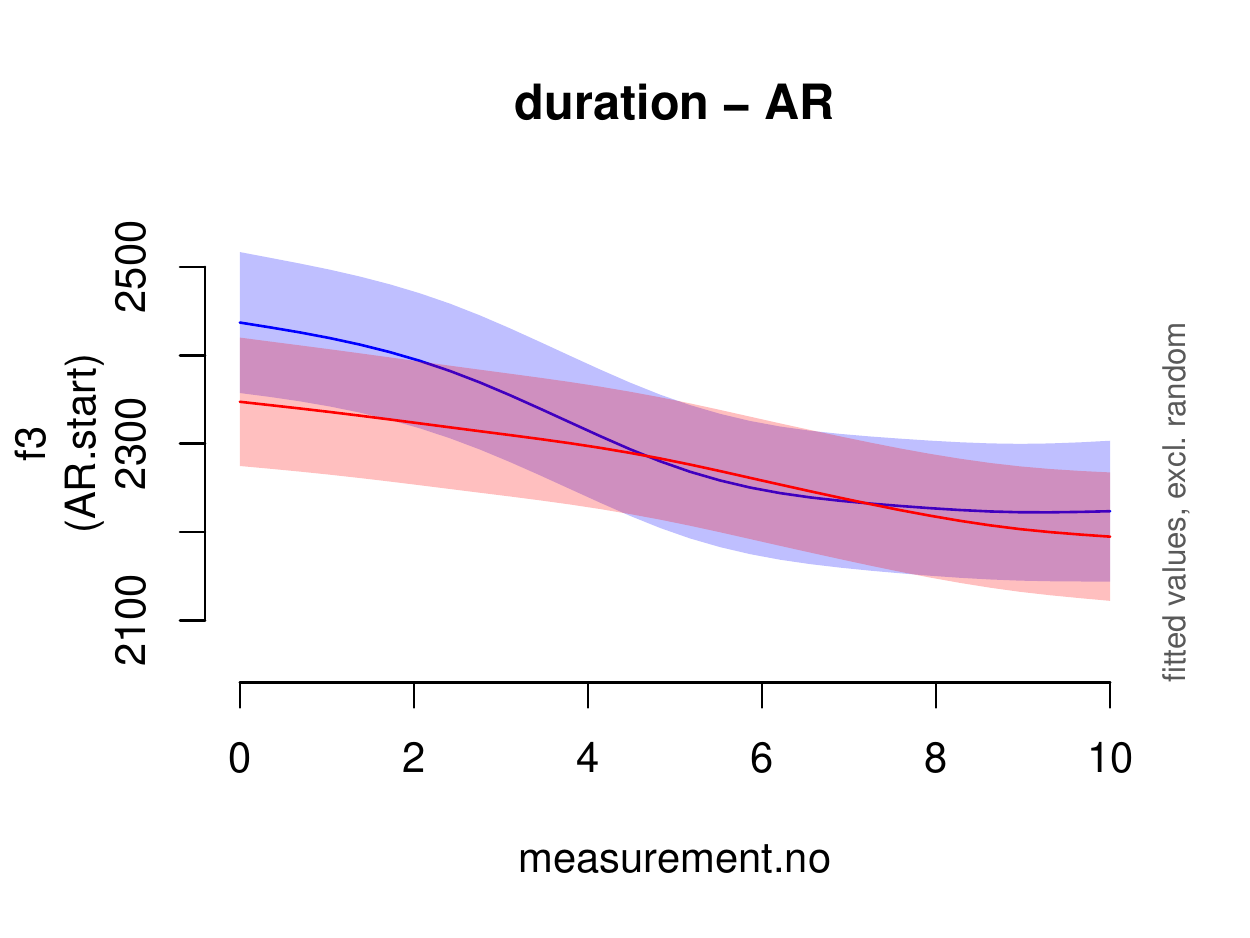} 

}

\end{knitrout}

\noindent Though the difference in smoothness seen for the previous graphs is also observed here, the general conclusions from the two models with respect to duration are very similar.

The plot below shows the estimated random smooths for different speakers from the model with random smooths. In order to get the graph to display the correct set of random smooths, we need to specify it using a number, which actually corresponds to its position in the smooth part of the model summary (i.e.\ it's on the 9th line in the smooth summary).
\newpage
\begin{knitrout}\footnotesize
\definecolor{shadecolor}{rgb}{0.969, 0.969, 0.969}\color{fgcolor}\begin{kframe}
\begin{alltt}
\hlkwd{inspect_random}\hlstd{(gl.r.gamm.traj,} \hlkwc{select}\hlstd{=}\hlnum{9}\hlstd{,} \hlkwc{lwd}\hlstd{=}\hlnum{3}\hlstd{,}
               \hlkwc{main}\hlstd{=}\hlstr{"random smooths by speaker"}\hlstd{)}
\end{alltt}
\end{kframe}

{\centering \includegraphics[width=0.495\textwidth]{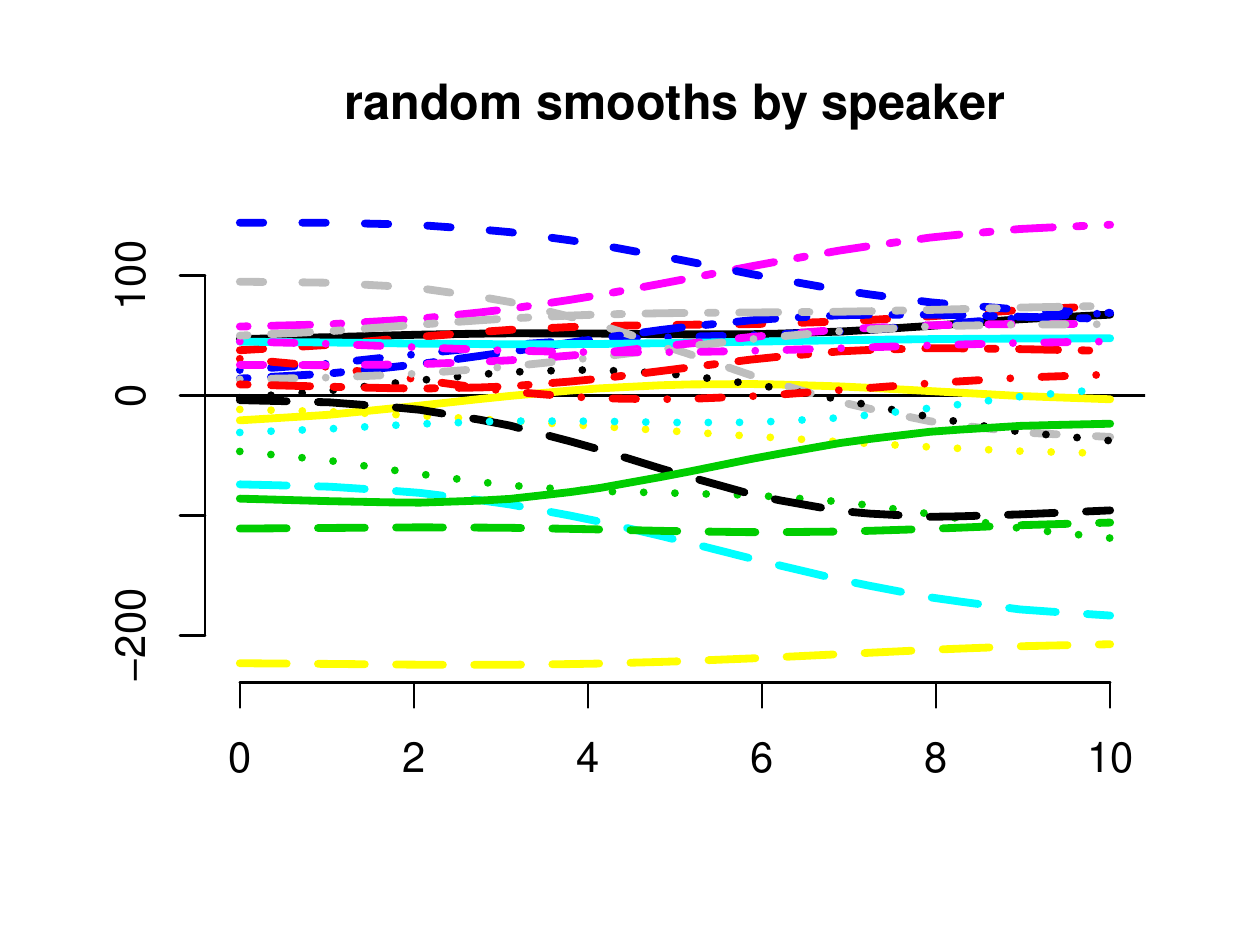} 

}

\end{knitrout}

\noindent Finally, we'll perform a model comparison to check whether the effect of decade is significant. The comparison is between the full GAMM with an AR model and a nested model that excludes \textit{all} terms with {\tt decade}. The comparison is shown below:

\begin{knitrout}\footnotesize
\definecolor{shadecolor}{rgb}{0.969, 0.969, 0.969}\color{fgcolor}\begin{kframe}
\begin{alltt}
\hlstd{gl.r.gamm.AR.0} \hlkwb{<-} \hlkwd{bam}\hlstd{(f3} \hlopt{~} \hlstd{stress} \hlopt{+}
                           \hlkwd{s}\hlstd{(measurement.no)} \hlopt{+} \hlkwd{s}\hlstd{(measurement.no,} \hlkwc{by}\hlstd{=stress)} \hlopt{+}
                           \hlkwd{s}\hlstd{(duration)} \hlopt{+} \hlkwd{ti}\hlstd{(measurement.no, duration)} \hlopt{+}
                           \hlkwd{s}\hlstd{(measurement.no, speakerStress,} \hlkwc{bs}\hlstd{=}\hlstr{"fs"}\hlstd{,} \hlkwc{m}\hlstd{=}\hlnum{1}\hlstd{,} \hlkwc{k}\hlstd{=}\hlnum{4}\hlstd{),}
                      \hlkwc{dat}\hlstd{=gl.r,} \hlkwc{method}\hlstd{=}\hlstr{"ML"}\hlstd{,}
                      \hlkwc{AR.start}\hlstd{=gl.r}\hlopt{$}\hlstd{start.event,} \hlkwc{rho}\hlstd{=gl.autocorr)}
\hlkwd{compareML}\hlstd{(gl.r.gamm.AR, gl.r.gamm.AR.0,} \hlkwc{print.output}\hlstd{=F)}\hlopt{$}\hlstd{table}
\end{alltt}
\begin{verbatim}
##            Model    Score Edf  Chisq     Df p.value Sig.
## 1 gl.r.gamm.AR.0 26391.20  13                           
## 2   gl.r.gamm.AR 26377.75  23 13.450 10.000   0.003  **
\end{verbatim}
\end{kframe}
\end{knitrout}

\noindent The difference between the models is significant, indicating that there is indeed a change in F3 as a function of time. This concludes our analysis of the Glasgow /r/ data set.

One remaining question to answer is whether we should prefer the GAMM with random smooths by trajectory or the one with an AR model -- or should we perhaps combine them? In this case, combining the two methods is probably not a good idea: though the random smooths introduce an artefactual correlation, an AR1 model wouldn't be able to deal with this issue appropriately, as it is at lag 2, not lag 1. There are no such artefacts in the AR1 version of the model, which tips the balance in favour of the latter model. Moreover, the AR1 version can be fitted using ML, which in turn makes model comparisons possible. In contrast, the version of the model with random smooths is too complex to be fitted using ML, so model comparison based on fixed effects is not an option. Finally, the AR1 version can be fitted in a much shorter time. This may not be a knock-down argument in the current case, but it is a very important consideration for data sets with a larger number of trajectories. Therefore, although the overall conclusions from the two models are similar, a GAMM with an AR1 error model seems preferable to a GAMM with random smooths by trajectory for the Glasgow /r/ data set.

It should be noted that while these arguments may hold for dynamic formant data of the type presented here, there is no one-size-fits-all solution for this issue, and the choice between the two options should be guided by careful consideration of the data and model criticism (e.g.\ inspection of residual autocorrelation plots). There may be also be cases where both types of structures are needed to account for patterns in the residuals. For more information on these issues, I strongly recommend \citet{baayenetal16} and \citet{baayenetal17}: both papers offer plenty of advice on using by-trajectory random smooths and autoregressive error models, and also illustrate their use through a wide variety of linguistic examples.

\section{Final comments}

The goal of this paper was to provide an introduction to GAMMs in the context of dynamic speech analysis. We have discussed a range of theoretical concepts and gone through two example data sets, covering many of the questions and issues that come up when working with these models. However, there are many more issues that I simply had to leave out in order to keep things short:

\begin{itemize}
  \item checking modelling assumptions: normal distribution of residuals, homoscedasticity
  \item dealing with violations of these assumptions
  \item other smoother types such as adaptive smoothers and cyclic smoothers (might be appropriate for pitch contours)
  \item GAMMs for two-dimensional spatial data
\end{itemize}

\subsection{Useful references}

Below is a short (and incomplete) list of useful GAMM references. I consulted many of these while putting this introduction together.

\begin{itemize}
  \item \citet{wood06}: The standard reference for GAMMs. It is an excellent book with a lot of detail, and can be very useful for finding out more about the methods and assumptions underlying GAMMs. Though many of the sections assume a strong background in mathematics, most of the example analyses and parts of the conceptual discussion are possible to follow without mathematical training.
  \item \href{https://arxiv.org/pdf/1601.02043v1.pdf}{\begin{NoHyper}\citet{baayenetal16}\end{NoHyper}}: A paper that focuses specifically on modelling autocorrelation using GAMMs, and presents analyses for three separate linguistic examples. It also includes some complicated GAMMs (even some with three-way interactions).
  \item \href{https://arxiv.org/abs/1511.03120}{\begin{NoHyper}\citet{baayenetal17}\end{NoHyper}}: A paper that looks at typical patterns of temporal autocorrelation in data obtained from humans. Provides advice on handling autocorrelation as well as discussion of other modelling issues (e.g.\ the difference between exploratory and confirmatory analyses).
  \item \href{https://rstudio-pubs-static.s3.amazonaws.com/24589_7552e489485b4c2790ea6634e1afd68d.html}{\begin{NoHyper}\citet{kelly14}\end{NoHyper}}: An online tutorial. This one covers not just GAMMs, but also a range of other methods for dealing with non-linearity, and makes various comparisons across these methods. Note that this tutorial relies on the {\tt gam} library, not {\tt mgcv}.
  \item \href{http://www.fromthebottomoftheheap.net/2014/05/09/modelling-seasonal-data-with-gam/}{\begin{NoHyper}\citet{simpson14a}\end{NoHyper}}: An interesting blog post that shows how to model seasonal data using GAMMs. It also explains basis functions in a clear and concise way.
  \item \href{http://www.fromthebottomoftheheap.net/2011/06/12/additive-modelling-and-the-hadcrut3v-global-mean-temperature-series/}{\begin{NoHyper}\citet{simpson14b}\end{NoHyper}}: Another great blog post with a lot of useful detail on autocorrelation components. It also has an interesting section on generating confidence intervals for the derivative of a smooth function.
  \item \href{http://www.sfs.uni-tuebingen.de/~jvanrij/Tutorial/GAMM.html}{\begin{NoHyper}\citet{vanrij15}\end{NoHyper}}: A great online tutorial that covers many of the technical aspects of fitting GAMMs. It also includes autocorrelation and model comparisons.
  \item \href{https://cran.r-project.org/web/packages/itsadug/vignettes/test.html}{\begin{NoHyper}\citet{vanrij16}\end{NoHyper}}: A brief R vignette that describes three different methods for significance testing using GAMMs. Note that some of the model comparisons are based on GAMMs fitted with fREML, which may lead to unreliable results (this is noted in the text as well).
  \item \href{https://academic.oup.com/jole/article/1/1/7/2281883/How-to-analyze-linguistic-change-using-mixed}{\begin{NoHyper}\citet{winterandwieling16}\end{NoHyper}}: Discusses GAMMs in the context of language change and evolution and provides a clear and concise introduction to GAMMs and mixed models in general. It also covers a number of topics that are not discussed here, such as autocorrelation and logistic / Poisson GAMMs. It comes with thoroughly commented example code that is also available at\\ \url{https://github.com/bodowinter/change_tutorial_materials}.
  \item \href{http://www.let.rug.nl/wieling/statscourse}{\begin{NoHyper}\citet{wieling17stats}\end{NoHyper}}: Lecture slides for a five-day workshop on advanced regression methods. Lectures 3--5 provide a very detailed introduction to GAMMs with tons of useful examples illustrating ways of dealing with autocorrelation, non-normally distributed residuals and also a range of different types of data (including ERP and articulatory data).
\end{itemize}

\bibliographystyle{apalike2}
\bibliography{/users/soskuthy/library/texmf/bibtex/bib/local/soskuthy_master}

\appendix

\section*{Appendix: R packages and functions for fitting GAMMs}\label{sec:packages}

There are a number of different functions that can be used for fitting GAMMs in R: \verb+gam()+, \verb+bam()+, \verb+gamm()+ (all three from the package {\tt mgcv}) and \verb+gamm4()+ (from the package {\tt gamm4}). These functions have slightly different strength and weaknesses. Some of the discussion below is fairly technical, and probably makes more sense to readers who already have a bit of experience with GAMMs. I have included it here as I think it may be useful for reference.

{\tt gam()/bam()} are the best documented and most reliable tools for fitting GAMMs, and most existing tutorials focus on these functions. Although these are separate functions, they can be used in very similar ways. \verb+bam()+ is in many ways a more advanced version of \verb+gam+ that can be much faster than \verb+gam()+ and uses less memory, so it is the preferred option for large or complex data sets. {\tt bam()} can also be run in parallel on multiple processor cores and includes an option for further performance gains (these can be accessed by setting {\tt discrete=TRUE} and specifying the number of parallel threads using the {\tt nthreads} option). {\tt bam()} also allows the inclusion of a simple autoregressive error model of order 1 (AR1), which can capture some patterns of autocorrelation in the residuals. \verb+gam()/bam()+ are generally more versatile than \verb+gamm()+ and \verb+gamm4()+, though the latter two can be a bit faster when the model contains random smooths with many levels. The package {\tt itsadug} has been developed primarily with \verb+gam()/bam()+ in mind, and many of its functions do not work properly with models fitted using \verb+gamm()+ or \verb+gamm4()+. Here is a brief list of the main features of \verb+gam()+ that are relevant for us (don't worry if some of these features are unclear at this point; many of them will be discussed soon):
\begin{itemize}
  \item can fit all smooth types, including traditional random effects and random smooths
  \item can fit crossed random smooths (e.g.\ random smooths by words and by speakers at the same time)
  \item fully compatible with \verb+itsadug+
  \item can fit smooth interactions (\verb+te()+ and \verb+ti()+), where the main effects and interaction terms are separable
  \item possible to compare models using \verb+anova()+ / \verb+compareML()+
  \item possible to model simple autocorrelation in the data when using \verb+bam()+ (but not \verb+gam()+)
  \item can be run on multiple processor cores (\verb+bam()+)
\end{itemize}
\verb+gamm()+ is superficially similar to \verb+gam()/bam()+, but it relies on an external package ({\tt nlme}) for estimating models. In practical terms, this means that \verb+gamm()+ can be faster than \verb+gam()/bam()+ when the model includes randoms smooths with many levels. In addition, \verb+gamm()+ differs from \verb+gam()/bam()+ in the following ways:
\begin{itemize}
  \item cannot fit crossed random smooths
  \item only partly compatible with \verb+itsadug+
  \item not possible to compare models using \verb+anova()+
  \item can include complex models of autocorrelation
  \item can deal with heteroscedascity in the data by using variance components
\end{itemize}
\verb+gamm4()+ is in many ways similar to \verb+gamm()+: it mostly uses the same syntax as \verb+gam()/bam()+, but performs model fitting with the help of the {\tt lme4} package. Like \verb+gamm()+, \verb+gamm4()+ is good at fitting models with random smooths, and possibly even faster than \verb+gamm()+. It also comes with its specific set of pros and cons compared to \verb+gam()/bam()+ and \verb+gamm()+:
\begin{itemize}
  \item can fit crossed random smooths
  \item only partly compatible with \verb+itsadug+
  \item possible to compare models using \verb+anova()+
  \item cannot deal with autocorrelation or heteroscedascity
  \item can only fit smooth interactions where the main effects and interaction terms are inseparable (so their significance cannot be evaluated separately)
\end{itemize}

\end{document}